\documentclass[journal]{vgtc}                     

\onlineid{0}

\vgtccategory{Research}

\vgtcpapertype{system}

\title{Raiven: LLM-Based Visualization Authoring via \\
Domain-Specific Language Mediation}

\author{%
  \authororcid{Alexandra Irger*}{0009-0004-1021-8887},
  \authororcid{Ella Hugie*}{0009-0009-1561-9897},
  \authororcid{Minghao Guo}{0000-0003-3408-4997},
  \authororcid{Simon Warchol}{0000-0001-9067-6888}, \\
  \authororcid{Kenneth Moreland}{0000-0002-7051-3288},
  \authororcid{David Pugmire}{0000-0003-0647-2634}, 
  \authororcid{Wojciech Matusik}{0000-0003-0212-5643},
  and 
  \authororcid{Hanspeter Pfister}{0000-0002-3620-2582}
}

\authorfooter{
  \item Alexandra Irger and Ella Hugie are co-first authors.
  \item Alexandra Irger, Ella Hugie and Hanspeter Pfister are with Harvard John A. Paulson School of Engineering and Applied Sciences. Email: airger|ehugie|pfister@g.harvard.edu
  \item Minghao Guo and Wojciech Matusik are with MIT Computer Science \& Artificial Intelligence Laboratory.
  \item Simon Warchol is with the Laboratory of Systems Pharmacology, Harvard Medical School. Email: simonwarchol@g.harvard.edu
  \item Kenneth Moreland and David Pugmire are with ORNL. Email: morelandkd|pugmire@ornl.gov
}

\abstract{Visualization is central to scientific discovery, yet authoring tools remain split between information and scientific visualization, and expertise in one rarely transfers to the other. Large Language Model (LLM) based systems promise to bridge this gap through natural language, but current approaches generate code non-deterministically, with no guarantee of correctness and no protection against silent data fabrication. We present Raiven, a conversational system that mediates visualization authoring through a formally defined domain-specific language. RaivenDSL unifies scientific and information visualization in a single representation spanning 2D, 3D, and tabular data. The LLM produces a compact RaivenDSL specification under schema-guided constraints, and a deterministic compiler translates it to executable D3 or VTK.js code. Because the LLM operates only on dataset metadata, outputs are deterministic, specifications are verifiable before execution, and data fabrication is impossible by construction. In a 100-task benchmark, Raiven achieves 100\% compilation, is up to six times faster and six times cheaper than state-of-the-art LLMs, while improving interaction quality, correctness, and data faithfulness. An expert user study shows that Raiven significantly reduces debugging effort and makes it easier to produce correct visualizations.
}

\keywords{Visualization Authoring, Natural Language Interfaces, Domain-Specific Languages, Visualization Systems}

\teaser{
  \centering
  \includegraphics[width=\linewidth, alt={
   Raiven overview. The user describes a visualization task in natural language; the LLM translates the request into a schema-constrained RaivenDSL specification, which the compiler deterministically transforms into an interactive visualization with a tailored control panel.
  }]{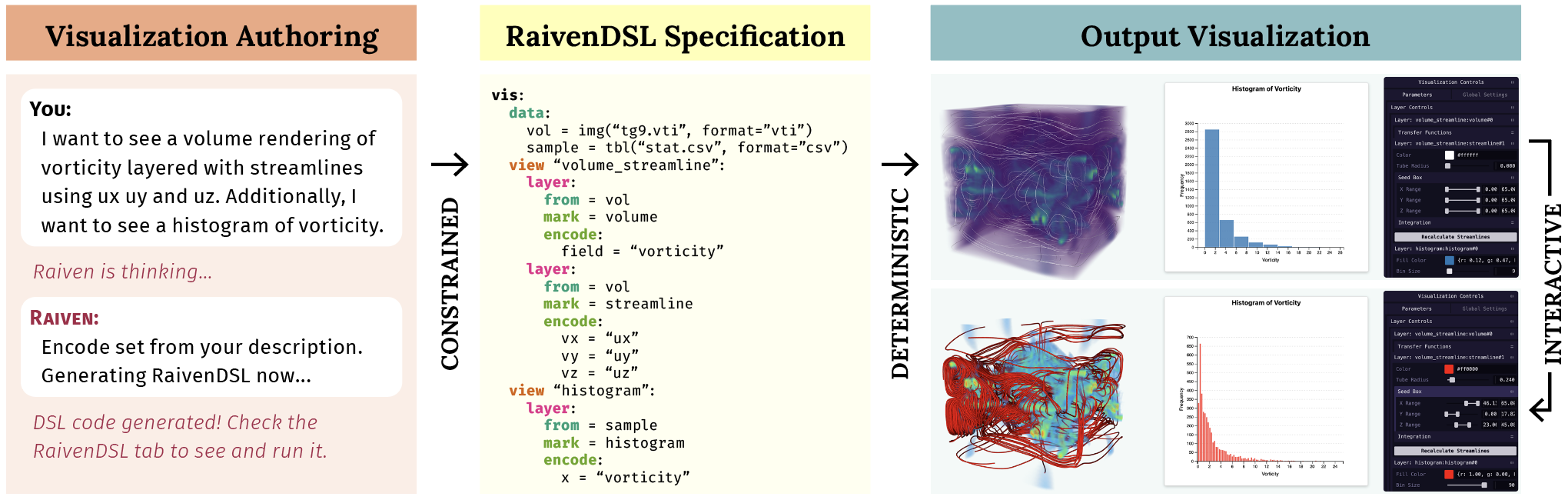}
  \caption{%
  \textbf{Raiven Overview.} The user describes a visualization task in natural language; the LLM translates the request into a schema-constrained RaivenDSL specification, which the compiler deterministically transforms into an interactive visualization and control panel.
  }
  \label{fig:teaser}
}

\graphicspath{{figs/}{figures/}{pictures/}{images/}{./}{figs/template}}

\usepackage{tabu}
\usepackage{booktabs}
\usepackage{amsmath}
\usepackage{multirow}
\usepackage{makecell}
\usepackage[table]{xcolor}
\usepackage{wasysym}
\usepackage{fancyvrb}
\usepackage{float}
\usepackage{pgfmath}
\usepackage{dblfloatfix}
\usepackage{subcaption}
\usepackage{soul}
\usepackage{xcolor}
\usepackage{tabularx}
\usepackage{listings}
\usepackage{xcolor}
\usepackage{tcolorbox}
\tcbuselibrary{breakable}
\usepackage{placeins}
\usepackage{mathptmx}

\newcommand{\para}[1]{\vspace{0em}\noindent\textbf{#1}}

\begin{document}

\firstsection{Introduction}

\maketitle

Despite decades of research and an ever-growing ecosystem of visualization tools, authoring visualizations remains surprisingly hard. This difficulty stems in part from fragmentation, both across domains and across tools. Scientific and information visualization have evolved as entirely separate disciplines, with distinct languages and conventions, meaning knowledge gained in one rarely carries over to the other. Even within a single domain, the landscape is crowded with libraries, frameworks, and applications each targeting a narrow use case. This challenge is compounded by the well-known tradeoff between expressiveness and ease of use, which forces users to choose between flexible but complex low-level libraries and simple but rigid turnkey applications~\cite{rautek2014vislang}.
Together, these challenges leave users with no single system that is both expressive enough to span domains and accessible enough for everyday use.\looseness=-1

Natural language interfaces promise to bridge this gap, but current Large Language Model (LLM) based approaches are opaque, costly, non-deterministic, and fragile. The LLM generates visualization code directly, with no guarantee of correctness and no enforcement of visualization best practices. Control over the visualization process lies entirely in large technology companies and their training data. Recent evaluations confirm these limitations: LLM-based generation struggles beyond simple chart types~\cite{wu2024automated, ribalta2026evaluating} and natural language interfaces remain incapable of supporting research-level visualization tasks~\cite{joseph2025astrovisbench}. These problems intensify for projects that span both scientific and information visualization, as no current LLM-based frameworks operate across both domains. Worse, because the model often operates directly on raw data values, LLMs can silently fabricate or substitute data, producing plausible-looking visualizations built on invented values.\looseness=-1

We argue that domain-specific languages (DSLs) are a solution to these problems.
DSLs occupy a middle ground between low-level libraries and simple turnkey applications~\cite{rautek2014vislang}.
A DSL defines a constrained vocabulary with formal semantics, enabling parsing, validation,
and compilation before rendering. A DSL inserted between the LLM and the visualization backend separates language interpretation from visualization execution. This separation provides \emph{verifiability}, because a DSL specification can be parsed and validated before compilation, and \emph{flexibility}, because a single specification can target multiple rendering backends, much as a shader language compiles to different GPU APIs. The visualization community itself has recognized the need for cross-domain tools: VAST, InfoVis, and SciVis merged into a single conference in 2021, and multifaceted scientific data increasingly demands both modalities, yet no existing visualization authoring system bridges this divide. Scientific visualization languages do not handle information visualization mark types, and information visualization grammars do not handle volumetric or flow data.\looseness=-1

In this paper, we present \emph{Raiven}, a system that mediates natural language visualization authoring through a pipeline that separates language interpretation from visualization execution. The user describes a visualization task in natural language, the LLM translates that request into a compact \emph{RaivenDSL} specification under schema-guided constraints, and a deterministic compiler translates the specification into executable D3 or VTK.js code. Because the LLM operates only on dataset metadata and never accesses raw data values, the architecture preserves data privacy by design, remains independent of context-window limitations that constrain direct code-generation approaches, and makes silent data fabrication impossible. In our evaluation, Raiven achieves 100\% compilation, runs up to six times faster, and costs up to six times less than direct code generation with state-of-the-art LLMs, while improving visualization correctness and data faithfulness.\looseness=-1

Our contributions are as follows. We present \emph{Raiven}, a conversational visualization authoring system with a type-checked LLM pipeline and a deterministic compiler, connected by \emph{RaivenDSL}, a backend-agnostic DSL spanning scientific and information visualization that provides a compact, verifiable specification structure designed for safe LLM-based visualization generation. We introduce \emph{VMPC} (Visualization Multi-view Prompt Compliance), an evaluation metric for multi-view visualization correctness that scores execution, data faithfulness, cross-view linking, and per-view compliance. We demonstrate Raiven's effectiveness through a novel \emph{benchmark} of 100 visualization tasks spanning scientific, information, and combined workflows, on which Raiven outperforms state-of-the-art large language models across all categories. All code and benchmarks will be released  upon acceptance.\looseness=-1
\begin{figure*}[!t]
    \centering
    \includegraphics[width=\linewidth]{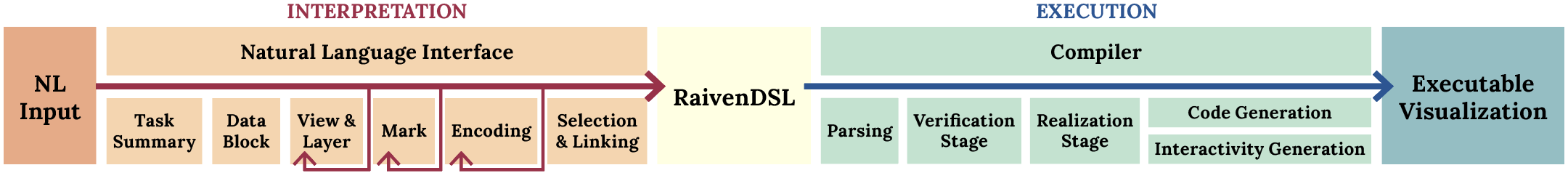}
    \caption{\textbf{Raiven pipeline.} Left: in the \textit{interpretation} phase,
    schema-mediated generation translates natural language into RaivenDSL
    through a sequence of validated stages. Right: in the \textit{execution} phase,
    the compiler parses, validates, resolves backend-specific defaults, and
    generates executable code and interactive controls. The user can
    interact with the system at three points: natural language input
    (Describe), direct DSL editing (Edit), and interactive controls on the
    rendered visualization (Explore).}
    \vspace{-15pt}
    \label{fig:pipeline}
\end{figure*}

\section{Related Work}

Prior work relevant to 
\emph{Raiven} spans three areas: direct natural language (NL) visualization authoring, domain-specific languages (DSL) for visualization, and systems that mediate natural language through an intermediate structured representation. Table~\ref{tab:related_work} summarizes these systems and their coverage.\looseness=-1

\subsection{Direct Natural Language Authoring}
An emerging line of research treats visualization authoring as a direct natural language-to-code generation problem.
In information visualization, Chat2VIS~\cite{maddigan2023chat2vis} generates visualization code directly from natural language, and LIDA~\cite{dibia2023lida} wraps code generation in a multi-stage pipeline that summarizes datasets before producing code. FlowSense~\cite{yu2019flowsense} takes a different approach, mapping utterances to dataflow graph operations.
In scientific visualization, ChatVis~\cite{mallick2024chatvis} generates ParaView Python scripts and iteratively repairs them, and Omega~\cite{royer2024omega} generates Python code for bioimage analysis tasks within napari. ParaView-MCP~\cite{liu2025paraview} and napari-mcp\footnote{\url{https://github.com/royerlab/napari-mcp}}, Omega's successor, replace free-form generation with structured tool invocation over their respective APIs.
All of these systems remain tied to specific libraries or execution environments, and direct code generation inherits the fragility of code synthesis, including syntax errors, execution failures, and malformed outputs.\looseness=-1

\begin{table}[ht]
    \centering
    \caption{Existing NL and DSL based visualization authoring systems. $S$ covers Scientific Visualization, $I$ covers Information Visualization.}
    \begin{tabular}{l | c | c c c c}
        System & Year & NL & DSL & $S$ & $I$ \\
        \hline \hline
        FlowSense \cite{yu2019flowsense} & 2020 & \CIRCLE & & & \CIRCLE \\
        Chat2Vis \cite{maddigan2023chat2vis} & 2023 & \CIRCLE & & & \CIRCLE \\
        LIDA \cite{dibia2023lida} & 2023 & \CIRCLE & & & \CIRCLE \\
        ChatVis \cite{mallick2024chatvis} & 2024 & \CIRCLE & & \CIRCLE & \\
        Omega \cite{royer2024omega} & 2024 & \CIRCLE & & \CIRCLE & \\
        ParaView-MCP \cite{liu2025paraview} & 2025 & \CIRCLE & & \CIRCLE & \\
        \hline
        Diderot \cite{chiw2012diderot} & 2012 & & Diderot & \CIRCLE & \\
        Vivaldi \cite{choi2014vivaldi} & 2014 & & Vivaldi & \CIRCLE & \\
        ViSlang \cite{rautek2014vislang} & 2014 & & \CIRCLE & \CIRCLE & \\
        Vega-Lite \cite{satyanarayan2016vega} & 2017 & & Vega-Lite & & \CIRCLE \\
        Shih et al. \cite{shih2018declarative} & 2019 & & \CIRCLE & \CIRCLE & \\
        DXR \cite{sicat2018dxr} & 2019 & & \CIRCLE & & \CIRCLE \\
        Scholz \cite{scholz2021modular} & 2021 & & \CIRCLE & \CIRCLE & \CIRCLE \\
        Harth et al. \cite{harth2023rapid} & 2023 & & \CIRCLE & & \CIRCLE \\
        GoFish \cite{pollock2025gofish} & 2026 & & GoFish & & \CIRCLE \\
        Gosling \cite{kouvril2025design} & 2026 & & Gosling & & \CIRCLE \\
        \hline
        NL4DV \cite{narechania2020nl4dv} & 2021 & \CIRCLE & Vega-Lite & & \CIRCLE \\
        NMT2Vis \cite{luo2021natural} & 2022 & \CIRCLE & VegaZero & & \CIRCLE \\
        NL2Viz \cite{wu2022nl2viz} & 2022 & \CIRCLE & \CIRCLE & & \CIRCLE \\
        FlowNL \cite{huang2022flownl} & 2023 & \CIRCLE & \CIRCLE & \CIRCLE & \\
        ChatModelling \cite{jia2024chat} & 2024 & \CIRCLE & \CIRCLE & \CIRCLE & \\
        YAC \cite{lange2025yac} & 2025 & \CIRCLE & \CIRCLE & & \CIRCLE \\
        ChartGPT \cite{tian2024chartgpt} & 2025 & \CIRCLE & \CIRCLE & & \CIRCLE \\
        NLI4VolVis \cite{ainli4volvis} & 2026 & \CIRCLE & \CIRCLE & \CIRCLE & \\
        VegaChat \cite{hostnik2026vegachat} & 2026 & \CIRCLE & Vega-Lite & & \CIRCLE \\
        \hline
        \textbf{Raiven} & 2026 & \CIRCLE & \textbf{RaivenDSL} & \CIRCLE & \CIRCLE
    \end{tabular}
    \vspace{-12pt}
    \label{tab:related_work}
\end{table}

\subsection{DSL Authoring}

We distinguish \emph{domain-specific languages} from \emph{visualization libraries} such as D3~\cite{bostock2011d3}, VTK~\cite{vtkBook}, Matplotlib~\cite{hunter2007matplotlib},
or ObservablePlot\footnote{\url{https://observablehq.com/plot/}}.
A library provides programmatic access to rendering primitives but imposes no structural constraints on how visualizations are specified.
A DSL, by contrast, defines a fixed vocabulary with formal semantics that can be parsed and validated prior to rendering.

The landscape of DSLs for scientific visualization is sparse. ViSlang~\cite{rautek2014vislang} composes procedural, declarative, and functional sub-languages for GPU-based scientific visualization. Shih et al.~\cite{shih2018declarative} introduce a declarative JSON grammar for volume visualization pipelines. Diderot~\cite{chiw2012diderot} is a compiled language for continuous field visualization, and Vivaldi~\cite{choi2014vivaldi} provides a procedural DSL for distributed volume processing. None of these target information visualization mark types.

DSL development is much more mature on the information visualization side. Vega-Lite~\cite{satyanarayan2016vega}, building on the Grammar of Graphics~\cite{wilkinson1999grammar}, is the dominant declarative grammar. GoFish~\cite{pollock2025gofish} formalizes Gestalt grouping as composable operators, Gosling~\cite{kouvril2025design} adapts the declarative model to genomic data, DXR~\cite{sicat2018dxr} targets immersive environments, and Harth et al.~\cite{harth2023rapid} coordinate heterogeneous web views through an interaction grammar. None of these support volumetric, flow, or spatially continuous scientific data. Scholz~\cite{scholz2021modular} spans both domains with a JSON-based DSL, but renders 3D content through Three.js, a general-purpose graphics library rather than a visualization backend. Draco\cite{moritz2018formalizing} encodes design knowledge as constraints over Vega-Lite and uses answer-set programming to rank valid designs; Dziban\cite{lin2020dziban} extends this framework with anchored recommendations to support incremental and iterative visualization design. Both, however, operate as recommendation engines rather than generation targets.

Across both communities, visualization DSLs have been designed for direct human authoring or programmatic construction, not as generation targets for language models.

\subsection{Natural Language Authoring via DSLs}
The line of research most similar to our approach combines natural language authoring with a DSL or other intermediate structured representation.
On the information visualization side, NL4DV~\cite{narechania2020nl4dv} (NLP-based) and VegaChat~\cite{hostnik2026vegachat} map natural language to Vega-Lite specifications. Vega-Zero~\cite{luo2021natural} introduces a seq2seq-friendly linearization of Vega-Lite, and ChartGPT~\cite{tian2024chartgpt} uses a thin intermediate chart representation that resolves to an established grammar. NL2Viz~\cite{wu2022nl2viz} and YAC~\cite{lange2025yac} introduce purpose-built intermediate languages, but both remain narrowly scoped to specific chart types or biomedical data exploration.

On the scientific visualization side, FlowNL~\cite{huang2022flownl} translates natural language into a declarative intermediate language for flow visualization. Chat Modeling~\cite{jia2024chat} mediates natural language through a JSON-based intermediate format for procedural modeling of biological structures. NLI4VolVis~\cite{ainli4volvis} uses function-calling agents to invoke volume visualization commands iteratively. All three are tightly scoped to particular tasks and domains.

No existing system combines natural-language authoring, a purpose-built backend-agnostic intermediate language, and support for both scientific and information visualization within a single typed language.

\section{System Overview}

Raiven mediates natural language visualization authoring through a structured pipeline that separates language interpretation from visualization execution (Figure~\ref{fig:pipeline}). The pipeline has two phases: \textit{interpretation}, which translates a natural language request into RaivenDSL, and \textit{execution}, which compiles that specification into an interactive browser-based visualization. 

In the interpretation phase (Section~\ref{sec:nli}), a user submits a visualization request through the chat interface. Rather than generating backend code directly, the LLM incrementally constructs a session schema, a structured representation that persists across conversational turns and records the evolving specification. The schema captures the task summary, typed data sources, and view structure, including layers, marks, encodings, styles, selections, and linked interactions. It is populated through a sequence of narrowly scoped prompts, each addressing a specific subproblem and followed by validation. When ambiguity remains, the system asks for clarification before proceeding. Once the schema is complete, a final prompt translates it into RaivenDSL.

The resulting RaivenDSL (Section~\ref{sec:dsl}) is returned directly to the user, who can inspect and edit the specification before compilation, render it immediately, or continue refining it through natural language. This transparency is deliberate: the DSL serves as a legible and verifiable mediator between user intent and compiler input.

In the execution phase (Section~\ref{sec:compiler}), the Raiven compiler parses the specification, resolves types, and assigns each view to an appropriate rendering backend. Raiven currently targets two rendering backends---D3 for 
information visualization and VTK.js for scientific visualization---both executed in the browser. This pairing was chosen 
deliberately: both are state-of-the-art libraries in their 
respective fields, together spanning both visualization 
domains. The compiler then assembles each view deterministically, resolving axes, color palettes, camera parameters, interaction defaults, and generated controls. Because this translation occurs per view within a single program, one RaivenDSL specification can produce coordinated multi-view visualizations that span both scientific and information visualization.

\section{RaivenDSL}
\label{sec:dsl}

\definecolor{data}{HTML}{1b9e77}
\definecolor{view}{HTML}{d95f02}
\definecolor{link}{HTML}{7570b3}
\definecolor{layer}{HTML}{e7298a}
\definecolor{mark}{HTML}{66a61e}

RaivenDSL specifies visualizations at the level of intent rather than execution. The user describes \emph{what} they want to see without concern for which backend renders it. 
The language restricts itself to mark types with established semantics across visualization practice, trading fine-grained control for lower specification burden, stronger validation, and backend portability. Although designed as a generation target for the natural language interface, RaivenDSL is also a human-authored language: users may write or edit specifications directly and compile them without going through the natural language pipeline.

We introduce the language through a concrete example before describing its design principles.

\subsection{A Running Example}
\label{sec:running-example}

Consider a neuroscientist exploring a CT scan of a human head. She wants a 3D volume rendering containing a 2D axial slice. In RaivenDSL, this visualization is a short program:

\begin{Verbatim}[commandchars=\\\{\},fontsize=\scriptsize]
vis:
  \textcolor{data}{data}:
    vol = img("head.vti", format="vti")
  \textcolor{view}{view} "volume\_slice":
    \textcolor{layer}{layer}:
      \textcolor{data}{from} = vol
      \textcolor{mark}{mark} = volume
    \textcolor{layer}{layer}:
      \textcolor{data}{from} = vol
      \textcolor{mark}{mark} = slice
      \textcolor{mark}{style}:
        axes = ["XY"]
\end{Verbatim}

\newcommand{\cul}[2]{{\setulcolor{#1}\setul{1pt}{2pt}\ul{\textbf{#2}}}}

Even this minimal program illustrates several design decisions. The \cul{data}{data} source carries a typed constructor: \texttt{img()} tells the compiler this is volumetric image data, not a table or network. The program declares a single \cul{view}{view}, a named container that holds one complete chart or 3D scene, containing two \cul{layer}{layers}, a volume rendering and a slice.
Each layer adds a distinct visual representation to the view. Here, the first layer renders the volume and the second adds a slice plane, both drawn in the same 3D scene because they share the same view. Each layer specifies a \cul{mark}{mark}, the primitive visual type to render (e.g., \texttt{volume}, \texttt{slice}, \texttt{scatter}, \texttt{bar}), and optionally an \cul{mark}{encoding} that binds data variables to the mark's visual channels (e.g., mapping columns to axes and colors).
The user specifies what to show; the compiler handles everything else, from axis construction to default color mappings.

\begin{figure}[H]
\vspace{-8pt}
    \centering    
    \includegraphics[width=0.75\linewidth]{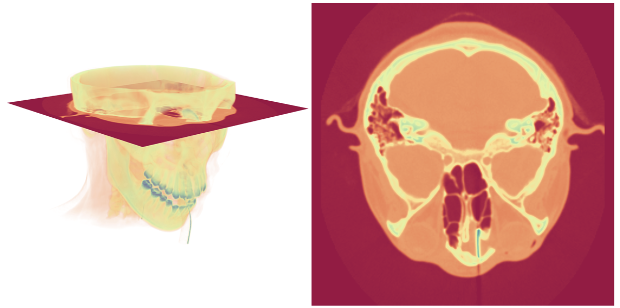}
    \caption{Output of the running example: a volume rendering and synchronized slice view of a CT head scan.}
    \vspace{-8pt}
\label{fig:dsl}
\end{figure}

Now suppose the neuroscientist wants a second linked view showing only the slice. The slice position gets synchronized with the first view and a shared color transfer function ensures that adjusting the mapping in one view updates the other (Figure~\ref{fig:dsl}).

\begin{Verbatim}[commandchars=\\\{\},fontsize=\scriptsize]
  \textcolor{view}{view} "slice\_xy":
    \textcolor{link}{link}(slice="xy\_link", axes=["XY"],
         views=["volume\_slice", "slice\_xy"])
    \textcolor{link}{link}(tf="head.vti\_shared",
         views=["volume\_slice", "slice\_xy"])
    \textcolor{layer}{layer}:
      \textcolor{data}{from} = vol
      \textcolor{mark}{mark} = slice
      \textcolor{mark}{style}:
        axes = ["XY"]
\end{Verbatim}

The two \cul{link}{link} declarations introduce cross-view coordination. The first synchronizes slice position: dragging the slice plane in one view moves it in the other. The second shares a color mapping between the slice layers so that adjustments propagate across views. The DSL expresses the \emph{intent} of the coordination, and the compiler generates the implementation, giving our neuroscientist a fully coordinated multi-view CT visualization in twenty-two lines of declarative code. The full RaivenDSL and corresponding visualization follow:

\begin{Verbatim}[commandchars=\\\{\},fontsize=\scriptsize]
vis:
  \textcolor{data}{data}:
    vol = img("head.vti", format="vti")
  \textcolor{view}{view} "volume\_slice":
    \textcolor{link}{link}(slice="xy\_link", axes=["XY"], 
         views=["volume\_slice", "slice\_xy"])
    \textcolor{link}{link}(tf="head.vti\_shared", 
         views=["volume\_slice", "slice\_xy"])
    \textcolor{layer}{layer}: ...                    \textit{\textcolor{gray}{// layers as in example 1}}
  \textcolor{view}{view} "slice\_xy": ...                 \textit{\textcolor{gray}{// view in example 2}}
\end{Verbatim}

\vspace{-4pt}

\subsection{Design Principles}
\label{sec:design-principles}

Four principles guided the design of RaivenDSL: typed data, separation of semantics from style, declarative coordination, and early rejection of invalid programs. 

\para{Typed data for backend routing.} A central challenge in unifying scientific and information visualization is that the two domains rely on fundamentally different rendering architectures. Volume data compiles to VTK.js; tabular data compiles to D3. Rather than forcing the user to choose a backend, we encode the distinction in the data declaration itself. Each data source carries a typed constructor that identifies its modality (image, table, network, or geospatial).
The compiler uses this data type to constrain which marks are valid using this data. Together with the mark type, it determines the appropriate rendering backend for each view and validates that the specified encodings are compatible with the data before generating code.
The running example illustrates this: the \texttt{img()} constructor restricts layers to marks such as \texttt{volume} and \texttt{slice}, which the compiler routes to VTK.js, while a \texttt{tbl()} constructor restricts layers to marks that compile to D3.

\para{Separating semantics from style.} A key design decision is the separation of encoding from styling. Encodings define the semantic content of the visualization: which data variables map to which visual channels. Style properties (color schemes, opacity, transfer function parameters) are optional and resolved by the compiler when omitted. This separation matters for two reasons. First, it reduces specification burden: the user communicates what the visualization means, and the compiler resolves how it looks. Second, it makes the DSL a better generation target for the LLM, because the model only needs to produce the semantically meaningful part of the specification. Stylistic defaults that would be unreliable and unpredictable if inferred by an LLM become deterministic when encoded in the compiler.

\para{Declarative coordination.} Cross-view coordination is one of the most complex aspects of multi-view visualization, typically requiring manual event wiring, shared state management, and backend-specific callback code. We chose to make coordination a first-class language construct: the user declares \emph{what} should be shared, and the compiler generates the synchronization logic. RaivenDSL currently supports five coordination types: brush-and-filter selections, point selection with cross-view highlighting, shared color mappings, synchronized color transfer functions, and linked slice indices. We deliberately made coordination explicit rather than inferred, because linked behaviors vary too widely across visualization tasks for any default to be reliable. A brush that filters in one context should highlight in another; a shared slice index is appropriate for medical imaging but not for a statistical dashboard. These decisions belong to the user, not the compiler. Currently, declared links only operate within a single backend (i.e., linking D3 views with D3 views, or VTK.js views with VTK.js views); cross-backend linking is a planned extension.

\para{Structural validation at parse time.} Because RaivenDSL is designed as a generation target for an LLM, catching errors early is essential. The grammar encodes structural constraints directly: a layer must declare a data source and a mark, selections must be defined before they can be linked, and backend-incompatible layers cannot share a view. Invalid specifications are rejected at parse time rather than discovered at runtime. This is a deliberate tradeoff against flexibility: the language does not permit arbitrary code or unconstrained composition. The benefit is that every syntactically valid RaivenDSL program is guaranteed to compile, which eliminates an entire class of errors that plague direct code generation.

The language supports two types of syntax: a brace-based syntax aligned with existing visualization grammars \cite{mcnutt2022no} and an indentation-based syntax familiar to Python users. Both parse to the same abstract syntax tree. The full grammar is provided in Appendix~\ref{app:grammar}.

\subsection{Expressiveness}
\label{sec:expressiveness}

RaivenDSL targets mark types that are well-established and broadly applicable across visualization practice, spanning both scientific visualization (volume rendering, isosurfaces, slices, streamlines, line integral convolution) and information visualization (statistical, distributional, relational, network, compositional, and geospatial marks). A complete list of supported marks with their corresponding encodings, style parameters, and generated controls is provided in Appendix~\ref{app:marks}.

The key observation is that both domains share the same specification structure: a data source, a mark type, mark-dependent channel-to-field encodings, and optional styling. A volume rendering and a scatter plot differ in their backend realization, but at the specification level they are both a mark applied to data, typically through an encoding. RaivenDSL capitalizes on this structural similarity to provide a single language that spans both domains.

\begin{table}[t]
\centering
\caption{Design-space coverage of RaivenDSL against the Tory \& M\"{o}ller~\cite{tory2004rethinking} taxonomy. All types and techniques are from the original classification. \CIRCLE~= supported,
\Circle~= not supported. Unsupported techniques in
\textcolor{lightgray}{gray}.}
\label{tab:tory-moller}
\begin{tabular}{llcl}
\toprule
Dim & Type & DSL & Techniques \\
\midrule
\multicolumn{4}{l}{\textbf{Continuous (given spatial layout)}} \\
\midrule
1D & Scalar & \CIRCLE & Line graph \\
2D & Scalar & \CIRCLE & Colour map, isolines \\
\multirow{2}{*}{3D} & Scalar & \CIRCLE & Volume rendering, isosurfaces \\
 & Tensor & \Circle & \textcolor{lightgray}{Tensor ellipsoids} \\
2D--3D & Vector & \CIRCLE & LIC, Particle traces, \textcolor{lightgray}{glyphs} \\
1D--3D & Multivar. & \CIRCLE & Combine scalar, vector \& \textcolor{lightgray}{tensor} \\
nD & --- & \CIRCLE & Multiple 1D, 2D, or 3D views \\
\midrule
\multicolumn{4}{l}{\textbf{Discrete (chosen / given layout)}} \\
\midrule
2D & Values & \CIRCLE & Scatter plot, bar chart \\
2D & Graphs & \CIRCLE & Node-link diagrams (2D) \\
3D & Values & \Circle & \textcolor{lightgray}{3D scatter plot, 3D bar chart} \\
3D & Graphs & \Circle & \textcolor{lightgray}{Node-link (3D)} \\
\multirow{2}{*}{nD} & \multirow{2}{*}{Values} & \multirow{2}{*}{\CIRCLE} & Charts + colour, multiple views, \\
 & & & \textcolor{lightgray}{glyphs}, parallel coordinates \\
\multirow{2}{*}{nD} & \multirow{2}{*}{Trees} & \multirow{2}{*}{\Circle} & \textcolor{lightgray}{Hierarchical graphs,} \\
 & & & \textcolor{lightgray}{Space-filling Mosaics} \\
\bottomrule
\end{tabular}
\vspace{-12pt}
\end{table}

\captionsetup[subfigure]{font={footnotesize,sf},position=top,justification=raggedright,singlelinecheck=false}

\begin{figure*}[t]
  \centering
  \begin{subfigure}{.32\linewidth}
    \begin{tikzpicture}
      \node[anchor=south west,inner sep=0] (img) {%
        \includegraphics[width=\linewidth,trim=0 0 310 0,clip]{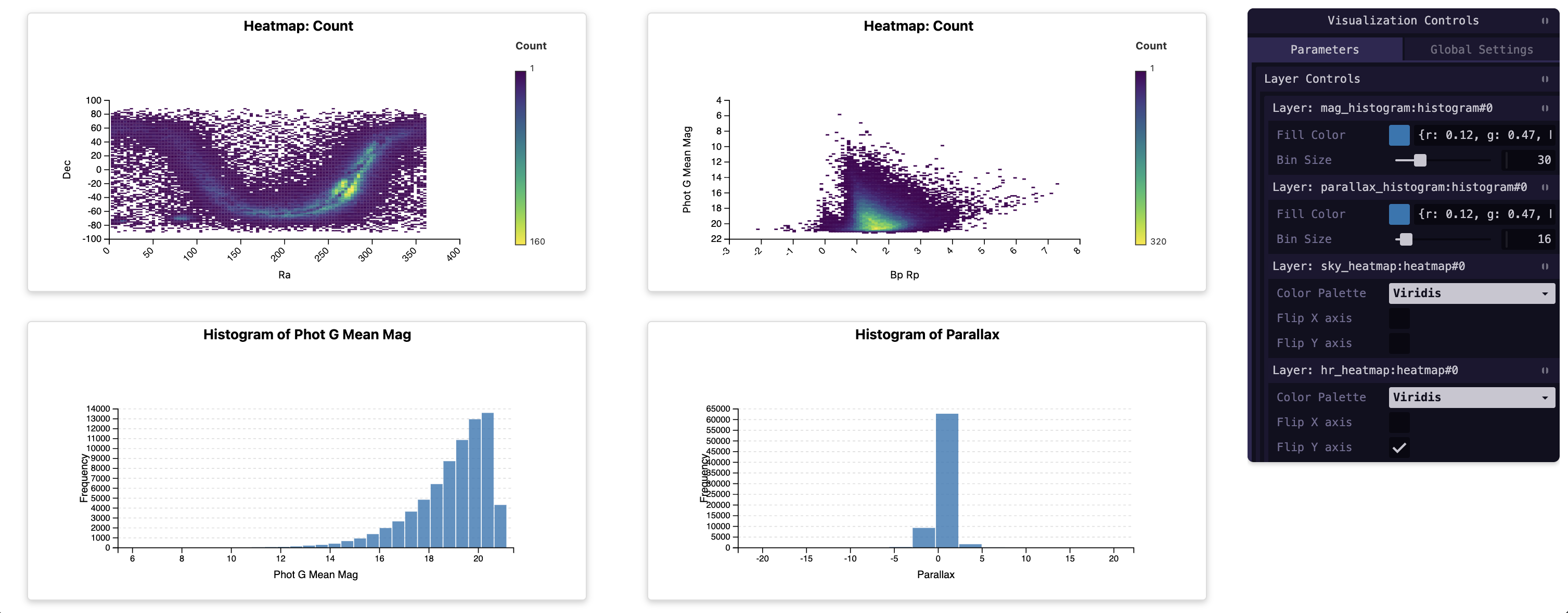}};
      \node[anchor=north west,font=\footnotesize\sffamily\bfseries,
            fill=white,opacity=0.85,text opacity=1,inner sep=2pt]
        at (img.north west) {(a)};
    \end{tikzpicture}
    \label{fig:mosaic}
  \end{subfigure}
  \hfill
  \begin{subfigure}{.32\linewidth}
    \begin{tikzpicture}
      \node[anchor=south west,inner sep=0] (img) {%
        \includegraphics[width=\linewidth,trim=0 0 310 0,clip]{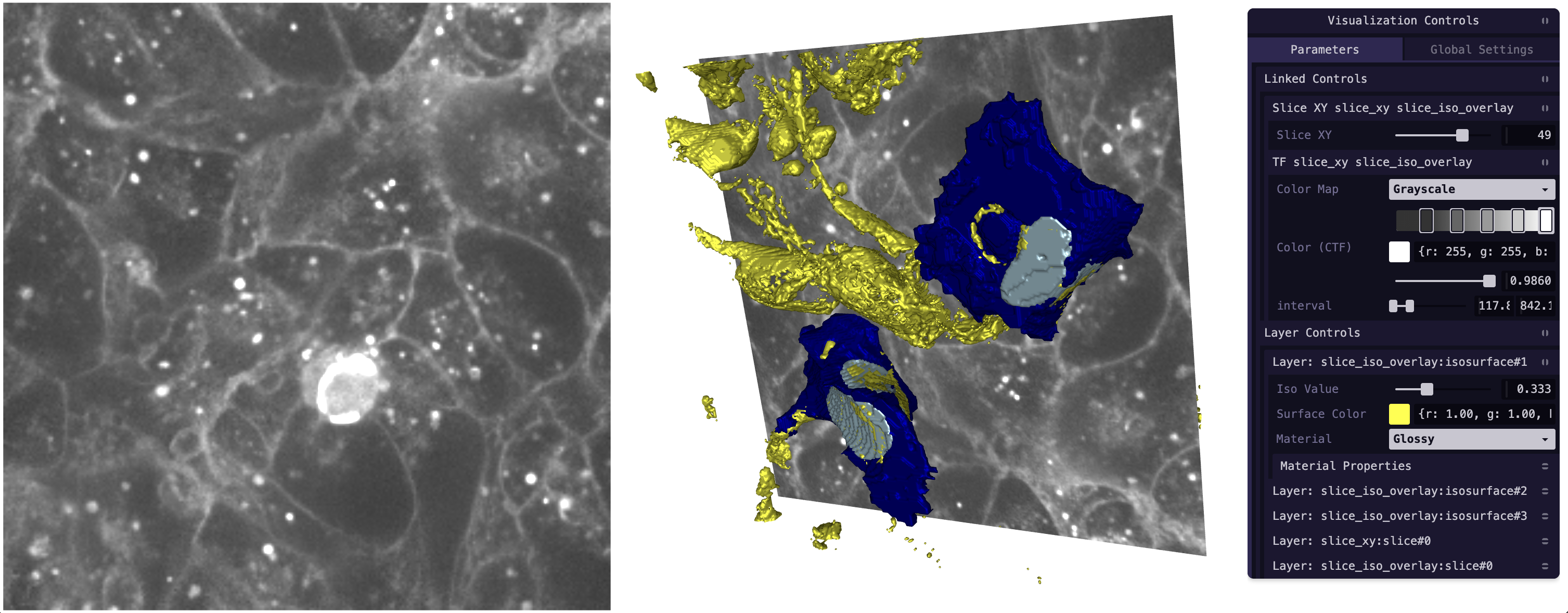}};
      \node[anchor=north west,font=\footnotesize\sffamily\bfseries,
            fill=white,opacity=0.85,text opacity=1,inner sep=2pt]
        at (img.north west) {(b)};
    \end{tikzpicture}
    \label{fig:neuroglancer}
  \end{subfigure}
  \hfill
  \begin{subfigure}{.32\linewidth}
    \begin{tikzpicture}
      \node[anchor=south west,inner sep=0] (img) {%
        \includegraphics[width=\linewidth,trim=0 0 310 0,clip]{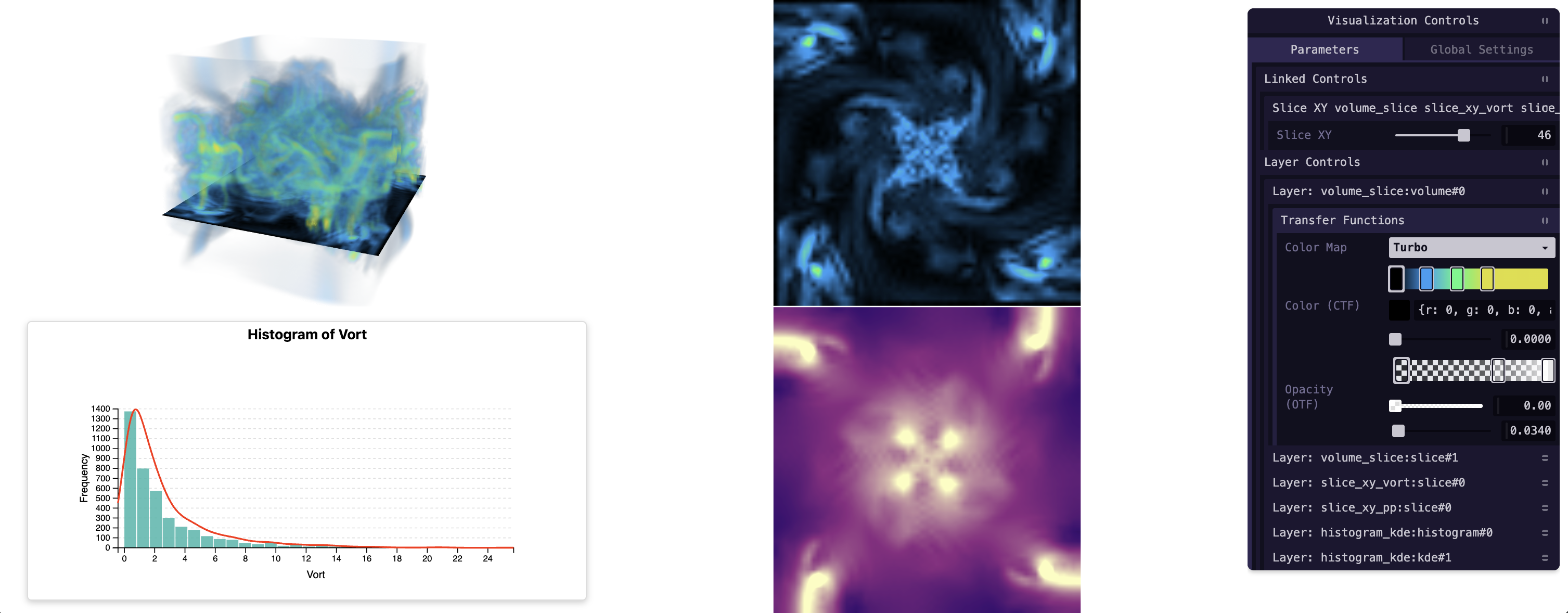}};
      \node[anchor=north west,font=\footnotesize\sffamily\bfseries,
            fill=white,opacity=0.85,text opacity=1,inner sep=2pt]
        at (img.north west) {(c)};
    \end{tikzpicture}
    \label{fig:polyphorm}
  \end{subfigure}
  \caption{Visualization types from published systems, recreated in
    RaivenDSL.
    (a)~Mosaic~\cite{heer2023mosaic}: linked heatmaps and histograms
    over a subset of the Gaia catalog~\cite{vallenari2023gaia}.
    (b)~Neuroglancer~\cite{maitin2021google} view recreated
    from~\cite{beyer2022survey}: 2D slice with 3D segmentations
    (Allen Cell Data~\cite{allencell2018}).
    (c)~Polyphorm~\cite{elek2020polyphorm} view recreated
    from~\cite{lan2021visualization}: volume rendering with linked
    slices and charts.}
    \vspace{-12pt}
  \label{fig:use_cases}
\end{figure*}

We evaluate RaivenDSL's coverage against the Tory and M\"{o}ller taxonomy~\cite{tory2004rethinking}, which classifies visualization techniques by whether their spatial layout is continuous or discrete and whether it is given by the data or chosen by the designer. This taxonomy is uniquely suited to our evaluation because it is the only high-level classification spanning both scientific and information visualization. Table~\ref{tab:tory-moller} summarizes the results. RaivenDSL covers 10 of 14 categories, including all scalar, vector, and multivariate continuous settings and all 2D discrete settings.

As a practical demonstration of this coverage, Figure~\ref{fig:use_cases} shows visualizations authored in RaivenDSL: recreations of key views from Mosaic~\cite{heer2023mosaic}, Neuroglancer~\cite{maitin2021google}, and Polyphorm~\cite{elek2020polyphorm}. We recreate the visualization types these systems produce, not their full functionality. The four unsupported categories in Table~\ref{tab:tory-moller} reflect deliberate scoping decisions. Tensors and hierarchical trees are left to future work. The discrete 3D categories (3D scatter plots, 3D node-link diagrams) are excluded because occlusion and depth ambiguity make them unreliable without careful view-dependent rendering, which conflicts with the language's goal of deterministic, intent-level specification.

\section{Natural Language Interface}
\label{sec:nli}

Rather than generating visualization code directly from a single 
prompt, Raiven first populates a structured intermediate schema 
that captures the user's intent, then compiles that schema into 
RaivenDSL only after validating it. The system is implemented as 
a FastAPI server that manages session state, orchestrates the 
multi-prompt pipeline, and communicates with the OpenAI API 
(ChatGPT~5.2) for all generation steps, exposing a browser-based 
chat interface to the user. We illustrate with a brief example 
before describing the mechanism.

\subsection{Revisiting the Running Example}
\label{sec:nl-example}
Returning to the running example from Sec~\ref{sec:running-example}: the neuroscientist uploads a VTI file of a CT head scan and types: ``Show me a volume rendering of the head data with a linked axial slice and a shared color transfer function.'' Raiven does not generate code directly from this sentence. Instead, the system proceeds through a sequence of targeted nodes, each populating one component of the session schema and validating it before the next step begins:

\begin{enumerate}

\item \textbf{Task summary.} The LLM extracts a high-level goal: \emph{volume rendering with linked axial slice of a 3D scalar dataset and shared transfer function.}

\item \textbf{Data block.} Raiven has already parsed the uploaded file. The schema records the source name, file path, format (\texttt{vti}), and variable descriptors (scalar field names, spatial dimensions), of which Raiven derives from the file header. The LLM never sees the raw voxel values.

\item \textbf{View and layer.} The LLM proposes two layers in a shared view: \texttt{volume} and \texttt{slice}. Raiven checks that at least one view with one layer exists, that every layer is assigned to a dataset, and that all layers in a view share a compatible backend.

\item \textbf{Mark type.} For each layer the LLM selects a primitive mark type: \texttt{volume} for the three-dimensional rendering and \texttt{slice} for the axial cut. Raiven normalizes the mark name and checks that it appears in the backend capability table.

\item \textbf{Encoding and Style.} The LLM updates the volumes field to the \texttt{vti}'s scalar variable 
and sets the slice axis to \texttt{XY}. Raiven checks that the mark and data types are compatible, that every referenced variable exists, and that all required channels are filled.

\item \textbf{Selection and Linking.} The LLM proposes a shared transfer function between the two views. Raiven records the link but enforces no additional validation.
\end{enumerate}

At each step, Raiven validates the schema before the pipeline advances. If the user had omitted the slice orientation, the system would have asked for it at the "Encoding and Style" stage rather than guessing. Only after all fields are populated and checked does a final prompt translate the schema into the RaivenDSL program shown in Section~\ref{sec:running-example}. The result is the same specification the user could have written by hand, but arrived at through conversation.

\subsection{Session Schema}
\label{sec:schema}

The example above illustrates the role of the session schema: a structured intermediate object that persists across conversational turns and tracks the accumulated specification. The schema records a task summary, a data block keyed by source name (including each source's type, path, and variable descriptors), and a views block that enumerates views, layers, marks, encodings, and styles. It tracks selections and linking definitions for interactive multi-view specifications.

A core challenge in natural-language visualization authoring is that users routinely underspecify their requests. Traditional LLM-based approaches respond to ambiguity by inferring missing details, which can produce hallucinated variable names, fabricated data sources, or visualizations that do not reflect user intent. Raiven addresses this by requiring a complete and validated specification before generating RaivenDSL. Users describe their intent through an interactive chat interface, refining and extending their request across multiple turns until the system has gathered a full specification. Rather than generating RaivenDSL from a single prompt, the LLM populates the schema incrementally through a sequence of targeted prompts, each responsible for a narrowly defined subproblem. The schema serves as a flattened intermediate representation of the eventual DSL, allowing the system to track what has been specified and what remains outstanding before compilation. Appendix~\ref{app:nli} describes the full schema structure 
and documents what triggers re-prompting at each pipeline stage, including the exact clarification messages presented to the user.

\subsection{Validation and Completion Checking}
\label{sec:validation}

The system advances through a series of sequential nodes that check each schema component before proceeding. When a session begins, Raiven immediately parses uploaded data files into the data block, grounding all subsequent generation in the actual dataset structure. Raiven exposes the LLM only to dataset metadata such as variable names and data types rather than raw values.

At each stage, Raiven validates proposed schema updates before committing them. Raiven checks mark types against its supported set, verifies referenced variables against the dataset, and checks encodings for compatibility with their data types. Crucially, the system will not proceed if required schema elements are missing or underspecified. If the user has not provided a mark type, the system asks for one. If encoding fields are absent or ambiguous, the system returns a message listing exactly what is still needed and waits for the user to respond. This continues until every required component is present and valid.

Only once the schema is fully populated and all checks pass does the system issue a final prompt that translates the schema into RaivenDSL. Because Raiven defers RaivenDSL generation until the schema is complete and validated, the translation step operates on well-formed input and avoids the hallucination and non-determinism that arise when LLMs generate visualization code directly from underspecified requests.

\section{Compiler}
\label{sec:compiler}

The Raiven compiler translates RaivenDSL specifications into executable visualization code deterministically.
Its design reflects a central architectural choice: all visualization logic, from default styling to interactive controls, is resolved by the compiler rather than the LLM.
This division means that the language captures what the user specified, and the compiler fills in what the user did not, with deterministic, reproducible results.

\subsection{Two-Stage Design}
\label{sec:compiler-design}

The compiler operates in two stages: a \emph{verification stage} that validates the specification independently of any rendering backend, and a \emph{realization stage} that assigns a rendering backend to each view, resolves defaults, and prepares the specification for code generation. We separate these stages so that the compiler catches structural errors (unknown data sources, invalid mark-encoding combinations, unresolved selection references) before making any backend-specific decisions. This ensures that validation logic and backend logic remain independent: adding a new backend requires only a new realization path, not changes to the validator.

Because a single RaivenDSL program can contain views that require different backends, for example volumetric views compiled to VTK.js alongside chart views compiled to D3, the realization stage assigns a backend independently to each view based on its mark types. This per-view routing is what makes cross-domain bridging work at the execution level: a single RaivenDSL program compiles to multiple backends, and the generated views are laid out together in the browser. We chose per-view rather than per-layer backend assignment because layers within a view must share a coordinate system and rendering context, a constraint that would be difficult to enforce across backends.

A validation pipeline runs at both stages. At the verification stage, the compiler checks for structural well-formedness. At the realization stage, it checks for backend compatibility. Because the specification has formal structure, the compiler can validate it before generating any code, which is not possible when an LLM emits arbitrary visualization-library programs.

\subsection{Code Generation}
\label{sec:codegen}

Code generation translates the realized specification into executable D3 or VTK.js code. 
For each view, the compiler produces the complete rendering logic required by the target backend.
The user specifies a mark type and, where applicable, an encoding; the compiler produces the full implementation deterministically.
The volume mark from the running example (Section~\ref{sec:running-example}) illustrates the scale of this expansion: two lines of RaivenDSL (a data source and a mark type) expand into a complete VTK.js pipeline that loads the data, constructs a transfer function, configures a volume mapper, and registers the actor with the renderer.

The complexity of the generated code varies substantially across mark types, and this asymmetry is a key part of the compiler's value. A D3 scatter plot requires scales, axes, gridlines, margins, and SVG structure. A scientific-visualization mark such as streamlines requires a multi-step pipeline: resolving velocity components from the source data, generating seed points, running numerical integration, and registering the result in the 3D renderer. In both cases, the user writes the same kind of specification, a mark type and an encoding, and the compiler absorbs the complexity that would otherwise fall on the user or the LLM.

The compiler is also responsible for defaults. When a specification leaves style properties unset, the compiler supplies deterministic values derived from the mark type and data schema: color palettes, bin sizes, iso-values, transfer function parameters. We encode defaults in the compiler rather than rely on the LLM to infer them because LLM-inferred styling varies across runs and cannot guarantee consistency. Users can adjust most defaults through the generated controls described below, so they can override defaults without editing the specification.

\subsection{Interactivity Generation}
\label{sec:interactivity}

Rather than requiring users to specify all interaction in RaivenDSL, the compiler generates interactive controls automatically. The alternative, producing static output with no controls, would force users to re-edit and recompile the specification for every parameter change. For exploratory tasks, this would make the system impractical. We therefore chose to have the compiler generate controls directly from the specification, so that users can adjust parameters in the UI without returning to RaivenDSL or the natural language interface.
Raiven generates interactive controls at three levels, each reflecting a different design rationale.

\para{Implicit controls.} The compiler provides standard navigation (rotation, zoom, pan) for all 3D views. We include these by default rather than require them in RaivenDSL, because free camera movement is necessary to explore any spatial visualization. Requiring users to declare navigation would add specification overhead without expressive benefit.

\para{Declared controls.} The compiler translates selections and links specified in the DSL into interactive behaviors. When a specification declares a brush selection linked across two views, the compiler generates the event listeners, data filtering logic, and visual highlighting needed to synchronize them. For example, a brush in a scatter plot that highlights corresponding points in a histogram requires the compiler to generate listener code in the source view, a filtering function that maps the brushed range to matching data points, and a highlight update in the target view. Shared transfer functions and synchronized slice indices work the same way: the compiler reads the declared link, determines which parameters must be shared, and generates the synchronization code. We chose to compile these from explicit declarations rather than infer them, consistent with the language design principle described in Section~\ref{sec:design-principles}.

\para{Mark-specific controls.} For each mark, the compiler determines which parameters are user-adjustable and generates the corresponding interface elements (Figure~\ref{fig:teaser}). A volume mark produces a transfer function editor; a force-directed graph produces controls for simulation strength and link distance; a streamline mark produces controls for seed region bounds, tube radius, and integration step size. We chose to generate these automatically because the mark type and the resolved specification fully determine the set of meaningful parameters, requiring no additional input from the user or the LLM.

Not all controls behave the same way at runtime. Changing a color, opacity value, tube radius, isosurface threshold, or force-graph simulation parameter takes effect immediately. Changing a seed region or integration step size for streamlines triggers a full numerical re-integration. Raiven applies these expensive controls only on user confirmation because the underlying computation is too costly to run on every slider movement. Appendix~\ref{app:marks} provides a complete list of per-mark-type controls.

Raiven implements the compiler in TypeScript; it runs entirely in the browser. ANTLR4 parses the brace-based syntax; the indentation-based syntax uses a separate parser that produces the same abstract syntax tree. The compiler implements validation, realization, and code generation as sequential pipeline stages. Generated D3 and VTK.js code executes in-browser, arranging views in a responsive grid and rendering interactive controls through Tweakpane. Compilation and rendering require no server-side computation: once Raiven generates RaivenDSL, the entire execution pipeline is client-side. We chose this split so that the only server dependency is the LLM API call; a user who writes RaivenDSL by hand can compile and render without any server at all. Vite handles self-contained HTML export, bundling all generated code, library dependencies, and data references in a single file. RaivenDSL supports four data formats: VTI (2D and 3D image data), CSV, JSON, and GeoJSON.
\section{Benchmark Evaluation}
\label{sec:benchmark}

To evaluate Raiven's end-to-end performance, we designed a benchmark of 100 visualization prompts spanning the full RaivenDSL design space. Each prompt is run once through Raiven and 3 alternate general-purpose LLMs, each in a single-shot setting, producing a self-contained HTML file scored for correctness and efficiency\footnote{All prompts, and outputs are available at \url{https://alexandrairger.github.io/RaivenBenchmark/benchmark/processed/gallery.html}.}.
No manual repair, iterative prompting, or task-specific scaffolding is permitted. Outputs that fail to produce executable HTML are treated as execution failures.

\para{Prompts.} The 100 prompts are grouped into three categories: InfoVis ($I$, 40 prompts), SciVis ($S$, 30 prompts), and Combined ($C$, 30 prompts), where Combined prompts reference both spatial and tabular data. The higher InfoVis count reflects the larger number of mark types in that category. Prompts vary in complexity from single-view, single-mark specifications to multi-view dashboards with cross-view linking. Each prompt is fully specified with respect to its accompanying dataset: given the prompt text and the dataset, the intended visualization is unambiguous, with no missing parameters or underspecified intent. All prompts are listed in Appendix~\ref{app:benchmark-full-results}.

\para{Datasets.} The benchmark uses real-world datasets across all three categories. SciVis prompts use 2D and 3D image data (VTI); InfoVis prompts use tabular (CSV), network (JSON), and geospatial (GeoJSON) data; Combined prompts reference both. CSV and JSON files are embedded in full in the model prompt, reflecting how a user would typically provide tabular data to an LLM; files exceeding 100K tokens are truncated, though this affects only a small subset of CSVs. VTI and GeoJSON files are too large to embed and are represented by their metadata headers only. Both are provided by URL, to avoid downsampling artifacts observed in early experiments. All files are also provided as URLs so that generated code can reference them directly at runtime. Full dataset descriptions, sources, and licenses are in Appendix~\ref{app:datasets}.

\para{Baselines.} We compare Raiven against three general-purpose LLMs: ChatGPT~5.4, Claude Opus~4.6, and Gemini~3.1~Pro. All models are called with temperature~0. Raiven itself uses ChatGPT~5.2 for all generation steps (Section~\ref{sec:nli}), a less capable model than the ChatGPT~5.4 baseline. Each baseline receives the same prompt and data as described above. Raiven, by contrast, receives only dataset metadata (variable names, data types, and source paths) and never sees raw data values, as described in Section~\ref{sec:nli}. All baselines share a system prompt (Appendix~\ref{app:prompts-benchmark}) instructing the model to produce a self-contained HTML file and to load data at runtime via the provided URLs. No additional scaffolding, examples, or few-shot demonstrations are provided. This is a system-level comparison: it evaluates the end-to-end authoring pipeline each approach provides, not model capability in isolation. Per-model API configurations are documented in Appendix~\ref{app:baseline-config}.

\definecolor{cell1}{HTML}{67a9cf}   % best
\definecolor{cell2}{HTML}{d1e5f0}
\definecolor{cell3}{HTML}{fddbc7}
\definecolor{cell4}{HTML}{ef8a62}  % worst

\definecolor{hc1}{HTML}{ef8a62}  % worst (soft red)
\definecolor{hc2}{HTML}{fddbc7}
\definecolor{hc3}{HTML}{f7f7f7}  % midpoint
\definecolor{hc4}{HTML}{d1e5f0}
\definecolor{hc5}{HTML}{67a9cf}  % best (soft blue)

\newcommand{\hc}[3]{%
  \edef\hcpct{\fpeval{
    abs(#2 - #3) < 0.0001 ? 50 : round((#1 - #3) / (#2 - #3) * 100, 0)
  }}%
  \ifnum\hcpct>75
    \edef\subpct{\fpeval{round((\hcpct - 75) * 4, 0)}}%
    \begingroup\edef\x{\endgroup\noexpand\cellcolor{hc5!\subpct!hc4}}\x
  \else\ifnum\hcpct>50
    \edef\subpct{\fpeval{round((\hcpct - 50) * 4, 0)}}%
    \begingroup\edef\x{\endgroup\noexpand\cellcolor{hc4!\subpct!hc3}}\x
  \else\ifnum\hcpct>25
    \edef\subpct{\fpeval{round((\hcpct - 25) * 4, 0)}}%
    \begingroup\edef\x{\endgroup\noexpand\cellcolor{hc3!\subpct!hc2}}\x
  \else
    \edef\subpct{\fpeval{round(\hcpct * 4, 0)}}%
    \begingroup\edef\x{\endgroup\noexpand\cellcolor{hc2!\subpct!hc1}}\x
  \fi\fi\fi
}

\begin{table}[t]
    \centering
    \caption{Benchmark results across InfoVis ($I$), SciVis ($S$), and
    Combined ($C$) categories reporting mean ($\mu$). Cell shading
    interpolates linearly per row from
    \colorbox{cell1}{best} to \colorbox{cell4}{worst}.}
    \begin{tabular}{cl|c|c|c|c}
        & & Gemini & ChatGPT & Claude & \multirow{2}{*}{Raiven} \\
        & & 3.1 Pro & 5.4 & Opus 4.6 & \\
        \hline \hline
        % Compile: higher is better → best=max, worst=min
        \multirow{4}{*}{\makecell{Compile\\($\%$)}}
         & $I$ &
         \hc{1.0000}{1.0}{0.9580}{1.0000} &
         \hc{1.0000}{1.0}{0.9580}{1.0000} &
         \hc{0.9580}{1.0}{0.9580}{0.9580} &
         \hc{1.0000}{1.0}{0.9580}{1.0000} \\
         & $S$ &
         \hc{0.6110}{1.0}{0.5110}{0.6110} &
         \hc{0.6670}{1.0}{0.5110}{0.6670} &
         \hc{0.5110}{1.0}{0.5110}{0.5110} &
         \hc{1.0000}{1.0}{0.5110}{1.0000} \\
         & $C$ &
         \hc{0.9440}{1.0}{0.8780}{0.9440} &
         \hc{0.8780}{1.0}{0.8780}{0.8780} &
         \hc{0.8780}{1.0}{0.8780}{0.8780} &
         \hc{1.0000}{1.0}{0.8780}{1.0000} \\
         & $\sum$ &
         \hc{0.8670}{1.0}{0.8000}{0.8670} &
         \hc{0.8630}{1.0}{0.8000}{0.8630} &
         \hc{0.8000}{1.0}{0.8000}{0.8000} &
         \hc{1.0000}{1.0}{0.8000}{1.0000} \\
\hline
        % Time: lower is better → best=min, worst=max
        \multirow{4}{*}{\makecell{Time in\\minutes\\ ($\mu$)}}
         & $I$ &
         \hc{1.432}{0.310}{1.432}{1.432} &
         \hc{0.857}{0.310}{1.432}{0.857} &
         \hc{0.890}{0.310}{1.432}{0.890} &
         \hc{0.310}{0.310}{1.432}{0.310} \\
         & $S$ &
         \hc{2.363}{0.342}{2.363}{2.363} &
         \hc{0.559}{0.342}{2.363}{0.559} &
         \hc{1.066}{0.342}{2.363}{1.066} &
         \hc{0.342}{0.342}{2.363}{0.342} \\
         & $C$ &
         \hc{2.118}{0.340}{2.118}{2.118} &
         \hc{0.986}{0.340}{2.118}{0.986} &
         \hc{1.365}{0.340}{2.118}{1.365} &
         \hc{0.340}{0.340}{2.118}{0.340} \\
         & $\sum$ &
         \hc{1.917}{0.329}{1.917}{1.917} &
         \hc{0.807}{0.329}{1.917}{0.807} &
         \hc{1.085}{0.329}{1.917}{1.085} &
         \hc{0.329}{0.329}{1.917}{0.329} \\
        \hline
        % Tokens: lower is better
        \multirow{4}{*}{\makecell{Total\\Tokens\\($\mu$)}}
         & $I$ &
         \hc{81221}{20880}{81221}{81221} &
         \hc{35211}{20880}{81221}{35211} &
         \hc{42960}{20880}{81221}{42960} &
         \hc{20880}{20880}{81221}{20880} \\
         & $S$ &
         \hc{19567}{3979}{19567}{19567} &
         \hc{3979}{3979}{19567}{3979} &
         \hc{6508}{3979}{19567}{6508} &
         \hc{18240}{3979}{19567}{18240} \\
         & $C$ &
         \hc{96363}{15906}{96363}{96363} &
         \hc{15906}{15906}{96363}{15906} &
         \hc{51719}{15906}{96363}{51719} &
         \hc{18824}{15906}{96363}{18824} \\
         & $\sum$ &
         \hc{67267}{19471}{67267}{67267} &
         \hc{20050}{19471}{67267}{20050} &
         \hc{34652}{19471}{67267}{34652} &
         \hc{19471}{19471}{67267}{19471} \\
        \hline
        % Cost: lower is better
        \multirow{4}{*}{\makecell{Estimated\\Cost in USD\\($\mu$)}}
         & $I$ &
         \hc{0.1761}{0.0487}{0.3018}{0.1761} &
         \hc{0.1361}{0.0487}{0.3018}{0.1361} &
         \hc{0.3018}{0.0487}{0.3018}{0.3018} &
         \hc{0.0487}{0.0487}{0.3018}{0.0487} \\
         & $S$ &
         \hc{0.0356}{0.0356}{0.1371}{0.0356} &
         \hc{0.0467}{0.0356}{0.1371}{0.0467} &
         \hc{0.1371}{0.0356}{0.1371}{0.1371} &
         \hc{0.0443}{0.0356}{0.1371}{0.0443} \\
         & $C$ &
         \hc{0.1957}{0.0471}{0.3928}{0.1957} &
         \hc{0.0925}{0.0471}{0.3928}{0.0925} &
         \hc{0.3928}{0.0471}{0.3928}{0.3928} &
         \hc{0.0471}{0.0471}{0.3928}{0.0471} \\
         & $\sum$ &
         \hc{0.1398}{0.0469}{0.2797}{0.1398} &
         \hc{0.0962}{0.0469}{0.2797}{0.0962} &
         \hc{0.2797}{0.0469}{0.2797}{0.2797} &
         \hc{0.0469}{0.0469}{0.2797}{0.0469} \\
        \hline \hline
        % VMPC: higher is better
        \multirow{4}{*}{\makecell{VMPC\\($\mu$)}}
         & $I$ &
         \hc{0.9738}{0.9900}{0.8847}{0.9738} &
         \hc{0.9722}{0.9900}{0.8847}{0.9722} &
         \hc{0.8847}{0.9900}{0.8847}{0.8847} &
         \hc{0.9900}{0.9900}{0.8847}{0.9900} \\
         & $S$ &
         \hc{0.3689}{0.9894}{0.3689}{0.3689} &
         \hc{0.4100}{0.9894}{0.3689}{0.4100} &
         \hc{0.3978}{0.9894}{0.3689}{0.3978} &
         \hc{0.9894}{0.9894}{0.3689}{0.9894} \\
         & $C$ &
         \hc{0.7109}{0.9851}{0.6823}{0.7109} &
         \hc{0.6966}{0.9851}{0.6823}{0.6966} &
         \hc{0.6823}{0.9851}{0.6823}{0.6823} &
         \hc{0.9851}{0.9851}{0.6823}{0.9851} \\
         & $\sum$ &
         \hc{0.7134}{0.9884}{0.6779}{0.7134} &
         \hc{0.7209}{0.9884}{0.6779}{0.7209} &
         \hc{0.6779}{0.9884}{0.6779}{0.6779} &
         \hc{0.9884}{0.9884}{0.6779}{0.9884} \\
    \end{tabular}
    \label{tab:benchmark_summary}
\vspace{-4pt}
\end{table}

\subsection{Efficiency}
\label{sec:efficiency}

We assess efficiency through generation time, token usage, and cost. Generation time is measured as wall-clock time from user request to produced visualization, reflecting the latency experienced during usage. Token usage is the total prompt and completion tokens consumed per task. Cost reflects the token-based API pricing applied to each task. Results are reported in Table~\ref{tab:benchmark_summary}.

\subsection{Visualization Correctness: VMPC}
\label{sec:vmpc}

Existing evaluation metrics for LLM-generated visualizations, such as VisEval~\cite{chen2024viseval} and VegaChat~\cite{hostnik2026vegachat}, are designed for single-chart tasks and do not address multi-view completeness, hallucinated data, or execution reliability. AstroVisBench~\cite{joseph2025astrovisbench} covers scientific visualization but uses coarse error categories insufficient for partially correct multi-view outputs. We introduce VMPC (Visualization Multi-View Prompt Compliance), a metric that extends these prior approaches to interactive multi-view visualization across both domains.

For each required view $v$, VMPC evaluates five binary criteria: view presence $V_v$ (whether the chart outline or container is rendered), mark correctness $M_v$ (whether the correct mark type is used), encoding correctness $E_v$ (whether the correct variables, color mappings, and channel assignments appear), data hallucination $H_v$ (whether the system used the original input data rather than fabricating or substituting values), and cross-view linking $L_v$ (whether requested coordination between views is present and functional). For single-view prompts or prompts that do not request linking, $L_v$ is set to~1. All criteria are binary: a criterion is either met or it is not. We chose binary scoring over partial-credit scales to eliminate subjectivity in grading. A global execution gate $X$ ensures that a system receives a score of zero if it fails to produce a runnable visualization. These per-view terms are averaged over all 5 criteria across the $N$ views specified in the prompt:
\begin{equation}
\mathrm{VMPC} = X \cdot \frac{1}{5N} \sum_{v=1}^{N}
\left( V_v + M_v + E_v + H_v + L_v \right) \in [0,1]
\label{eq:vmpc}
\end{equation}

\begin{figure}[t]
    \centering
    \begin{subfigure}[b]{0.48\linewidth}
        \begin{tikzpicture}
          \node[anchor=south west,inner sep=0] (img) {%
            \includegraphics[width=\linewidth]{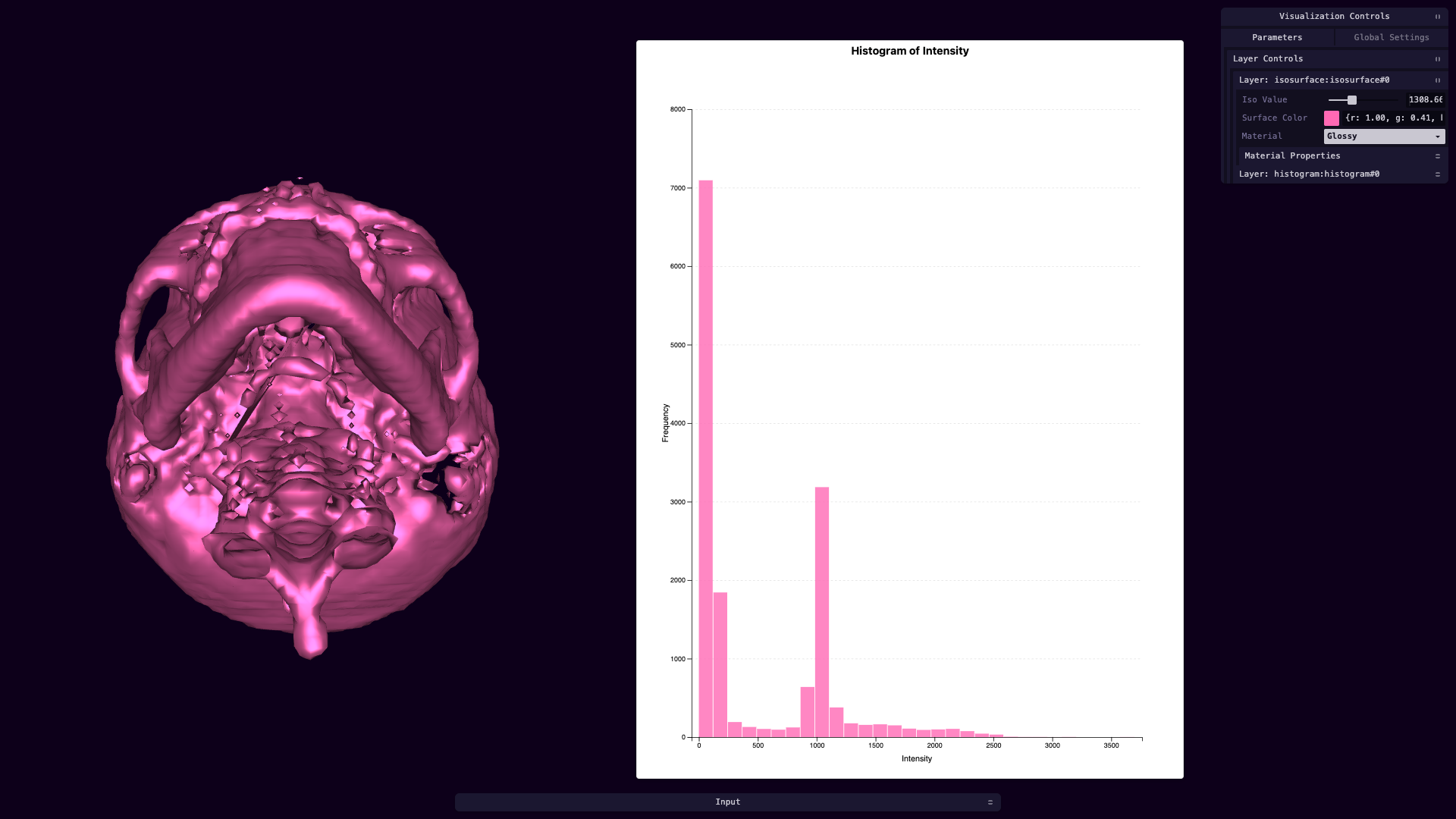}};
          \node[anchor=north west,font=\footnotesize\sffamily\bfseries,
                fill=white,opacity=0.85,text opacity=1,inner sep=2pt]
            at (img.north west) {(a)};
        \end{tikzpicture}
        \phantomsubcaption\label{fig:c73_raiven}
    \end{subfigure}\hfill
    \begin{subfigure}[b]{0.48\linewidth}
        \begin{tikzpicture}
          \node[anchor=south west,inner sep=0] (img) {%
            \includegraphics[width=\linewidth]{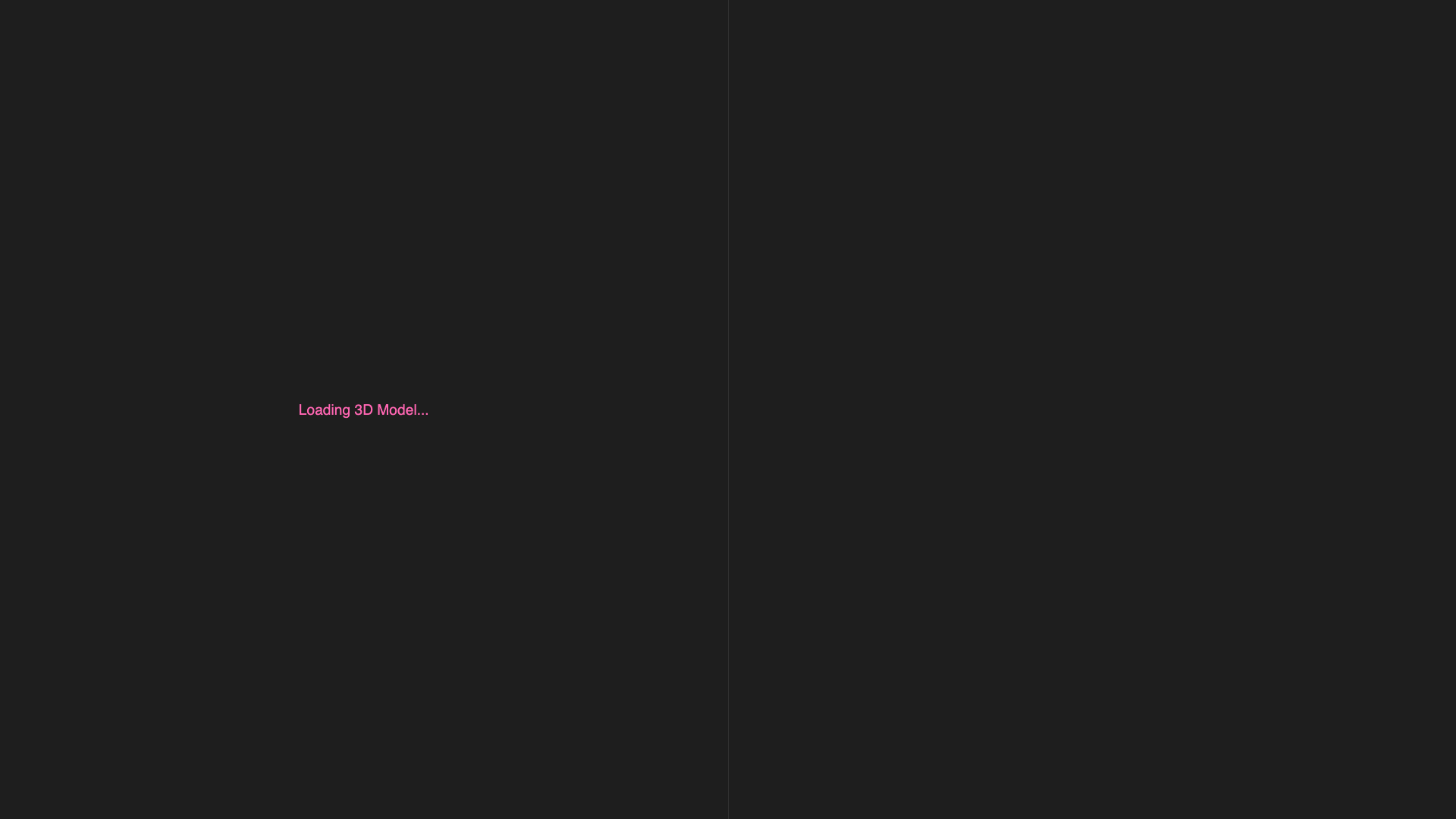}};
          \node[anchor=north west,font=\footnotesize\sffamily\bfseries,
                fill=white,opacity=0.85,text opacity=1,inner sep=2pt]
            at (img.north west) {(b)};
        \end{tikzpicture}
        \phantomsubcaption\label{fig:c73_gemini}
    \end{subfigure}\\[4pt]
    \centering
    \footnotesize
    \setlength{\tabcolsep}{4pt}
    \renewcommand{\arraystretch}{1.05}
\begin{tabular}{l|c|ccccc|ccccc|c}
    \toprule
    System & $X$ 
      & $V_1$ & $M_1$ & $E_1$ & $H_1$ & $L_1$ 
      & $V_2$ & $M_2$ & $E_2$ & $H_2$ & $L_2$ 
      & VMPC \\
    \midrule
    Raiven  & 1 & 1 & 1 & 1 & 1 & 1 & 1 & 1 & 1 & 1 & 1 & 1.00 \\
    Gemini  & 1 & .67 & 0 & 0 & 1 & 1 & .67 & 0 & 0 & 1 & 1 & 0.53 \\
    \bottomrule
\end{tabular}

\vspace{-6pt}
    \caption{VMPC scoring for benchmark prompt~\#73: \textit{``Render the CT isosurface in pink from \texttt{head.vti}, then summarize \texttt{head\_sample.csv} using a histogram of intensity colored pink.''} (a)~Raiven output (VMPC\,=\,1.00). (b)~Gemini output (VMPC\,=\,0.53):  marks and encodings are missing. Fractional scores (e.g. $V$ = 0.67) reflect disagreement among graders. The table shows per-view component scores averaged across three human graders; subscripts denote view~1 (isosurface) and view~2 (histogram). Both systems execute successfully ($X$ = 1).}
    \vspace{-12pt}
    \label{fig:grading_examples}
\end{figure}

Figure~\ref{fig:grading_examples} illustrates VMPC scoring on a two-view prompt. Raiven produces both views correctly (VMPC=1.00); Gemini renders both view containers but with missing marks and encodings (VMPC=0.53).
Because VMPC awards partial credit for incomplete outputs, the baseline scores reported in this paper represent generous lower bounds; the practical gap from a user's perspective is larger than the numbers suggest. Full VMPC grading example breakdowns are provided in Appendix~\ref{app:vmpc-scoring}.

\subsection{Scoring Validation}
\label{sec:scoring}

Each visualization is scored independently by three human graders, and the final VMPC score is taken as the average across all three assessments. Inter-rater reliability across the three human graders was high (Krippendorff's $\alpha = 0.93$), indicating strong consistency in how graders applied the rubric. Because all graders are authors of this paper, we additionally validated human scores against an independent vision-language model (GPT-4.1), which received the benchmark prompt, rendered visualization, and VMPC rubric. We observe strong agreement between human and VLM evaluations (Pearson $r = 0.924$, Spearman $\rho = 0.881$, $n = 400$), confirming that the rubric is sufficiently well-specified to be applied objectively. All scores reported in this paper are human-assigned; VLM scores serve only as a validation instrument.

\subsection{Results}
\label{sec:benchmark-results}

The benchmark reveals a clear and consistent story. As shown in Table~\ref{tab:benchmark_summary}, Raiven scores near-perfect across all prompt groups and components (overall VMPC $= 0.988$, averaged across three human graders), while the three general-purpose baselines range between $0.678$ and $0.721$ on the same aggregate. Notably, Raiven achieves these results using ChatGPT~5.2, a less capable model than the ChatGPT~5.4 baseline, suggesting that the gains stem from the DSL-mediated architecture rather than from the underlying language model.

The performance gap is driven almost entirely by scientific visualization ($S$) prompts, which expose simultaneous failures across multiple VMPC components in all baselines. On $S$ prompts, general-purpose models fail at the most basic level, with execution rates suggesting a substantial portion of all prompts either crash or produce no visible output, and mark correctness and encoding correctness collapse among those that do. InfoVis ($I$) prompts tell a different story: execution is strong across all models and ChatGPT and Gemini nearly match Raiven on overall VMPC, with the remaining gap concentrated in mark and encoding correctness. Combined ($C$) prompts sit in between: execution recovers relative to $S$, but overall VMPC remains substantially below Raiven's, as the $S$ components within these prompts continue to drag down mark correctness, encoding correctness, and cross-view coordination for all baselines.

Across the full benchmark, data hallucination affects $4\%$ of Claude outputs, $2\%$ of ChatGPT and Gemini outputs, and $0\%$ of Raiven outputs. A visualization that silently fabricates values is not an aesthetic failure but a correctness one, with real consequences ranging from flawed research findings to misdiagnoses in clinical settings. 
We note that VMPC does not score interaction, which understates Raiven's practical advantage: the compiler generates interactive controls for every visualization, including parameter sliders, transfer function editors, and cross-view coordination widgets, while the baselines produced little to no interactivity.

Raiven never fails to compile or hallucinates data across any prompt or category, not because it is a more capable language model but because its architecture makes these failure modes structurally impossible. However, this same property is also a limitation: because the LLM never directly inspects the data, it cannot apply reasoning to catch logical inconsistencies. For example, timestep labels such as \texttt{t1}, \texttt{t10}, \ldots, \texttt{t19}, \texttt{t2} were sorted alphabetically by the compiler rather than numerically. Similarly, Raiven plots all data as provided, with no automatic filtering of null or zero values, which can produce visually misleading results. 
\section{User Study}
\label{sec:userstudy}

To validate whether the benchmark gains translate into measurable improvements in practice, we conducted an expert user study comparing Raiven against participants' own LLM-assisted visualization workflows.

\subsection{Study Design}
\label{sec:study-design}

We recruited seven visualization experts (2 female, 5 male), consisting of three PhD students and four postdoctoral researchers whose research focuses on visualization. Participants reported an average visualization expertise of 4.43 on a 1--5 self-assessment scale. Each study session lasted approximately 60--90 minutes, and participants were compensated \$20 for their participation. All participants provided informed consent. The study protocol was approved by the Harvard University Institutional Review Board (IRB25-1536) and the Department of Energy (ORAU001356).

Participants completed three visualization authoring tasks spanning information visualization, scientific visualization, and a combined workflow. Each task was presented as a printed image of a target visualization dashboard to replicate, simulating a common workflow in which a practitioner begins with a visualization concept and must recreate it using available tools. Typed instructions were intentionally avoided to prevent participants from directly copying task descriptions into either system. 

Each participant completed all three tasks using both Raiven and a baseline workflow, with system order alternating across tasks and the starting assignment counterbalanced between participants. For their baseline, participants used their own preferred LLM-assisted tools: six of seven chose Claude in some form, spanning chat interfaces, notebook environments, and IDE-integrated agentic workflows (Claude web, Claude Code, Cursor, Antigravity, and Google Colab). All tools represented the highest-tier models available at the time of the study. To control for unequal access, we provided a standardized workstation with unlimited access to ChatGPT, Claude, Claude Code, Gemini, Codex, Cursor, Antigravity, VS Code, and Google Colab. The comparison is therefore between two LLM-mediated authoring strategies: Raiven's structured DSL pipeline versus free-form code generation through general-purpose models.

Each task began with a comprehension phase in which participants examined the printed target dashboard and wrote prompts or notes in a document. Participants then completed the task using the assigned system. We recorded comprehension time, task completion time, and the number of interaction iterations required to produce the visualization, and collected all generated outputs. After completing all tasks, participants filled out a post-study survey consisting of Likert-scale questions and open-ended feedback. Full task descriptions, target dashboards, and study materials are provided in Appendix~\ref{app:tasks}.

% Likert score colors:
\definecolor{color1}{HTML}{b2182b}
\definecolor{color2}{HTML}{ef8a62}
\definecolor{color3}{HTML}{E8E8E8}
\definecolor{color4}{HTML}{67a9cf}
\definecolor{color5}{HTML}{2166ac}

\begin{figure}[t]
    \centering
    \includegraphics[width=\linewidth]{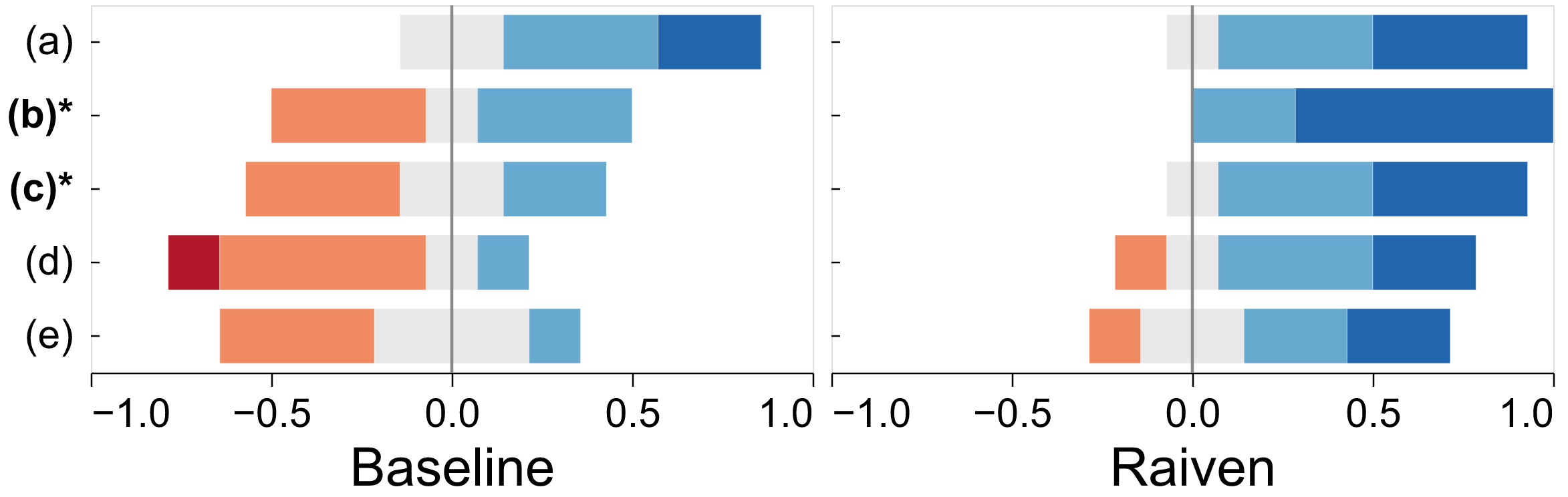}
    \caption{Fraction of participants at each Likert level. Rows (a)--(e) index paired items in survey order: (a)~ease of use; \textbf{(b)*}~task efficiency; \textbf{(c)*}~ease of creating correct visualizations; (d)~trust in visualizations; (e)~perceived control. Colors indicate Likert score \colorbox{color1}{\textcolor{white}{1}}\colorbox{color2}{\textcolor{black}{2}}\colorbox{color3}{\textcolor{black}{3}}\colorbox{color4}{\textcolor{black}{4}}\colorbox{color5}{\textcolor{white}{5}} (1 = strongly disagree, 5 = strongly agree). *$p < 0.05$.}
    \label{fig:user_study_results_1}
    \vspace{-12pt}
\end{figure}

\subsection{Raiven reduces debugging effort and iteration time}
\label{sec:efficiency-results}

\begin{figure}[t]
    \centering
    \includegraphics[width=\linewidth]{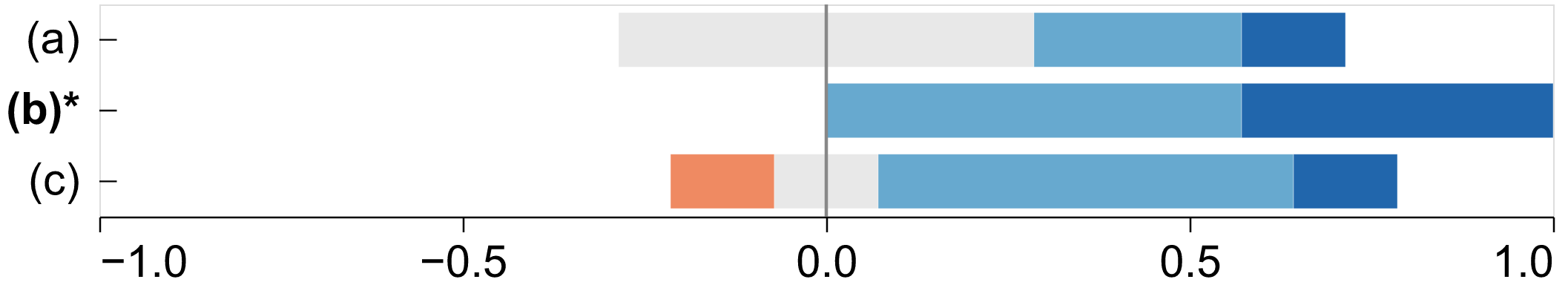}
    \caption{Comparing Raiven with the Baseline: (a)~required less mental effort; \textbf{(b)*}~reduced debugging effort; (c)~preferred for future tasks. Colors indicate Likert scores \colorbox{color1}{\textcolor{white}{1}}\colorbox{color2}{\textcolor{black}{2}}\colorbox{color3}{\textcolor{black}{3}}\colorbox{color4}{\textcolor{black}{4}}\colorbox{color5}{\textcolor{white}{5}} (1 = strongly disagree, 5 = strongly agree). *$p < 0.05$.}
    \vspace{-12pt}
    \label{fig:user_study_results_2}
\end{figure}

Raiven demonstrated clear efficiency advantages over baseline workflows across all three tasks. Participant ratings reflected this directly: Raiven scored significantly higher than baseline on task completion efficiency (Wilcoxon signed-rank, $p = 0.031$) and ease of creating correct visualizations (Wilcoxon signed-rank, $p = 0.031$), Figure~\ref{fig:user_study_results_1}. The strongest result in the study was for debugging effort, where Raiven was rated as significantly reducing debugging compared to baseline (Wilcoxon signed-rank vs.\ neutral, $p = 0.016$), Figure~\ref{fig:user_study_results_2}. Ease of use did not differ significantly between systems ($p = 0.75$), likely reflecting the high usability floor inherent to natural language interfaces generally. Raiven's gains were specific to output reliability and iteration speed rather than general usability.

Task completion patterns favored Raiven across all three tasks. Task~3 (combined) was the most consistent for Raiven, with nearly every participant completing it correctly on the first or second attempt. Task~2 (scientific visualization) proved the most challenging for baseline workflows, with several participants spending upwards of 10--15 minutes with limited success. Baseline failures were varied: several participants received non-interactive static outputs when tasks required linked interactive views, one participant's map was generated for only a small geographic subset of the data, and another encountered repeated compile errors across two different tools and was unable to produce a working output at all.

Participants echoed these results in open-ended feedback, citing speed as Raiven's single biggest advantage: ``by far the speed,'' ``speed, better result in initial try,'' and ``easy to iterate'' were representative responses. One participant noted that with Raiven it was ``extremely straightforward to get the correct result,'' attributing this to how the DSL clarified the mental model: ``I need data, I need to know a visualization type, I need to know encodings --- the system prompting for decisions makes this really clear, without implementation details.''

Terminology and specification challenges were a recurring issue across both systems. Several participants used imprecise vocabulary, asking for a ``bar chart'' when the target was a histogram, or ``volume'' when they meant isosurface. Both systems took these descriptions literally, although baseline systems frequently missed parts of the specifications even when stated correctly. One of the most commonly noted limitations of Raiven was that its clarifying prompts could have been clearer and more intuitive, though all participants were ultimately able to answer them and allow Raiven to proceed.

\subsection{Baseline systems silently fabricate data; Raiven's DSL makes outputs verifiable}
\label{sec:fabrication-trust}

The most consequential baseline failure was silent data fabrication, which surfaced repeatedly in Task~2. Despite VTI being a standard volumetric data format, several baseline systems failed to parse it correctly and silently substituted fabricated or simulated datasets, producing plausible-looking visualizations built on invented data. One participant's Cursor-based workflow produced a visualization with no real data at all; another spent over 20 minutes with Antigravity before realizing the same issue; a third participant received a synthetic skull volume generated with a Gaussian filter rather than the actual file. One participant caught the issue mid-task and added ``don't mess with my data'' explicitly to their prompt. In most cases, the output looked superficially correct, requiring active verification to detect.

Silent data fabrication of this kind is architecturally impossible in Raiven: data is handled entirely by deterministic code, and the LLM never directly accesses or processes the underlying data. Raiven handled Task~2 well, with most participants finishing in under 4 minutes using the actual data. Where Raiven did fall short, such as missing volume rendering components, it did so transparently rather than silently substituting data. This distinction between \emph{honest failure} and \emph{silent failure} has real consequences, particularly for users without the data familiarity and domain knowledge to recognize that the underlying data has been replaced entirely.

Trust emerged as a central theme across both qualitative feedback and behavioral observations. Baseline trust concerns centered on process opacity: ``I had no idea what type of output it would produce, what software architecture or visualization libraries it would use. I was not confident it would read the data formats and understand their contents as expected.'' Raiven fostered trust primarily through the readability of its RaivenDSL output and confidence in data integrity. The RaivenDSL specification serves as a legible mediator between user intent and compiler input: participants can verify what the system understood before any code is generated. As one participant noted, ``I could quickly read its RaivenDSL output,'' and another emphasized that ``I could trust that the system was not manipulating my data.'' Participant ratings reflected this pattern, with trust in visualizations trending toward Raiven though falling short of significance (Wilcoxon signed-rank, $p = 0.063$), likely reflecting limited statistical power at $n=7$ rather than a genuine null effect.

\subsection{Experts prefer Raiven for prototyping and unfamiliar data, baselines for production workflows}
\label{sec:use-cases}

Participants drew a clear distinction between the types of tasks each system was best suited for. Raiven was preferred for demos, prototyping, quick data exploration, standalone dashboards, and tasks involving niche or unfamiliar data formats, particularly those combining 3D volumetric data with information visualization. Baseline was preferred for production projects requiring many iterations, tasks with specific library or language requirements, extending existing codebases, and workflows demanding heavy verification and fine-grained control. One participant captured the baseline use case succinctly: ``if I want to be a micromanager.'' When asked about overall system preference, five of seven participants reported preferring Raiven (one definitely, four probably), two reported no preference, and none preferred the baseline.

\section{Conclusions and Future Work}
\label{sec:conclusion}

A central result of this work is that an LLM does not need to generate backend code to support effective visualization authoring. In Raiven, the model operates only over RaivenDSL, while the compiler handles implementation logic deterministically. This separation reduces fragility, keeps visualization behavior under system control, and makes the pipeline easier to verify. By externalizing visualization knowledge into the language and compiler rather than relying on implicit model knowledge of libraries and conventions, Raiven makes data faithfulness and structural correctness easier to enforce than in direct code-generation pipelines. Our findings also highlight the value of a unified representation spanning both scientific and information visualization. Prior systems remain split between chart-focused information visualization and narrowly scoped scientific-visualization workflows, even though many real analysis tasks require both. A cross-domain DSL provides a practical way to bridge this divide. More broadly, the results suggest that visualization authoring systems should be evaluated not only by syntactic validity or visual plausibility, but also by data faithfulness, view completeness, and cross-view structure. VMPC is one step toward that broader evaluation framework.

The current limitations of Raiven fall along a natural boundary: some constrain only the implementation, while others reflect deliberate choices in the language design. On the implementation side, datasets are loaded entirely into the browser with a 100~MB cap, excluding large-scale data and streaming workflows. Animation, cross-backend linking, and unstructured mesh geometry are not yet supported but require only new compiler capabilities, not language changes. On the language side, RaivenDSL provides one rendering implementation per mark type, prioritizing determinism over algorithmic choice, and deliberately excludes data preprocessing to remain declarative. Tensor and hierarchical mark types would extend the design space. Adding new rendering backends such as Matplotlib or a WebGPU-based renderer requires only a new code-generation path, since backend routing already supports per-view assignment. The conversational flow assumes users with some visualization vocabulary; novice users would benefit from more guided clarification strategies. Fine-tuning a model specifically on RaivenDSL specifications is a tractable next step that would further improve generation reliability.

Raiven demonstrates that DSL-mediated generation is a practical foundation for trustworthy visualization authoring via natural language, and its architecture is designed to remain relevant as models continue to improve. The mediation layer, languages, compilers, and validation pipelines that sit between language models and rendering backends, is where the visualization community can most effectively assert its own standards of correctness, faithfulness, and design. Raiven is one step in building that foundation.

\section{Acknowledgements}
\label{sec:acknowledgements}

This work was supported by the U.S. Department of Energy, Office of Science, Office of Advanced Scientific Computing Research's Computer Science Competitive Portfolios program under Contract No. DE-AC05-00OR22725.

\bibliographystyle{abbrv-doi-hyperref}

\bibliography{bibliography}

\appendix

\clearpage
\newpage
\appendix

\section{RaivenDSL}
\label{app:dsl}
This section provides the complete RaivenDSL language reference: the formal grammar (Section~\ref{app:grammar}), the language structure including data constructors, layer properties, link types, and selection types (Section~\ref{app:language-structure}), and per-mark-type specifications (Section~\ref{app:marks}).

\subsection{Grammar}
\label{app:grammar}
As referenced by Section~\ref{sec:design-principles}.
RaivenDSL supports two surface syntaxes that parse to the same abstract syntax tree. Listings~\ref{lst:pythonic} and~\ref{lst:brace} show the same specification in both forms. The brace-based syntax is defined by the ANTLR4 grammar in Section~\ref{app:grammar-brace}; the indentation-based variant is described in Section~\ref{app:grammar-pythonic}.

\lstdefinestyle{raiven}{
  basicstyle=\ttfamily\scriptsize,
  columns=fullflexible,
  keepspaces=true,
  showstringspaces=false,
  breaklines=true,
  escapeinside={(*}{*)},
}

\begin{lstlisting}[style=raiven,
  caption={Indentation-based syntax for the Teaser Figure~\ref{fig:teaser} example.},
  label={lst:pythonic}]
vis:
  (*\textcolor{data}{data}*):
    vol = img("taylorgreen_9.vti", format="vti")
    sample = tbl("tg9_sample.csv", format="csv")
  (*\textcolor{view}{view}*) "volume_streamline":
    (*\textcolor{layer}{layer}*):
      (*\textcolor{data}{from}*) = vol
      (*\textcolor{mark}{mark}*) = volume
      (*\textcolor{mark}{encode}*):
        field = "vorticity"
    (*\textcolor{layer}{layer}*):
      (*\textcolor{data}{from}*) = vol
      (*\textcolor{mark}{mark}*) = streamline
      (*\textcolor{mark}{encode}*):
        vx = "ux"
        vy = "uy"
        vz = "uz"
  (*\textcolor{view}{view}*) "histogram":
    (*\textcolor{layer}{layer}*):
      (*\textcolor{data}{from}*) = sample
      (*\textcolor{mark}{mark}*) = histogram
      (*\textcolor{mark}{encode}*):
        x = "vorticity"
\end{lstlisting}
\begin{lstlisting}[style=raiven,
  caption={Brace-based syntax for the same specification. Both listings produce an identical AST.},
  label={lst:brace}]
vis {
  (*\textcolor{data}{data}*) {
    vol: img("taylorgreen_9.vti", format: "vti");
    sample: tbl("tg9_sample.csv", format: "csv");
  }
  (*\textcolor{view}{view}*) "volume_streamline" {
    (*\textcolor{layer}{layer}*) {
      (*\textcolor{data}{from}*): vol;
      (*\textcolor{mark}{mark}*): volume;
      (*\textcolor{mark}{encode}*): { field: "vorticity" };
    }
    (*\textcolor{layer}{layer}*) {
      (*\textcolor{data}{from}*): vol;
      (*\textcolor{mark}{mark}*): streamline;
      (*\textcolor{mark}{encode}*): { vx: "ux", vy: "uy", vz: "uz" };
    }
  }
  (*\textcolor{view}{view}*) "histogram" {
    (*\textcolor{layer}{layer}*) {
      (*\textcolor{data}{from}*): sample;
      (*\textcolor{mark}{mark}*): histogram;
      (*\textcolor{mark}{encode}*): { x: "vorticity" };
    }
  }
}
\end{lstlisting}

\lstdefinestyle{antlr}{
  basicstyle=\ttfamily\footnotesize,
  keywordstyle=\bfseries,
  commentstyle=\itshape\color{gray},
  stringstyle=\color{brown},
  morekeywords={grammar, fragment},
  morecomment=[l]{//},
  morecomment=[s]{/*}{*/},
  columns=fullflexible,
  keepspaces=true,
  showstringspaces=false,
  breaklines=true,
  xleftmargin=1em,
  frame=none,
  tabsize=2,
}

\subsubsection{Classic Brace-Based Syntax---ANTLR4 Grammar}
\label{app:grammar-brace}
\begin{lstlisting}[style=antlr]
    grammar VisDSL;

program    : 'vis' '{' topStmt* '}' ;

topStmt    : dataBlock
           | viewDecl
           | selectionsBlock
           ;

dataBlock  : 'data' '{' (dataDecl ';')* '}' ;
dataDecl   : (IDENT | 'geo') ':' dataCtor ;

dataCtor   : 'img' '(' STRING (',' argList)? ')'
           | 'tbl' '(' STRING (',' argList)? ')'
           | 'net' '(' STRING (',' argList)? ')'
           | 'geo' '(' STRING (',' argList)? ')'
           | 'func' '(' argList? ')'
           ;

argList    : namedArg (',' namedArg)* ;
namedArg   : IDENT ':' value ;

value      : STRING
           | NUMBER
           | 'true' | 'false' | 'null'
           | obj | arr
           ;

obj        : '{' (namedArg (',' namedArg)*)? '}' ;
arr        : '[' (value (',' value)*)? ']' ;

viewDecl   : 'view' STRING '{' (layerDecl | linkDecl | interactionsBlock)* '}' ;

layerDecl  : 'layer' '{' layerStmt* '}' ;

layerStmt  : 'from'     ':' layerIdent  ';'
           | 'geo'      ':' layerIdent  ';'
           | 'mark'     ':' IDENT  ';'
           | 'encode'   ':' obj    ';'
           | 'style'    ':' obj    ';'
           | 'where'    ':' expr   ';'
           ;

selectionsBlock : 'selections' '{' selectionDecl* '}' ;
selectionDecl   : 'select' '(' argList ')' ';' ;

linkDecl   : 'link' '(' argList ')' ';' ;

interactionsBlock : 'interactions' '{' interactionDecl* '}' ;
interactionDecl   : 'on' '(' STRING ')' '{' action* '}' ;
action            : 'bind' '(' STRING (',' argList)? ')' ';' ;

expr       : ;

IDENT      : [A-Za-z_][A-Za-z0-9_]* ;
NUMBER     : [+-]?([0-9]+('.'[0-9]+)?)([eE][+-]?[0-9]+)? ;
STRING     : '"' ( '\\' . | ~["\\] )* '"' | '\'' ( '\\' . | ~['\\] )* '\'' ;
WS         : [ \t\r\n]+ -> skip ;
COMMENT    : '//' ~[\r\n]* -> skip ;
BLOCKCOMM  : '/*' .*? '*/' -> skip ;

\end{lstlisting}

\subsubsection{Pythonic Indentation-Based Syntax}
\label{app:grammar-pythonic}

The indentation-based syntax replaces braces with colon-plus-indentation blocks, semicolons with newlines, and uses \texttt{=} for property assignment within layers (e.g., \texttt{from = ct}) rather than \texttt{:}. Data constructors and function-call forms (\texttt{link(...)}, \texttt{select(...)}, \texttt{bind(...)}) retain their parenthesized syntax unchanged. The \texttt{encode} and \texttt{style} properties additionally accept a block form with indented key-value pairs as an alternative to inline objects. A hand-written tokenizer emits \texttt{INDENT}/\texttt{DEDENT} tokens from leading whitespace; the parser is recursive descent and produces an identical AST to the ANTLR-based parser.

\subsection{Language Structure}
\label{app:language-structure}

\label{app:language-structure}

\definecolor{selection}{HTML}{e6ab02}

This section provides the complete language reference for RaivenDSL's compositional rules: top-level constructs, data constructors, layer properties, link types, and selection types. Per-mark-type specifications follow in Section~\ref{app:marks}.

\subsubsection{Top-Level Constructs}

A RaivenDSL program is a single \texttt{vis} block containing data declarations, view definitions, and an optional selections block.

\begin{table}[H]
\centering
\footnotesize
\caption{Top-level constructs in a RaivenDSL program.}
\begin{tabular}{lll}
\toprule
Construct & Contains & Cardinality \\
\midrule
\texttt{vis}        & \texttt{\textcolor{data}{data}}, \texttt{\textcolor{view}{view}}, \texttt{\textcolor{selection}{selections}} & 1 per program \\
\texttt{\textcolor{data}{data}}       & named data declarations    & 1{+} sources \\
\texttt{\textcolor{view}{view}}       & \texttt{\textcolor{layer}{layer}}, \texttt{\textcolor{link}{link}}, \texttt{\textcolor{selection}{interactions}} & 1--9 views \\
\texttt{\textcolor{selection}{selections}} & \texttt{\textcolor{selection}{select(...)}} declarations & 0--1 blocks \\
\bottomrule
\end{tabular}
\label{tab:top-level}
\end{table}

\subsubsection{\textcolor{data}{Data Constructors}}

Each data source carries a typed constructor that determines which marks and backends are valid. The constructor name identifies the data modality; the compiler uses it together with the mark type to route each view to the appropriate rendering backend.

\begin{table}[H]
\centering
\footnotesize
\caption{Data constructors. All constructors except \texttt{func()} take a path string as their first positional argument.}
\begin{tabular}{lllp{3.2cm}}
\toprule
Constructor & Format & Backend & Optional Args \\
\midrule
\texttt{img()} & VTI         & VTK.js & \texttt{format} \\
\texttt{tbl()} & CSV, JSON   & D3     & \texttt{format} \\
\texttt{net()} & JSON        & D3     & \texttt{format} \\
\texttt{geo()} & GeoJSON     & D3     & \texttt{format}, \texttt{crs} \\
\texttt{func()}& procedural  & VTK.js & \texttt{equations}, \texttt{dims}, \texttt{bounds}, \texttt{range}, \texttt{params} \\
\bottomrule
\end{tabular}
\label{tab:data-constructors}
\end{table}

\subsubsection{\textcolor{layer}{Layer Properties}}

Each layer declares a data source and a mark type; encodings, styles, and geographic references are optional at the language level but may be required by specific marks.

\begin{table}[H]
\centering
\footnotesize
\caption{Layer properties. \CIRCLE~= required, \Circle~= optional. Encoding keys and style keys are mark-dependent (Section~\ref{app:marks}).}
\begin{tabular}{llcl}
\toprule
Property & Type & Req. & Description \\
\midrule
\texttt{\textcolor{data}{from}}   & identifier & \CIRCLE & Data source reference \\
\texttt{\textcolor{mark}{mark}}   & identifier & \CIRCLE & Visual mark type \\
\texttt{\textcolor{mark}{encode}} & object     & \Circle & Channel-to-field mappings \\
\texttt{\textcolor{mark}{style}}  & object     & \Circle & Appearance overrides \\
\texttt{\textcolor{data}{geo}}    & identifier & \Circle & Geographic data reference \\
\bottomrule
\end{tabular}
\label{tab:layer-props}
\end{table}

\subsubsection{\textcolor{link}{Link Types}}

Links declare cross-view coordination. Each link specifies exactly one coordination type (\texttt{tf}, \texttt{slice}, or \texttt{selection}) and the participating views.

\begin{table}[H]
\centering
\footnotesize
\caption{Link parameters. A link must specify exactly one of \texttt{selection}, \texttt{tf}, or \texttt{slice}. \texttt{selection} links operate within D3; \texttt{tf} and \texttt{slice} links are within VTK.js.}
\begin{tabular}{lll}
\toprule
Parameter & Type & Required With \\
\midrule
\texttt{\textcolor{selection}{selection}} & brush-and-filter    & \texttt{target} or \texttt{\textcolor{view}{views}} \\
\texttt{tf}        & shared transfer fn. & \texttt{\textcolor{view}{views}} \\
\texttt{slice}     & synced slice index  & \texttt{\textcolor{view}{views}}, \texttt{axes} \\
\texttt{axes}      & slice orientation   & \texttt{slice} \\
\texttt{\textcolor{view}{views}}     & view ID array       & all \texttt{\textcolor{link}{link}} types \\
\texttt{target}    & single view ID      & \texttt{\textcolor{selection}{selection}} only \\
\bottomrule
\end{tabular}
\label{tab:link-types}
\end{table}

\subsubsection{\textcolor{selection}{Selections and Interactions}}

Selections declare named handles for cross-view coordination; the \texttt{interactions} block binds events to those handles. Brush-based selection is driven by the DSL; point selection with cross-view highlighting is generated automatically by the compiler for \texttt{force\_graph} and \texttt{heatmap} marks. All selection-based coordination currently operates within D3 views only.

\begin{table}[H]
\centering
\footnotesize
\caption{Selection and interaction constructs.}
\begin{tabular}{lll}
\toprule
Construct & Required & Effect \\
\midrule
\texttt{\textcolor{selection}{select}(\textit{name})} & \texttt{name} & Registers a named selection \\
\texttt{on("brush")}  & event = \texttt{"brush"} & Installs a rectangular D3 brush \\
\texttt{bind(\textit{sel.})} & selection name & Publishes brush to named selection \\
\bottomrule
\end{tabular}
\label{tab:selection-types}
\end{table}

\FloatBarrier
\subsection{\textcolor{mark}{Marks}}

\label{app:marks}
As referenced by Sections~\ref{sec:expressiveness} and~\ref{sec:interactivity}.
RaivenDSL supports 22 mark types: 5 for scientific visualization (rendered via VTK.js) and 17 for information visualization (rendered via D3). Tables~\ref{tab:scivis-marks} and~\ref{tab:infovis-marks} list each mark with its encoding channels, style parameters, and compiler-generated controls. All style fields are optional; when omitted, the compiler resolves defaults deterministically. Encoding channels define the semantic content of the visualization (which data variables map to which visual channels), while style parameters control appearance and can be adjusted at runtime through the generated controls without re-editing the specification. All SciVis views additionally receive implicit navigation controls (rotation, zoom, pan).

\subsubsection{Scientific Visualization Marks}

All scientific visualization marks accept \texttt{img} or \texttt{func} data sources. All encode and style fields are optional; when omitted, the compiler selects the first available scalar or vector field and resolves data-derived ranges.

Table~\ref{tab:scivis-marks} lists each mark's encoding channels, style parameters, and compiler-generated controls. Table~\ref{tab:scivis-meta} summarizes data type constraints, layering compatibility, implicit controls, and linking options.

\para{Slice rendering modes.} The slice mark's rendering mode and camera are determined by the view configuration. A single slice layer with a single axis renders as a 2D image with pan, zoom, and scroll-to-index navigation. Any other configuration---multiple axes (e.g.\ \texttt{["XY","XZ","oblique"]}), or any additional layer in the same view---renders in 3D with a trackball camera. Oblique slices follow the same rule, with one exception: a single oblique slice defaults to 2D but can be forced to 3D by setting \texttt{is3DPlane: true}.

\begin{table*}[t]
\centering
\footnotesize
\caption{Scientific visualization mark specifications. \textcolor{gray}{Gray} encode fields are semantically important but syntactically optional (the compiler infers them from the data when omitted). All style fields are optional. Controls marked with $\dagger$ are runtime-only (generated by the compiler, not settable in the specification).}
\label{tab:scivis-marks}
\begin{tabular}{llll}
\toprule
Mark & Encode & Style & Generated Controls \\
\midrule

\texttt{volume}
  & \textcolor{gray}{\texttt{field}: scalar array name}
  & \texttt{sample\_distance}: ray step size (default 0.7)
  & Color transfer function editor \\
  & \quad \textcolor{gray}{defaults to first array} & \texttt{palette}: named colormap (e.g.\ \texttt{"viridis"})
  & Opacity transfer function editor \\
  & & \texttt{ctf}: color TF stops (overrides \texttt{palette})
  & Palette dropdown \\
  & & \texttt{otf}: opacity TF stops
  & \\

\addlinespace
\midrule
\addlinespace

\texttt{isosurface}
  & \textcolor{gray}{\texttt{field}: scalar array name}
  & \texttt{iso\_value}: threshold (default $1/3$ of range)
  & Iso-value slider \\
  & \quad \textcolor{gray}{defaults to first array} & \texttt{color}: hex color
  & RGBA color picker with opacity \\
  & & \texttt{opacity}: 0--1
  & Material preset$^\dagger$ (matte/glossy/metallic/custom) \\
  & & & Specular, diffuse, ambient sliders$^\dagger$ \\

\addlinespace
\midrule
\addlinespace

\texttt{slice}
  & \textcolor{gray}{\texttt{field}: scalar array name}
  & \texttt{axes}: \texttt{"XY"}, \texttt{"XZ"}, \texttt{"YZ"}, \texttt{"oblique"},
  & \textit{Axis-aligned} (per axis in \texttt{axes}): \\
  & \quad \textcolor{gray}{defaults to first array} & \quad or array (e.g.\ \texttt{["XY","XZ"]}); default \texttt{"XY"}
  & \quad Slice index slider \\
  & & \texttt{palette}: named colormap
  & \quad Visibility toggle$^\dagger$ \\
  & & \texttt{ctf}: color TF stops (overrides \texttt{palette})
  & Color range interval slider$^\dagger$ \\
  & & \texttt{quaternion}: orientation (oblique only)
  & Color transfer function editor \\
  & & \texttt{offset}: position (oblique only)
  & Palette dropdown \\
  & & \texttt{is3DPlane}: oblique rendering mode
  & \textit{Oblique} (when \texttt{"oblique"} in \texttt{axes}): \\
  & & \quad (auto-inferred; see text)
  & \quad 3D rotation gizmo$^\dagger$ \\
  & & & \quad Offset slider$^\dagger$ \\

\addlinespace
\midrule
\addlinespace

\texttt{streamline}
  & \texttt{vx}, \texttt{vy}, \texttt{vz}: velocity
  & \texttt{seed\_bounds}: $[x_0,x_1,y_0,y_1,z_0,z_1]$
  & Seed region$^\dagger$: dual-handle slider per axis \\
  & \quad component array names
  & \texttt{seed\_count}: number of seeds (default 100)
  & Seed count slider \\
  & & \texttt{integration\_step}: step size (default 0.5)
  & Integration step slider \\
  & & \texttt{max\_steps}: max steps (default 1000)
  & Max steps slider \\
  & & \texttt{color}: hex color
  & Color picker \\
  & & \texttt{tube\_radius}: streamtube radius
  & Tube radius slider \\
  & & & Recalculate button$^\dagger$ \\

\addlinespace
\midrule
\addlinespace

\texttt{lic}
  & \texttt{vx}, \texttt{vy}: velocity
  & \texttt{number\_of\_steps}: integration steps (default 50)
  & Step count slider \\
  & \quad component array names
  & \texttt{step\_size}: step length (default 1.0)
  & Step size slider \\
  & & \texttt{enhanced\_lic}: contrast enhancement (default \texttt{true})
  & Enhanced LIC toggle \\
  & & \texttt{lic\_intensity}: blending weight (default 0.8)
  & Intensity slider \\

\bottomrule
\end{tabular}
\end{table*}

\begin{table}[H]
\centering
\footnotesize
\caption{Scientific visualization mark metadata. All SciVis marks except for LIC can be layered together within a single view. Implicit controls are provided automatically based on rendering mode.}
\label{tab:scivis-meta}
\begin{tabular}{llll}
\toprule
Mark & Layerable With & Implicit Controls & Linking \\
\midrule
\texttt{volume}      & iso., slice, stream. & 3D trackball camera   & --- \\
\texttt{isosurface}  & vol., slice, stream. & 3D trackball camera   & --- \\
\texttt{slice}       & vol., iso., stream.  & 2D or 3D (see text)   & \texttt{slice}, \texttt{tf} \\
\texttt{streamline}  & vol., iso., slice    & 3D trackball camera   & --- \\
\texttt{lic}         & --- & 2D image camera      & --- \\
\bottomrule
\end{tabular}
\end{table}

\subsubsection{Information Visualization Marks}

Most information visualization marks consume tabular data (\texttt{tbl} with
CSV or an inline table). Force-directed graph layers use a network source
(\texttt{net}, JSON). Chord, Sankey, and most chart marks expect tables;
choropleth layers typically pair a geographic source (\texttt{geo}, GeoJSON
or similar) with table columns that supply region keys and values. Required
encode channels are per mark (for example \texttt{x} and \texttt{y} for
scatter plots, \texttt{dimensions} for parallel coordinates, \texttt{source}
and \texttt{target} for flows and networks); optional channels such as
\texttt{color} are listed in the schema when allowed. Style parameters are
optional throughout. When optional encodings or styles are omitted, the
compiler and renderer apply defaults (for example named color schemes, bin
counts, or layout parameters). Table~\ref{tab:infovis-marks} lists each
mark's encoding channels, style parameters, and compiler- or UI-generated
controls. Table~\ref{tab:infovis-meta} summarizes data-type constraints,
layering compatibility, implicit controls, and linking options.

\medskip
\noindent\textbf{Multi-view interaction and layout.}\quad
Views that share a linked selection use \texttt{selections},
\texttt{link(selection:\,\ldots)}, and \texttt{interactions} (for example
brush-to-filter) so brushing in one chart updates linked views; parallel
coordinates additionally expose per-axis brushes in the runtime UI. Network
and flow marks use dedicated layouts (force simulation, chord diagram,
Sankey) rather than shared Cartesian axes; Cartesian marks (points, lines,
bars, distributions, heatmaps, hexbins, etc.) are drawn on scales derived
from the data unless overridden in the view or layer specification.

\begin{table}[H]
\centering
\footnotesize
\setlength{\tabcolsep}{4pt}
\caption{Information visualization mark metadata. Single-view layering is
supported only for the combinations listed (see text). Brush and linking are
enabled when the program declares \texttt{selections},
\texttt{link(selection:\,\ldots)}, and \texttt{interactions\,on("brush")}.}
\label{tab:infovis-meta}
\begin{tabular}{>{\raggedright\arraybackslash}p{1.6cm}
                >{\raggedright\arraybackslash}p{2.0cm}
                >{\raggedright\arraybackslash}p{1.0cm}
                >{\raggedright\arraybackslash}p{2.5cm}}
\toprule
Mark & Layerable With & Implicit & Linking \\
\midrule
\texttt{points} / \texttt{bubble} & hexbin, choropleth & --- & brush (emit/follow) \\
\hline
\texttt{hexbin}        & points, choropleth & --- & --- \\
\hline
\texttt{line}          & line, band         & --- & --- \\
\hline
\texttt{band}          & line               & --- & --- \\
\hline
\texttt{histogram}     & KDE, histogram     & --- & brush (follow) \\
\hline
\texttt{kde}           & histogram          & --- & brush (follow) \\
\hline
\texttt{heatmap}       & ---                & --- & brush (emit/follow),\newline point (follow) \\
\hline
\texttt{bar}           & ---                & --- & --- \\
\hline
\texttt{boxplot}       & ---                & --- & --- \\
\hline
\texttt{violin}        & ---                & --- & --- \\
\hline
\texttt{ridgeline}     & ---                & --- & --- \\
\hline
\texttt{parallel}\newline\texttt{coordinates} & --- & --- & brush (emit/follow) \\
\hline
\texttt{pie}           & ---                & --- & --- \\
\hline
\texttt{chord}         & ---                & --- & --- \\
\hline
\texttt{sankey}        & ---                & --- & --- \\
\hline
\texttt{force\_graph}  & ---                & Node dragging & point (emit) \\
\hline
\texttt{choropleth}    & points, hexbin     & --- & --- \\
\bottomrule
\end{tabular}
\end{table}

\begin{table*}[t]
\centering
\footnotesize
\setlength{\tabcolsep}{4pt}
\caption{Information visualization mark specifications. \textcolor{gray}{Gray} encode fields are semantically important but syntactically optional. All style fields are optional. Controls marked with \textsuperscript{$\dagger$} are runtime-only (generated by the compiler, not settable in the specification). \texttt{bubble} is treated as \texttt{points} with a \texttt{size} encoding in the generated IR.}
\label{tab:infovis-marks}
\begin{tabular}{>{\raggedright\arraybackslash}p{2.0cm}
                >{\raggedright\arraybackslash}p{4.2cm}
                >{\raggedright\arraybackslash}p{5.2cm}
                >{\raggedright\arraybackslash}p{4.0cm}}
\toprule
Mark & Encode & Style & Generated Controls \\
\midrule

\texttt{points}
  & \texttt{x}, \texttt{y};\newline \textcolor{gray}{\texttt{color}: category or value}
  & \texttt{radius}, \texttt{fill\_color}
  & Fill color; Categories if \texttt{color}-encoded; brush interval\textsuperscript{$\dagger$} \\

\addlinespace\midrule\addlinespace

\texttt{hexbin}
  & \texttt{x}, \texttt{y};\newline \textcolor{gray}{\texttt{color}: value field}
  & \texttt{radius}, \texttt{color\_scheme}
  & Color palette dropdown \\

\addlinespace\midrule\addlinespace

\texttt{heatmap}
  & \texttt{x}, \texttt{y};\newline \textcolor{gray}{\texttt{color}: cell value}
  & \texttt{color\_scheme}
  & Color palette; flip x/y toggles; brush\textsuperscript{$\dagger$} (numeric density); point selection\textsuperscript{$\dagger$} (adjacency) \\

\addlinespace\midrule\addlinespace

\texttt{histogram}
  & \texttt{x} \quad (count implicit)
  & \texttt{bins}, \texttt{fill\_color}, \texttt{stroke\_color}
  & Fill color; bin size slider (5--100); brush follow\textsuperscript{$\dagger$} \\

\addlinespace\midrule\addlinespace

\texttt{kde}
  & \texttt{x}
  & \texttt{bandwidth}, \texttt{stroke\_width}, \texttt{stroke\_color}
  & Stroke color; brush follow\textsuperscript{$\dagger$} \\

\addlinespace\midrule\addlinespace

\texttt{boxplot}
  & \texttt{x}, \texttt{y};\newline \textcolor{gray}{\texttt{color}: group field}
  & \texttt{width}, \texttt{fill\_color}, \texttt{stroke\_color}, \texttt{stroke\_width}
  & Fill color / Categories \\

\addlinespace\midrule\addlinespace

\texttt{violin}
  & \texttt{x}, \texttt{y};\newline \textcolor{gray}{\texttt{color}: group field}
  & \texttt{bandwidth}, \texttt{fill\_color}, \texttt{stroke\_color}, \texttt{stroke\_width}, \texttt{show\_median}
  & Fill color / Categories \\

\addlinespace\midrule\addlinespace

\texttt{ridgeline}
  & \texttt{x}, \texttt{y} (categorical group);\newline \textcolor{gray}{\texttt{color}: group field}
  & \texttt{bandwidth}, \texttt{fill\_color}, \texttt{stroke\_color}, \texttt{stroke\_width}, \texttt{overlap}, \texttt{height}
  & Fill color / Categories \\

\addlinespace\midrule\addlinespace

\texttt{line}
  & \texttt{x}, \texttt{y} (\texttt{y} may be array);\newline \textcolor{gray}{\texttt{color}: series field}
  & \texttt{stroke\_width}, \texttt{stroke\_color}
  & Stroke color; Categories if \texttt{color}-encoded; linked selection dims opacity\textsuperscript{$\dagger$} \\

\addlinespace\midrule\addlinespace

\texttt{band}
  & \texttt{x}, \texttt{y0}, \texttt{y1};\newline \textcolor{gray}{\texttt{color}, \texttt{opacity}: optional}
  & \texttt{fill\_color}, \texttt{fill\_opacity}, \texttt{stroke\_color}, \texttt{stroke\_width}
  & Fill color \\

\addlinespace\midrule\addlinespace

\texttt{bar}
  & \texttt{x}, \texttt{y};\newline \textcolor{gray}{\texttt{color}: stacking or grouping}
  & \texttt{fill\_color}, \texttt{stroke\_color}
  & Fill color / Categories \\

\addlinespace\midrule\addlinespace

\texttt{pie}
  & \texttt{label}, \texttt{value};\newline \textcolor{gray}{\texttt{color}: label field}
  & \texttt{inner\_radius}, \texttt{outer\_radius}
  & Categories \\

\addlinespace\midrule\addlinespace

\texttt{chord}
  & \texttt{source}, \texttt{target}, \texttt{value};\newline \textcolor{gray}{\texttt{group}: optional}
  & \texttt{pad\_angle}, \texttt{inner\_radius}, \texttt{outer\_radius}
  & Categories \\

\addlinespace\midrule\addlinespace

\texttt{sankey}
  & \texttt{source}, \texttt{target}, \texttt{value};\newline \textcolor{gray}{\texttt{node}: optional}
  & \texttt{node\_width}, \texttt{node\_padding}, \texttt{link\_opacity}, \texttt{align}, \texttt{link\_color}
  & Link color mode (static / source / target / interpolate) \\

\addlinespace\midrule\addlinespace

\texttt{force\_graph}
  & \texttt{source}, \texttt{target};\newline \textcolor{gray}{\texttt{value}, \texttt{color}: optional}
  & \texttt{node\_radius}, \texttt{link\_distance}, \texttt{link\_strength}, \texttt{charge\_strength}, \texttt{stroke\_width}, \texttt{stroke\_opacity}, \texttt{fill\_color}, \texttt{stroke\_color}, \texttt{color\_scheme}
  & Node/link sliders; path source/target fields\textsuperscript{$\dagger$}; Categories if \texttt{color}-encoded; click-to-select\textsuperscript{$\dagger$} \\

\addlinespace\midrule\addlinespace

\texttt{choropleth}
  & \texttt{region}, \texttt{value};\newline \textcolor{gray}{\texttt{color}: value field}
  & \texttt{color\_scheme}, \texttt{stroke\_color}, \texttt{stroke\_width}, \texttt{projection}
  & Color palette per layer \\

\addlinespace\midrule\addlinespace

\shortstack[l]{\texttt{parallel}\\\texttt{coordinates}}
  & \texttt{dimensions}: list of columns;\newline \textcolor{gray}{\texttt{color}: category field}
  & \texttt{stroke\_width}, \texttt{stroke\_opacity}, \texttt{color\_scheme}
  & Fill color / Categories; per-axis brush\textsuperscript{$\dagger$} \\

\bottomrule
\end{tabular}
\end{table*}

\FloatBarrier
\clearpage
\section{Natural Language Interface}
As referenced by Section~\ref{sec:schema}.

\label{app:nli}

\subsection{Schema Structure}
\label{app:schema}

The schema (Fig.~\ref{fig:vis-schema}) is a structured representation used for Natural Language to RaivenDSL generation. It consists of the following components (fields shown in gray in Fig.~\ref{fig:vis-schema} denote conditionally included components, which appear only when required by the data type or visualization):

\begin{itemize}
    \item \textbf{\texttt{task\_summary}} (string): Natural-language description of the visualization goal.

    \item \textbf{\texttt{data}} (object): Mapping from source name to data source definition.
    \begin{itemize}
        \item \texttt{type}: One of \texttt{tbl}, \texttt{img}, \texttt{net}, \texttt{geo}, \texttt{func}
        \item \texttt{path}: File path or URL
        \item \texttt{args}: Optional parameters
        \item \texttt{variables}: List of variable definitions
        \begin{itemize}
            \item \texttt{name}: Field/column name used in encodings
            \item \texttt{data\_type}: Primitive type (e.g., number, string, date)
            \item \texttt{semantic\_type}: \texttt{quantitative} or \texttt{qualitative} (optional)
            \item \texttt{role\_hint}: Optional role (e.g., category, value, id)
        \end{itemize}
        \item \texttt{dimensions}: Dimensions of image/volume data (e.g., VTI); required for volumetric sources and omitted otherwise
    \end{itemize}

    \item \textbf{\texttt{views}} (array): List of view specifications.
    \begin{itemize}
        \item \texttt{view\_id}: Unique view identifier
        \item \texttt{layers}: List of layers
        \begin{itemize}
            \item \texttt{from}: Data source reference
            \item \texttt{geo}: Geographic reference (if applicable)
            \item \texttt{mark}: Visual mark type
            \item \texttt{encode}: Channel-to-field mapping
            \item \texttt{style}: Optional styling parameters
        \end{itemize}
        \item \texttt{links\_out}: Target view IDs for linking (optional)
        \item \texttt{interactions}: View-specific interaction configuration (optional)
    \end{itemize}

    \item \textbf{\texttt{selections}} (array): Interaction definitions.
    \begin{itemize}
        \item \texttt{name}: Selection identifier (optional)
        \item \texttt{type}: \texttt{interval} or \texttt{point} (optional)
        \item \texttt{bind\_view}: View where interaction occurs (optional)
        \item \texttt{bind\_channels}: Channels or fields used in selection (optional)
    \end{itemize}

    \item \textbf{\texttt{linking}} (object): Multi-view coordination.
    \begin{itemize}
        \item \texttt{shared\_data\_source}: Data source used by linked views (optional)
        \item \texttt{linked\_view\_ids}: Views participating in linking (optional)
        \item \texttt{selection\_name}: Selection used for coordination (optional)
        \item \texttt{link\_style}: Linking mode (optional)
    \end{itemize}

    \item \textbf{\texttt{slice\_linking}} (array): Slice-based interaction groups for volume visualizations, included only when slice interactions are present.
    \begin{itemize}
        \item \texttt{linked\_view\_ids}: Views participating in slice linking
        \item \texttt{axes}: Slice orientation (e.g., \texttt{XY}, \texttt{XZ}, \texttt{YZ}, or oblique)
        \item \texttt{slice\_link\_id}: Identifier for synchronizing slice position across views
        \item \texttt{tf\_link\_id}: Identifier linking slice interaction to transfer-function interaction
    \end{itemize}

    \item \textbf{\texttt{tf\_linking}} (array): Transfer-function linking groups, included only when transfer-function interactions are present.
    \begin{itemize}
        \item \texttt{linked\_view\_ids}: Views participating in transfer-function linking
        \item \texttt{tf\_link\_id}: Identifier for shared transfer-function interaction
    \end{itemize}
\end{itemize}

\lstdefinestyle{jsonschema}{
  basicstyle=\ttfamily\scriptsize,
  frame=none,
  columns=fullflexible,
  keepspaces=true,
  showstringspaces=false,
  breaklines=true,
  moredelim=[is][\bfseries]{@@}{@@},
  moredelim=[is][\color{gray}]{~~}{~~}
}

\phantomsection
\label{fig:vis-schema}

\begin{lstlisting}[style=jsonschema]
{
  "@@task_summary@@": "",
  "@@data@@": {
    "<source_name>": {
      "type": "tbl|img|net|geo|func",
      "path": "",
      "args": {},
      "variables": [
        {
          "name": "",
          "data_type": "",
          ~~"semantic_type": "quantitative|qualitative",~~
          ~~"role_hint": ""~~
        }
      ],
      ~~"dimensions": [x, y, z]~~
    }
  },
  "@@views@@": [
    {
      "view_id": "",
      "layers": [
        {
          "from": "",
          "geo": "",
          "mark": "",
          "encode": {},
          "style": {}
        }
      ],
      "links_out": [],
      "interactions": {}
    }
  ],
  @@"selections"@@: [
    {
      ~~"name": "",~~
      ~~"type": "interval|point",~~
      ~~"bind_view": "",~~
      ~~"bind_channels": []~~
    }
  ],
  @@"linking"@@: {
    ~~"shared_data_source": "",~~
    ~~"linked_view_ids": [],~~
    ~~"selection_name": "",~~
    ~~"link_style": "views"~~
  },
  ~~"slice_linking": [
    {
      "linked_view_ids": [],
      "axes": "XY|list of XY/XZ/YZ/oblique",
      "slice_link_id": "",
      "tf_link_id": ""
    }
  ],~~
  ~~"tf_linking": [
    {
      "linked_view_ids": [],
      "tf_link_id": ""
    }
  ]~~
}
\end{lstlisting}

\noindent\textbf{Validation and Progression:}
The schema is constructed incrementally through a sequence of nodes. Each node must satisfy structural and semantic constraints before the workflow advances. Validation enforces correct data references, valid mark types, compatibility between variables and encoding channels, and completeness of required fields. If any constraint fails, the schema is not updated and the workflow remains on the current node, prompting for clarification until the requirements are satisfied.

\begin{enumerate}
    \item \textbf{Task Definition}
\begin{itemize}
    \item Requirement: Non-empty textual summary derived from the user's initial description.
    \item Behavior: Generates \texttt{task\_summary} by summarizing the user's first message; always advances with no validation constraints.
\end{itemize}

    \item \textbf{Data}
    \begin{itemize}
        \item Requirement: Valid data sources with parsed variables.
        \item Behavior: Enforced at session start through file upload. The workflow cannot begin unless a data file is successfully uploaded and parsed; unsupported or invalid files prevent initialization.
    \end{itemize}

    \item \textbf{View \& Layer}
\begin{itemize}
    \item Requirements:
    \begin{itemize}
        \item Each view must define a unique \texttt{view\_id}.
        \item Each layer must reference a valid data source (\texttt{from $\in$ data}).
        \item When only one data source is present, all layers must resolve to that source.
    \end{itemize}
    \item Behavior: 
    \begin{itemize}
        \item Uniqueness of \texttt{view\_id} and single-source assignment are enforced programmatically.
        \item Missing or invalid layer-to-data assignments trigger re-prompting, requiring the user to explicitly specify the data source for each layer.
    \end{itemize}
\end{itemize}

    \item \textbf{Mark}
\begin{itemize}
    \item Requirement: Each layer must specify a valid mark from the RaivenDSL mark set.
    \item Behavior: If no visualization type is specified or inferable for all layers, the user is re-prompted before proceeding; invalid marks block progression and require correction.
\end{itemize}

    \item \textbf{Encode}
    \begin{itemize}
        \item Requirements:
        \begin{itemize}
            \item Encoded variables must exist in the associated data source.
            \item Variable types must be compatible with the selected mark and channel.
        \end{itemize}
        \item Behavior: Non-existent or incompatible variables trigger re-prompting, and the workflow remains at this step until all required channels are assigned valid, compatible variables.
    \end{itemize}

    \item \textbf{Selections \& Linking}
    \begin{itemize}
        \item Requirement: None explicitly enforced.
        \item Behavior: Inferred programmatically from the user description; always completes.
    \end{itemize}
\end{enumerate}

\noindent\textbf{LLM Output Requirements:}
For LLM-driven nodes, progression additionally requires:
\begin{itemize}
    \item Valid JSON output conforming to the expected schema structure
    \item A signal that the provided information is sufficient (e.g., ``Enough'')
\end{itemize}

\noindent If these conditions are not met, the schema is not updated and the workflow remains at the current node until valid output is produced.

\subsection{Prompt Templates}
\label{app:prompts}

The following prompts define the LLM interactions at each stage of the schema construction pipeline. Each node uses a task-specific prompt to elicit structured outputs that progressively populate the schema.

\subsubsection{Task Summary Node Prompt}

\noindent\footnotesize\{user\} denotes the user’s natural language input describing the visualization task.

\newcommand{\fullprompt}[1]{%
  \noindent\rule{\linewidth}{0.4pt}
  \begin{quote}\ttfamily\small #1\end{quote}
  \noindent\rule{\linewidth}{0.4pt}}

\noindent\rule{\linewidth}{0.4pt}
\begin{quote}\small

You are given a user's description of a visualization task. Summarize in a short paragraph what this visualization is mainly about.

\medskip

\textbf{Output format:}
\begin{itemize}
    \item Return a markdown block containing the summary:
\end{itemize}

\begin{lstlisting}[basicstyle=\ttfamily\scriptsize]
```markdown
<your answer>
```
\end{lstlisting}

\textbf{Input:}
\begin{itemize}
\item User description: {user}
\end{itemize}

\end{quote}
\noindent\rule{\linewidth}{0.4pt}

\normalsize
\subsubsection{Data Node Prompt}
\noindent\footnotesize\{user\} denotes the user’s natural language input describing the visualization task.

\noindent\rule{\linewidth}{0.4pt}
\begin{quote}\small

You are an assistant that helps structure user descriptions into a predefined schema.  
The current task is about the data block (data sources for the visualization).

\medskip

\textbf{Schema:}
\begin{lstlisting}[basicstyle=\ttfamily\scriptsize]
{
  "data": {
    "source_name": {
      "type": "",
      "path": "",
      "args": {},
      "variables": []
    },
    ...
  }
}
\end{lstlisting}

\textbf{Data source types:}
\begin{itemize}
    \item tbl (table/CSV)
    \item img (image/volume)
    \item net (network)
    \item geo (geographic)
    \item func (computed)
\end{itemize}

\textbf{Variable extraction (tbl):}
\begin{itemize}
    \item When the user mentions variables, add each to \texttt{variables}.
    \item Use the exact names provided by the user.
    \item Do not introduce new variable names.
    \item Leave \texttt{variables = []} if none are specified.
\end{itemize}

\textbf{Inference rule (volume / slice / isosurface):}
\begin{itemize}
    \item If the user requests volume, slice, isosurface, triplanar, or oblique slice:
    \begin{itemize}
        \item set \texttt{type = img}
        \item set \texttt{path} to the provided file/URL
        \item assign a short \texttt{source\_name}
    \end{itemize}
    \item Return \texttt{Enough}; do not ask for missing type/path.
\end{itemize}

\textbf{Instructions:}
\begin{enumerate}
    \item Read the user's description: \{user\}
    
    \item Fill the schema:
    \begin{itemize}
        \item \texttt{source\_name}: short identifier
        \item \texttt{type}: one of tbl, img, net, geo, func
        \item \texttt{path}: exact file path or URL
        \item \texttt{args}: optional arguments (used for func)
    \end{itemize}
    
    \item Insert any provided file path or URL into \texttt{path}, then decide Enough vs Not Enough.
    
    \item Provide two outputs:
    \begin{itemize}
        \item Schema output (JSON)
        \item Feedback (Markdown)
        \begin{itemize}
            \item If complete: first line \texttt{Enough}
            \item If incomplete: first line \texttt{Not Enough}, followed by a specific request
        \end{itemize}
    \end{itemize}
\end{enumerate}

Ensure both outputs are always produced.

\end{quote} \noindent\rule{\linewidth}{0.4pt}

\normalsize
\subsubsection{View \& Layer Node Prompt}
\noindent\footnotesize\{user\}, \{data\_ref\}, and \{data\_ref\_with\_types\} denote runtime-injected values from the current schema and user input.

\nopagebreak

\noindent\rule{\linewidth}{0.4pt}
\begin{quote}\small

You are an assistant that helps structure user descriptions into a predefined schema.
The current task is about \textbf{view(s) and layer} — which view(s) to create and which data source each uses.

\medskip

\textbf{Available data sources (name and type):} {data\_ref\_with\_types}

\medskip

\textbf{You may return either:}

\begin{itemize}
\item \textbf{A) Single view} (when the user asks for one chart):
\end{itemize}

\begin{lstlisting}[basicstyle=\ttfamily\scriptsize]
{
"view_name": "",
"layer_from": "",
"geo": ""
}
\end{lstlisting}

\begin{itemize}
\item \textbf{B) Multiple views} (when the user asks for more than one chart, or one view with multiple layers):
\end{itemize}

\begin{lstlisting}[basicstyle=\ttfamily\scriptsize, breaklines=true]
{
"views": [
{ "view_id": "", "layer_from": "", "geo": "" },
{ "view_id": "", "layers": [ { "layer_from": "", "geo": "" }, { "layer_from": "", "geo": "" } ] },
...
]
}
\end{lstlisting}

\textbf{Field rules:}
\begin{itemize}
\item \texttt{view\_id} / \texttt{view\_name}: short identifier for the view. Only use \texttt{volume\_triplanar} when the user explicitly asked for triplanar slices, three slices, or XY/XZ/YZ slices. If the user said only \textit{a slice}, \textit{volume with a slice}, or \textit{volume and a slice}, use \texttt{volume\_slice} or \texttt{combined}, not \texttt{volume\_triplanar}.
\item \texttt{layer\_from}: the name of a data source --- use the exact name from the list above (e.g., {data\_ref}). Do not use generic names like \texttt{data} or \texttt{sample}. When there are multiple sources of the same type (e.g., two \texttt{img}), set \texttt{layer\_from} only for the view(s) the user explicitly specified. If the user said \textit{volume with a slice from vtk and a streamline}, set \texttt{vtk} only for the volume/slice view and leave the streamline source empty.
\item \texttt{layers}: use when one view contains multiple layers (e.g., volume and slice in the same view). Each layer may use the same data source. Omit \texttt{layer\_from} at the view level when using \texttt{layers}.
\item \texttt{geo}: optional; for geographic layers only. Leave empty if not applicable.
\end{itemize}

\textbf{Inference rules:}
\begin{itemize}
\item \textbf{Single data source:} when there is only one data source ({data\_ref}), use it for \texttt{layer\_from} for every view and return \texttt{Enough}.
\item \textbf{One img and one tbl:} when there is exactly one \texttt{img} and one \texttt{tbl}, and the user asked for both a volume/slice/isosurface view and a chart view, infer \texttt{layer\_from}: assign the \texttt{img} source to volume-like view(s) and the \texttt{tbl} source to chart view(s). Return \texttt{Enough}.
\item \textbf{Multiple data sources:} only output \texttt{Not Enough} when there are multiple sources of the same type and it is unclear which source each view/layer should use.
\end{itemize}

\textbf{Volume + slice(s): combined vs.\ separate views}
\begin{itemize}
\item \textbf{``With'' means one view:} phrases such as \textit{volume with streamline}, \textit{volume with slice}, or more generally \textit{X with Y} mean one view with two layers from the same data source.
\item \textbf{Layered means one view:} if the user says \textit{layered}, return one view with multiple layers. Examples include layered volumes, layered streamlines, or mixed layered volumes and streamlines.
\item \textbf{Multichannel volume:} if the user says \textit{multichannel volume} or \textit{volume using layers A, B, C}, return one view with one layer per named field, all from the same source.
\item \textbf{Volume + triplanar + streamline:} if the user explicitly asks for triplanar or three slices and a separate streamline, return two views:
\begin{itemize}
\item \texttt{volume\_triplanar} with volume + one slice layer
\item \texttt{streamline}
\end{itemize}
\item \textbf{Volume + slice(s) + streamline (separate):} if volume+slice(s) and streamline are listed separately, return two views:
\begin{itemize}
\item one combined view with volume + one slice layer
\item one streamline view
\end{itemize}
\item \textbf{Combined volume + slice(s):} if the user says volume with slice(s) only, return one view with two layers:
\begin{itemize}
\item first layer = volume (or isosurface)
\item second layer = slice
\end{itemize}
Use \texttt{volume\_triplanar} only if triplanar is explicitly requested; otherwise use \texttt{volume\_slice} or \texttt{combined}. Even if the user says \textit{two slices} or \textit{N slices}, still use one slice layer; slice planes are chosen later in the encode step.
\item \textbf{Combined with triplanar + oblique:} if the user says volume with xy, yz, xz, and oblique slice, return one combined view with three layers:
\begin{itemize}
\item volume
\item slice for triplanar
\item slice for oblique
\end{itemize}
If the user also asks for separate slice views, add four additional slice views: \texttt{slice\_xy}, \texttt{slice\_oblique}, \texttt{slice\_xz}, \texttt{slice\_yz}.
\item \textbf{Separate views:} if the user lists volume and slice(s) as separate items (without \textit{with} or \textit{layered}), return separate views, one per item.
\end{itemize}

\textbf{General rules:}
\begin{itemize}
\item If the user describes two or more distinct charts/views, return multiple views.
\item Treat \textit{interactive scatterplot} and \textit{interactive chart} as meaning a brushable scatterplot: one view only.
\item Only when the user explicitly asks for linked views should multiple linked views be returned.
\item If the user wants only multiple slice planes and no volume, use one view with one slice layer; slice axes are set later.
\item A \textit{linked} slice is a separate slice view using the same data source as the linked source view.
\end{itemize}

\textbf{Map / choropleth rules:}
\begin{itemize}
\item If the user wants a map and both a GeoJSON and CSV source are available:
\begin{itemize}
\item set \texttt{layer\_from} to the table (CSV)
\item set \texttt{geo} to the geographic (GeoJSON) source
\end{itemize}
\item \textbf{Map always includes a choropleth layer:}
\begin{itemize}
\item \textit{map} only $\rightarrow$ one choropleth layer
\item \textit{map with hexbin}, \textit{map with bubble}, or \textit{map with points} $\rightarrow$ choropleth + overlay layer
\item \textit{map with hexbin and bubble} $\rightarrow$ choropleth + hexbin + bubble
\end{itemize}
\end{itemize}

\textbf{Histogram with KDE:}
\begin{itemize}
\item If the user says \textit{histogram with kde}, \textit{histogram and kde}, \textit{kde over histogram}, or similar, return one view with two layers:
\begin{itemize}
\item histogram
\item kde
\end{itemize}
Both layers use the same table source.
\end{itemize}

\textbf{Instructions:}
\begin{enumerate}
\item Read the user's description: {user}. If it says \textit{interactive scatterplot} or \textit{interactive chart}, treat that as a brushable scatterplot: one view only, no second view.
\item If the user asked for multiple views/charts, fill \texttt{views} with one object per view. If the user asked for one chart, fill \texttt{view\_name}, \texttt{layer\_from}, and \texttt{geo}.
\item When there are multiple data sources:
\begin{itemize}
\item if there is one \texttt{img} and one \texttt{tbl} and the user asked for both volume/slice and a chart type, infer \texttt{layer\_from} and return \texttt{Enough}
\item only when there is real ambiguity (e.g., two \texttt{tbl} or two \texttt{img} sources) output \texttt{Not Enough} and ask which dataset each view/layer uses
\end{itemize}
\item Provide two outputs:
\begin{itemize}
\item \textbf{Schema output} (inside \texttt{json ...}): set \texttt{layer\_from} only when (a) there is one data source, or (b) the user explicitly said which dataset a view uses. If there are two or more sources of the same type and the user specified only one dataset, leave the others empty.
\item \textbf{Feedback} (inside \texttt{markdown ...}): use exact dataset names (e.g., {data\_ref}). First line \texttt{Enough} when at least one view is defined and each \texttt{layer\_from} matches a data source; otherwise \texttt{Not Enough} with a specific clarification request.
\end{itemize}
    \end{enumerate}

Make sure to always output both parts.

\end{quote} \noindent\rule{\linewidth}{0.4pt}

\normalsize
\subsubsection{Mark Node Prompt}
\noindent\footnotesize\{user\}, \{schema\_desc\}, and \{MARK\_TYPES\_STR\} denote runtime-injected values from the current schema and user input.

\nopagebreak

\noindent\rule{\linewidth}{0.4pt}
\begin{quote}\small

You are an assistant that helps structure user descriptions into a predefined schema.
The current task is about the \textbf{mark type(s)} (chart/graph type) for the visualization.

\medskip

\textbf{Valid mark types (you MUST use only these):}
{MARK\_TYPES\_STR}

\medskip

\textbf{Aliases:}
\begin{itemize}
\item If the user says \textbf{map} (e.g. \textit{I want a map}, \textit{create a map}), use mark type \textbf{choropleth}.
\item \textbf{Map with overlay:} when the user said \textbf{map with hexbin} (or \textit{map with bubble}, \textit{map with points}) and the schema has \textbf{one view with two layers}, return \texttt{view\_marks} \textbf{[ ["choropleth", "hexbin"] ]} (or \textbf{["choropleth", "bubble"]} / \textbf{["choropleth", "points"]}) --- the first layer must be \textbf{choropleth}; do not return only \textbf{["hexbin"]} or put geo on hexbin.
\item When the user said \textbf{map with hexbin and bubble} and the schema has one view with \textbf{three} layers, return \textbf{["choropleth", "hexbin", "bubble"]}.
\item If the user says \textbf{bubble} or \textbf{bubble chart}, use mark type \textbf{bubble} (x, y, and size required; color optional).
\item If the user says \textbf{time series}, \textbf{timeseries}, \textbf{time series plot}, or similar (e.g. \textit{show over time}, \textit{trend over time}), use mark type \textbf{line} (line plot).
\item For an \textbf{oblique slice} (single plane at an angle), use mark \textbf{slice}; axes will be set to \texttt{"oblique"}.
\item For \textbf{triplanar}, \textbf{two slices}, \textbf{three slices}, or \textbf{multiple slice planes} in one view, use a \textbf{single} mark \textbf{slice} for that view --- one layer, not multiple. The axes (e.g. \textbf{["XY", "YZ"]} or \textbf{["XY", "XZ", "YZ"]}) are set in the encode step; do not create multiple slice layers for \textit{two slices} or \textit{three slices}.
\item \textbf{Volume with triplanar + oblique in one view:} when the schema has one view with \textbf{three} layers (volume, slice, slice), use \texttt{view\_marks} so that view gets \textbf{["volume", "slice", "slice"]} --- first slice layer = triplanar (XY, XZ, YZ), second slice layer = oblique.
\item \textbf{Histogram with KDE:} when the user said \textbf{histogram with kde}, \textbf{histogram and kde}, \textbf{kde over histogram}, or \textbf{histogram with density} and the schema has \textbf{one view with two layers}, return \texttt{view\_marks} \textbf{[ ["histogram", "kde"] ]} --- first layer \textbf{histogram}, second layer \textbf{kde} (KDE curve overlaid on the same variable as the histogram).
\end{itemize}

\textbf{Do not assume a chart type.} If the user did not specify what kind of visualization they want (e.g. they only said \textit{show variable b} or \textit{visualization of x} without saying histogram, scatterplot, heatmap, bar chart, etc.), output \texttt{{ "mark": "" }} (or empty \texttt{"marks"} array) and in Feedback say \texttt{Not Enough} and ask them to specify the type of visualization, listing the valid mark types above.

\medskip

\textbf{The schema is defined as follows:}
{schema\_desc}

\medskip

\textbf{Instructions:}
\begin{enumerate}
\item Read the user's description: \{user\}

\item Only if the user clearly specified a chart/visualization type (e.g. histogram, scatter, heatmap, bar, line, pie, volume, slice), choose the matching mark type(s) from the list above.

\begin{itemize}
    \item \textbf{Layered (one view, multiple layers):} When the schema has one view with multiple layers from \textit{layered} (e.g. two volume layers + one streamline layer), return one mark per layer in order (e.g. \texttt{view\_marks: [ ["volume", "volume", "streamline"] ]}).

    \item \textbf{Multichannel volume:} When the user specifies multiple fields (e.g. \textit{volume using layers X, Y, Z}), return one volume mark per layer (e.g. \texttt{["volume", "volume", ...]}).

    \item \textbf{Volume + slice(s):} When one view contains volume and slice, return two marks: first \texttt{volume} (or \texttt{isosurface}), second \texttt{slice}.

    \item \textbf{"Two slices" or "three slices":} Still use one slice layer (e.g. \texttt{["volume", "slice"]}), not multiple slice layers.

    \item \textbf{Three-layer case:} If the schema has volume + triplanar slice + oblique slice, return \texttt{["volume", "slice", "slice"]}.

    \item \textbf{Separate views:} If volume and slices are in separate views, return one mark per view (e.g. \texttt{["volume"], ["slice"], ["slice"]}).

    \item \textbf{Triplanar in one view:} Use a single \texttt{slice} mark; do not return multiple slice marks.

    \item \textbf{Linked slice:} A linked slice is a separate view with mark \texttt{"slice"}.

    \item \textbf{Single-layer view:} If a view has one layer, return \texttt{"mark"} as a string.
\end{itemize}

\item Provide \textbf{two outputs}:
\begin{itemize}
    \item \textbf{Schema output} (inside \texttt{json ...})
    \item \textbf{Feedback} (inside \texttt{markdown ...})
    \begin{itemize}
        \item First line \texttt{Enough} when valid marks are provided
        \item Otherwise \texttt{Not Enough} with a request to specify the visualization type
    \end{itemize}
\end{itemize}

\end{enumerate}

Make sure to always output both parts.

\end{quote} \noindent\rule{\linewidth}{0.4pt}

\normalsize
\subsubsection{Encode Node Prompt}
\noindent\footnotesize
\{user\_intent\_block\}, \{data\_info\}, and \{refinement\_context\_block\} denote runtime-injected values from the current schema and user input. \\
\{no\_encode\_marks\} specifies mark types that do not require encoding.

\noindent\rule{\linewidth}{0.4pt}
\begin{quote}\small

You are an assistant that helps structure user descriptions into a predefined schema.
The current task is about the \textbf{encode} block — which variables (fields) the user wants on which visual channels.

\medskip

\textbf{Barrier rules:}
\begin{itemize}
\item You may ONLY use variable names from the allowed list above. If the list is empty, output \texttt{{ "encode": {} }} and say \texttt{Not Enough}.
\item Do not infer or guess encode from the data. Only fill encode when the user has explicitly said which variables to use (e.g. \textit{use c and h}, \textit{x is sales, y is revenue}, \textit{plot column A vs B}).
\item If the user only described the chart type (e.g. \textit{I want a scatterplot}) without naming which columns/variables to use, output \texttt{{ "encode": {} }} and say \texttt{Not Enough}.
\item In your JSON output, include ONLY encode keys for channels the user explicitly assigned.
\item Example: if the user said \textit{x is sales, y is revenue} and \textit{sales}, \textit{revenue} are allowed, output \texttt{{ "encode": { "x": "sales", "y": "revenue" } }}. Do not add \texttt{color} or any other key they did not specify.
\end{itemize}

\textbf{User intent block (when provided):}
{user\_intent\_block}

\medskip

\textbf{Context:}
{data\_info}

\medskip

\textbf{Refinement context (when provided):}
{refinement\_context\_block}

\medskip

\textbf{Exception — encode step can be bypassed for these marks:}
{no\_encode\_marks}

These display the whole dataset and do not map table columns to x/y/color. For these mark types, output \texttt{Enough} with an empty encode object.

\medskip

\textbf{Output format:}
\begin{itemize}
\item Single view with one layer needing encode:
\texttt{{ "encode": { <only channels user assigned> } }}
\item Single view with multiple layers needing encode, or multiple views:
\texttt{{ "encodes": [ [ \{ ... \}, \{ ... \} ], ... ] }}
\begin{itemize}
\item one element per view
\item each element is an array of encode objects
\item one encode object per layer that needs encode, in layer order
\end{itemize}
\item Example: one view with 2 histogram layers 
→
→ \texttt{"encodes": [ [ \{ "x": "s" \}, \{ "x": "t" \} ] ]}
\item For a view whose layers are all {no\_encode\_marks}, use \texttt{[ \{\} ]}.
\end{itemize}

\textbf{Instructions:}
\begin{enumerate}
\item If single view and mark type is one of {no\_encode\_marks}, output empty encode and first line \texttt{Enough}.
\item If multiple views: for each view with mark in {no\_encode\_marks}, use \texttt{{}} for that view's encode. Only request or fill encode for views that need variable mapping (points, bar, line, etc.).
\item If allowed variable names is empty and at least one view needs encode, output empty encodes for those views and first line \texttt{Not Enough}; ask the user to specify which columns/variables to use (only for the view(s) that need encode, not for volume/slice/isosurface views).
\item If the user did not specify which variable goes on which channel for a view that needs encode, do not guess and do not pick from the allowed list — output empty encode for that view and say \texttt{Not Enough}, asking the user to choose which variables to use (only for that view).
\item \textbf{Underspecification:} If a view needs multiple variables (e.g. scatterplot/points needs x and y) but the user only specified one variable for that view, output empty encode for that view and say \texttt{Not Enough}, naming which view is underspecified, what is required (e.g. x, y), and what is optional for this mark if any (e.g. color). Example: \textit{Scatterplot view: required encodings are x and y; optional encodings are color. You specified one variable. Please specify the other required variable(s).}
\item Use only the required and optional channels listed in Context for this mark. Include optional channels (e.g. color, size, opacity for scatterplot) only if the user explicitly assigned them.
\item Provide \textbf{two outputs}: Schema (inside \texttt{json ...}) and Feedback (inside \texttt{markdown ...}). First line of feedback: \texttt{Enough} when every view that needs encode has all required channels filled with names from the allowed list (and volume/slice/isosurface views have empty encode); otherwise \texttt{Not Enough}.
\end{enumerate}

Make sure to always output both parts.

\end{quote} \noindent\rule{\linewidth}{0.4pt}

\normalsize
\subsubsection{Selections \& Linking Node Prompt}
\noindent\footnotesize\{user\}, \{view\_ids\}, and \{data\_sources\} denote runtime-injected values from the current schema and user input.

\nopagebreak
\normalfont

\noindent\rule{\linewidth}{0.4pt}
\begin{quote}\small

You are an assistant that helps structure user descriptions into a predefined schema.
The current task is about \textbf{selections and linking} — whether the user wants to link views so that selecting in one view updates another.

\medskip

\textbf{View IDs from previous step:} {", ".join(view\_ids) or "(none)"}
\textbf{Data sources:} {", ".join(data\_sources) or "(none)"}

\medskip

\textbf{Interpretation:} Treat \textbf{interactive scatterplot} and \textbf{interactive chart} as meaning \textbf{brushable scatterplot} (one brushable view).

\medskip

If the user asked to \textbf{link} views, \textbf{brush/select}, or an \textbf{interactive} (brushable) chart so that selecting in one view updates another, fill in the schema below. Use the exact \texttt{view\_id} names and the exact field names the user said for \texttt{bind\_channels}.

\medskip

\textbf{Linking works by data rows, not by variables:} Linked views can use different axes (e.g. one scatter a vs b, another scatter c vs d) as long as they use the same data source. The selection identifies rows; the other view highlights/filters those same rows even if it plots different columns. So \texttt{shared\_data\_source} is the key; \texttt{bind\_channels} are only for the view where the user brushes (\texttt{bind\_view}).

\medskip

If the user did NOT ask for linking, brushing, or an interactive/brushable chart, return:

\begin{lstlisting}[basicstyle=\ttfamily\scriptsize,breaklines=true]
{ "selections": [], "linking": {} }
\end{lstlisting}

\textbf{Schema when linking is requested:}

\begin{lstlisting}[basicstyle=\ttfamily\scriptsize,breaklines=true]
{
"selections": [
{
"name": "<selection_name>",
"type": "interval",
"bind_view": "<view_id where user brushes>",
"bind_channels": ["<x_field>", "<y_field>"]
}
],
"linking": {
"shared_data_source": "<data source name used by all linked views>",
"linked_view_ids": ["<view_id1>", "<view_id2>"],
"selection_name": "<same as selections[0].name>"
}
}
\end{lstlisting}

\textbf{Instructions:}
\begin{enumerate}
\item Read the user's description: {user}

\item If they want linked views with brush/selection: fill \texttt{selections} (name, type \texttt{"interval"}, \texttt{bind\_view}, \texttt{bind\_channels} for that view only) and \texttt{linking} (\texttt{shared\_data\_source}, \texttt{linked\_view\_ids}, \texttt{selection\_name}). Linked views may use different encode channels (e.g. scatter a vs b and scatter c vs d from the same table) — same data source is sufficient. Use exact \texttt{view\_id} and field names from the task.

\item If they did not ask for linking: return \texttt{\{ "selections": [], "linking": \{\} \}} and first line \texttt{Enough}.

\item Provide \textbf{two outputs}:
\begin{itemize}
    \item \textbf{Schema output} (inside \texttt{json ...})
    \item \textbf{Feedback} (inside \texttt{markdown ...}). First line \texttt{Enough} in both cases.
\end{itemize}

\end{enumerate}

Make sure to always output both parts.

\end{quote} \noindent\rule{\linewidth}{0.4pt}

\subsection{Agent Clarification Messages}
\label{app:clarification-messages}

When Raiven cannot proceed with a workflow step due to missing or 
ambiguous information, it pauses and returns a clarification message 
to the user rather than making an uninformed assumption. Below we 
document what a user will see at each stage of the pipeline when 
clarification is needed.

\medskip
\noindent\textbf{Task Definition.} This stage always completes 
successfully and never requests clarification from the user.

\medskip
\noindent\textbf{Data Block.} If the dataset to use is unclear, 
the user sees:
\begin{quote}
{\ttfamily\small Please specify the dataset (file path or URL) 
you want to use.}
\end{quote}

\medskip
\noindent\textbf{View \& Layer.} If the user has multiple datasets 
and has not specified which dataset each view should use:
\begin{quote}
{\ttfamily\small You have multiple datasets: [names]. Each view 
uses exactly one dataset; multiple views can share the same 
dataset. Please specify which dataset each view uses.}
\end{quote}
With a single dataset:
\begin{quote}
{\ttfamily\small Please specify which dataset this view should 
use (or multiple views can share the same dataset).}
\end{quote}
If ambiguity persists after the LLM has inferred views, Raiven 
provides a structured per-view breakdown:
\begin{quote}
{\ttfamily\small You have multiple image data sources: [names].\newline
Which dataset should each view use?\newline
- View 'spatial' (volume + triplanar slices): ct\_1.vti}
\end{quote}

\medskip
\noindent\textbf{Mark.} If the user has not specified a 
visualization type:
\begin{quote}
{\ttfamily\small You must specify the type of visualization 
(chart type) you want. Supported types: [supported types]. 
For example: ``I want a histogram of b'', ``make a scatterplot 
of x and y'', ``show a heatmap of a, b, and c''.}
\end{quote}
If a chart type was specified but is not supported:
\begin{quote}
{\ttfamily\small [LLM explanation]. The mark type must be one 
of: [supported types].}
\end{quote}

\medskip
\noindent\textbf{Encode.} The encoding stage produces the most 
detailed clarification messages, as it must resolve all 
variable-to-channel assignments. If the mark and data 
combination is invalid:
\begin{quote}
{\ttfamily\small The variable encoding underspecified check is 
not satisfied.\newline
[per-view error details]}
\end{quote}
or:
\begin{quote}
{\ttfamily\small This combination isn't allowed:\newline
[per-view error details]}
\end{quote}
When encoding information is incomplete, Raiven returns a 
structured per-view breakdown of what has been specified and 
what is still needed:
\begin{quote}
{\ttfamily\small I need a few more details to complete.\newline
View 'main' (line): Already specified: y (all numerical 
variables). Still needed: x (numerical or categorical). 
Optionally: color (categorical). Please specify the x 
variable (e.g.\ 'along d', 'plot them along x', or 'x is d').}
\end{quote}
Available variable names from the loaded dataset are listed 
directly below each view block. For choropleth views, 
additional guidance on providing a matching GeoJSON boundary 
file is included.

\medskip
\noindent\textbf{Selections \& Linking.} This stage always 
completes successfully and never requests clarification from 
the user, as linking is inferred programmatically from the 
visualization structure.

\subsection{NL to Schema to RaivenDSL}
\label{app:nli-example}

We illustrate the end-to-end pipeline from natural language input to schema construction and final RaivenDSL generation. Given the datasets \texttt{stats.csv} and \texttt{tg9.vti}, the user provides the following request:

\begin{quote}
"I want to see a volume rendering of vorticity layered with streamlines using ux, uy, and uz. Additionally, I want to see a histogram of vorticity."
\end{quote}

This input corresponds to the visualization shown in Fig.~\ref{fig:teaser}. It is first translated into the intermediate schema representation (Fig.~\ref{fig:example-schema}), which captures the data sources, view structure, and encodings. The corresponding RaivenDSL program generated from this schema is shown in Listing~\ref{lst:brace}.

\phantomsection
\label{fig:example-schema}

\begin{lstlisting}[style=jsonschema]
{
  "@@task_summary@@": "The visualization focuses on exploring fluid dynamics by displaying a volume rendering of vorticity. It also includes a histogram of vorticity values to show the distribution and magnitude of rotational intensity within the dataset.",

  "@@data@@": {
    "stats.csv": {
      "type": "tbl",
      "path": "/var/folders/n4/vzm78dy505g2qkxwbtf2bh780000gp/T/vis_upload_ef_a9qzq/stats.csv",
      "args": { "format": "csv" },
      "variables": [
        { "name": "ux", "data_type": "number" },
        { "name": "uy", "data_type": "number" },
        { "name": "uz", "data_type": "number" },
        { "name": "vorticity", "data_type": "number" },
        { "name": "pp", "data_type": "number" },
        { "name": "critq", "data_type": "number" }
      ]
    },
    "tg9.vti": {
      "type": "img",
      "path": "/var/folders/n4/vzm78dy505g2qkxwbtf2bh780000gp/T/vis_upload_ef_a9qzq/tg9.vti",
      "args": { "format": "vti" },
      "variables": [
        { "name": "critq", "data_type": "number" },
        { "name": "pp", "data_type": "number" },
        { "name": "ux", "data_type": "number" },
        { "name": "uy", "data_type": "number" },
        { "name": "uz", "data_type": "number" },
        { "name": "vorticity", "data_type": "number" }
      ],
      "dimensions": [65, 65, 65]
    }
  },

  "@@views@@": [
    {
      "view_id": "volume",
      "layers": [
        {
          "from": "tg9.vti",
          "geo": "",
          "mark": "volume",
          "encode": {
            "field": "vorticity"
          },
          "style": {}
        }
      ],
      "links_out": [],
      "interactions": {}
    },
    {
      "view_id": "histogram",
      "layers": [
        {
          "from": "stats.csv",
          "geo": "",
          "mark": "histogram",
          "encode": {
            "x": "vorticity"
          },
          "style": {}
        }
      ],
      "links_out": [],
      "interactions": {}
    }
  ],

  "@@"selections"@@": [],
  "@@"linking"@@": {}
}
\end{lstlisting}

\clearpage
\newpage
\section{Compiler}
\label{app:compiler}
\label{app:compiler}

This section details the two-stage compiler pipeline through the running example from Figure~\ref{fig:teaser}. The compiler translates a parsed RaivenDSL AST into executable visualization code through a verification stage (Section~\ref{app:compiler-verification}) and a realization stage (Section~\ref{app:compiler-realization}).

\subsection{Verification Stage}
\label{app:compiler-verification}

The verification stage transforms the raw AST into a typed \texttt{ProgramSpec}, validating structural constraints independently of any rendering backend. The key transformations are:

\begin{itemize}
    \item Data constructors are resolved to typed kinds: \texttt{img} $\to$ \texttt{ImageData}, \texttt{tbl} $\to$ \texttt{Table}, \texttt{net} $\to$ \texttt{Network}, \texttt{geo} $\to$ \texttt{GeoJSON}.
    \item Each layer receives a unique identifier of the form \texttt{viewId:markType\#index}.
    \item Data source references (\texttt{from}) are validated against the data block.
    \item Mark types are checked against the supported set.
    \item Selection and link declarations are validated for referential integrity.
\end{itemize}

\noindent For the running example, the AST:

\begin{lstlisting}[style=jsonschema]
{
  "@@kind@@": "Program",
  "@@data@@": {
    "vol":    { "ctor": "img", "args": { "path": "taylorgreen_9.vti", "format": "vti" }},
    "sample": { "ctor": "tbl", "args": { "path": "tg9_sample.csv", "format": "csv" }}
  },
  "@@views@@": [
    { "kind": "View", "id": "volume_streamline",
      "layers": [
        { "kind": "Layer", "from": "vol", "mark": "volume",
          "encode": { "field": "vorticity" }},
        { "kind": "Layer", "from": "vol", "mark": "streamline",
          "encode": { "vx": "ux", "vy": "uy", "vz": "uz" }}
      ]},
    { "kind": "View", "id": "histogram",
      "layers": [
        { "kind": "Layer", "from": "sample", "mark": "histogram",
          "encode": { "x": "vorticity" }}
      ]}
  ]
}
\end{lstlisting}

\noindent is lowered to the following \texttt{ProgramSpec}:

\begin{lstlisting}[style=jsonschema]
{
  "@@data@@": {
    "vol":    { "ctor": "DataLoader", ~~"kind": "ImageData"~~,
               "args": { "path": "taylorgreen_9.vti", "format": "vti" }},
    "sample": { "ctor": "DataLoader", ~~"kind": "Table"~~,
               "args": { "path": "tg9_sample.csv", "format": "csv" }}
  },
  "@@views@@": [
    { "id": "volume_streamline",
      "layers": [
        { ~~"id": "volume_streamline:volume#0"~~,
          ~~"kind": "volume"~~, "from": "vol",
          "encode": { "field": "vorticity" }},
        { ~~"id": "volume_streamline:streamline#1"~~,
          ~~"kind": "streamline"~~, "from": "vol",
          "encode": { "vx": "ux", "vy": "uy", "vz": "uz" }}
      ]},
    { "id": "histogram",
      "layers": [
        { ~~"id": "histogram:histogram#0"~~,
          ~~"kind": "histogram"~~, "from": "sample",
          "encode": { "x": "vorticity" }}
      ]}
  ]
}
\end{lstlisting}

\noindent Gray fields highlight the verification stage's contributions: data kinds are resolved from constructors, and each layer receives a unique identifier. The user's specification is otherwise preserved unchanged. At this stage, the compiler has confirmed that all data references resolve, all marks are valid, and all encodings are structurally well-formed---without knowing which backend will render each view.

\subsection{Realization Stage}
\label{app:compiler-realization}

The realization stage assigns a rendering backend to each view, resolves all defaults, and produces the \texttt{RenderIR}: the final, fully specified representation consumed by code generation. The key transformations are:

\begin{itemize}
    \item Backend assignment: views are routed to \texttt{vtkjs} or \texttt{d3} based on their layer mark types. When a program contains views targeting both backends, the top-level backend is set to \texttt{multi}.
    \item Default resolution: color transfer functions, opacity profiles, sample distances, palettes, bin counts, fill colors, and all other style properties are resolved to concrete values.
    \item Control generation: for each mark, the compiler produces the set of runtime-adjustable parameters with their ranges and defaults.
    \item Data URLs are resolved to their runtime paths.
\end{itemize}

\noindent For the running example, the \texttt{ProgramSpec} is realized to:

\begin{lstlisting}[style=jsonschema]
{
  ~~"backend": "multi"~~,
  "@@views@@": [
    {
      "viewId": "volume_streamline",
      ~~"backend": "vtkjs"~~,
      "layers": [
        { "type": "volume",
          "id": "volume_streamline:volume#0",
          "field": "vorticity",
          "url": "/data/teaser/taylorgreen_9.vti",
          ~~"sampleDistance": 0.7~~,
          ~~"range": [0, 28.82]~~,
          ~~"ctf": [{"r":0.27,"g":0.00,"b":0.33,"s":0}, ...]~~,
          ~~"otf": [{"a":0,"s":0}, {"a":0.3,"s":10.09},
                  {"a":0.9,"s":28.82}]~~,
          ~~"_palette": "viridis"~~
        },
        { "type": "streamline",
          "id": "volume_streamline:streamline#1",
          "encode": { "vx": "ux", "vy": "uy", "vz": "uz" },
          "url": "/data/teaser/taylorgreen_9.vti",
          ~~"integrator": { "step": 0.5, "max_steps": 1000 }~~,
          ~~"seedSpec": { "n": 100,
                        "region": { "type": "box" }}~~,
          ~~"style": { "color": null, "tubeRadius": null }~~
        }
      ],
      ~~"controls": {~~
        ~~"sampleDistance": { "min":0.1, "max":2,~~
                           ~~"default":0.7, "step":0.01 },~~
        ~~"palette": "viridis",~~
        ~~"ctf_stops": [ ... ],~~
        ~~"otf_stops": [ ... ],~~
        ~~"layerControls": {~~
          ~~"volume_streamline:streamline#1": {~~
            ~~"streamline": {~~
              ~~"color": "#ffffff",~~
              ~~"count": 100,~~
              ~~"integrationStep": 0.5,~~
              ~~"maxSteps": 1000,~~
              ~~"tubeRadius": 0,~~
              ~~"seedBoxX": { "min":0, "max":65 },~~
              ~~"seedBoxY": { "min":0, "max":65 },~~
              ~~"seedBoxZ": { "min":0, "max":65 }~~
            ~~}~~
          ~~}~~
        ~~}~~
      ~~}~~    
    },
    {
      "viewId": "histogram",
      ~~"backend": "d3"~~,
      "layers": [
        { "type": "histogram",
          "id": "histogram:histogram#0",
          "data": "/data/teaser/tg9_sample.csv",
          "encoding": { "x": "vorticity" }
        }
      ],
      ~~"controls": {~~
        ~~"colors": {~~
          ~~"histogram:histogram#0": {~~
            ~~"bins": 30,~~
            ~~"fill_color": "#1f77b4"~~
          ~~}~~
        ~~},~~
        ~~"palette": "viridis"~~
      ~~}~~
    }
  ]
}
\end{lstlisting}

\noindent Gray fields highlight the realization stage's contributions: backend assignment, resolved defaults, and generated control specifications. The scalar range is extracted from the dataset (\texttt{range: [0, 28.82]} for vorticity) and used to position CTF and OTF stops within the data domain. Seed region control ranges are derived from the image dimensions (\texttt{seedBox: \{min:0, max:65\}} per axis), though the actual seed bounds are resolved at mount time after data loading. The volume layer's two-line DSL specification---a data source and a mark type---has expanded to a full rendering configuration; the streamline layer has acquired integrator and seed parameters; the histogram has received a default bin count and fill color. Bold fields denote user-specified values that pass through unchanged. This is the final representation consumed by the D3 and VTK.js code generators.

\clearpage
\newpage
\section{Benchmark}
\label{app:benchmark}

\subsection{Datasets}
\label{app:datasets}

As referenced by Section~\ref{sec:benchmark}.
The benchmark uses real-world datasets across all three prompt categories. Tables~\ref{tab:datasets-scivis}, \ref{tab:datasets-combined}, and~\ref{tab:datasets-infovis} summarize each dataset's format, dimensions, and source. CSV and JSON files are embedded in full in the model prompt; VTI and GeoJSON files are represented by metadata headers only and provided by URL for runtime loading.

\subsubsection{Scientific Visualization Datasets}

Four volumetric datasets provide the spatial data for SciVis and Combined prompts.

\begin{table}[H]
\centering
\scriptsize
\caption{Scientific visualization datasets (VTI format).}
\label{tab:datasets-scivis}
\begin{tabular}{lll}
\toprule
Dataset & Dimensions & Source \\
\midrule
\texttt{head.vti}
  & $64\!\times\!64\!\times\!93$, 1 var.
  & Kitware\footnotemark \\
\texttt{cells.vti}
  & $256\!\times\!256\!\times\!71$, 6 vars.
  & Allen Inst.\footnotemark \\
\texttt{divcurl\_\{t\}.vti}
  & $2048\!\times\!1024\!\times\!1$, 4 vars., 20 t
  & ORNL \\
\texttt{taylorgreen\_\{t\}.vti}
  & $65\!\times\!65\!\times\!65$, 6 vars., 10 t
  & ORNL \\
\bottomrule
\end{tabular}
\end{table}

\addtocounter{footnote}{-1}
\footnotetext{Kitware, Inc.\ (2016). VTK.js sample data. \url{https://github.com/Kitware/vtk-js}. BSD-3-Clause.}
\stepcounter{footnote}
\footnotetext{Allen Institute for Cell Science (2018). hiPSC Single-cell Image Dataset [AICS-10\_8]. \url{https://allencell.org/3d-cell-viewer}.}

\subsubsection{Combined Datasets}

Combined prompts pair volumetric data with tabular CSV extracts derived from the same sources. Each CSV provides a statistical summary or random sample of the parent volume, enabling InfoVis marks alongside SciVis marks in coordinated dashboards.

\begin{table}[H]
\centering
\scriptsize
\caption{Combined tabular datasets (CSV), derived from the volumetric datasets in Table~\ref{tab:datasets-scivis}.}
\label{tab:datasets-combined}
\begin{tabular}{lrl}
\toprule
File & Rows & Description \\
\midrule
\multicolumn{3}{l}{\textit{From \texttt{head.vti}}} \\
\texttt{head\_sample.csv}        & 15k & Random CT intensities \\
\texttt{head\_region\_stats.csv}  & 27  & $3\!\times\!3\!\times\!3$ block stats \\
\addlinespace
\multicolumn{3}{l}{\textit{From \texttt{divcurl\_\{t\}.vti}}} \\
\texttt{divcurl\_sample.csv}     & 4k  & Raw field values (t\,=\,10) \\
\texttt{divcurl\_stats.csv}      & 20  & Per-timestep summaries \\
\texttt{divcurl\_binned\_2d.csv}  & \raise.17ex\hbox{$\scriptstyle\sim$}2.5k & $50\!\times\!50$ spatial bins \\
\addlinespace
\multicolumn{3}{l}{\textit{From \texttt{taylorgreen\_\{t\}.vti}}} \\
\texttt{tg\_sample.csv}          & 4k  & Raw field values (t\,=\,9) \\
\texttt{tg\_stats.csv}           & 10  & Per-timestep summaries \\
\texttt{tg\_t9\_binned.csv}       & 512 & $8^3$ block averages \\
\addlinespace
\multicolumn{3}{l}{\textit{From \texttt{cells.vti}}} \\
\texttt{cells\_channel\_sample.csv} & 15k & Co-sampled channels \\
\texttt{cells\_seg\_stats.csv}      & 6   & Segmentation stats \\
\texttt{cells\_per\_seg\_sample.csv} & \raise.17ex\hbox{$\scriptstyle\sim$}6k & Per-compartment samples \\
\bottomrule
\end{tabular}
\end{table}

\subsubsection{Information Visualization Datasets}

InfoVis prompts use tabular, network, and geospatial datasets drawn from public sources.

\para{Country statistics.} The primary tabular data source is a merged dataset combining World Bank Development Indicators\footnote{The World Bank: World Development Indicators. \url{https://data.worldbank.org}. CC~BY~4.0.} with UN Population Division median age estimates.\footnote{United Nations, Dept.\ of Economic and Social Affairs, Population Division (2024). World Population Prospects 2024. \url{https://population.un.org/dataportal}. CC~BY~3.0~IGO.} Three tabular views are provided: \texttt{countries\_timeseries.csv} (13{,}760 rows; one row per country per year, 1960--2023), \texttt{countries\_latest.csv} (215 rows; most recent snapshot per country), and aggregated files at continent, UN sub-region, and World Bank trade-region levels. An energy breakdown file (\texttt{countries\_energy.csv}, 645 rows) provides long-format renewable/nuclear/fossil shares. A world life-expectancy aggregate (\texttt{world\_life\_exp.csv}, 64 rows) provides population-weighted global averages from 1960--2023.

\para{Trade flows.} Regional trade data (\texttt{trade\_edges.csv}, 56 rows) reports flows between seven World Bank regions, sourced from the World Integrated Trade Solution (WITS).\footnote{World Bank WITS. \url{https://wits.worldbank.org}. World Bank license.}

\para{Border network.} Land border data (\texttt{borders.csv}, 203 rows; \texttt{border\_edges.csv}, 313 edges) is derived from Wikipedia's list of land borders.\footnote{Wikipedia: List of countries and territories by number of land borders. CC~BY-SA~4.0.} Network JSON files (\texttt{border\_network.json}, \texttt{trade\_network.json}) provide node-link representations for force-directed graph and Sankey visualizations.

\para{Capital cities.} Coordinates and populations for primary capital cities are from SimpleMaps World Cities.\footnote{SimpleMaps. \url{https://simplemaps.com/data/world-cities}. Basic license.}

\para{Geospatial.} GeoJSON boundary files are derived from Natural Earth Admin-0 boundaries,\footnote{Natural Earth. \url{https://naturalearthdata.com}. Public domain.} split into per-continent, per-region, and per-trade-region subsets. A full world boundary file (\texttt{world\_iso3.geojson}) is also provided.

\para{Gaia star catalog.} A random sample of $\sim$74{,}300 sources from ESA Gaia DR3\footnote{ESA Gaia mission. \url{https://www.cosmos.esa.int/web/gaia}. CC~BY-NC~3.0~IGO. This work has made use of data from the European Space Agency (ESA) mission Gaia, processed by the Gaia Data Processing and Analysis Consortium (DPAC).} provides right ascension, declination, G-band magnitude, parallax, and BP--RP colour for scatter plots and sky maps.

\begin{table}[H]
\centering
\scriptsize
\caption{Information visualization datasets.}
\label{tab:datasets-infovis}
\begin{tabular}{llrl}
\toprule
File & Fmt & Rows & Source \\
\midrule
\texttt{countries\_timeseries} & CSV & 13.8k & WB, UN \\
\texttt{countries\_latest}     & CSV & 215   & WB, UN \\
\texttt{world\_life\_exp}       & CSV & 64    & Derived \\
\texttt{countries\_energy}      & CSV & 645   & WB \\
\texttt{*\_latest/ts/energy}    & CSV & var.  & Aggreg. \\
\texttt{borders}               & CSV & 203   & Wikipedia \\
\texttt{border\_edges}          & CSV & 313   & Wikipedia \\
\texttt{trade\_edges}           & CSV & 56    & WITS \\
\texttt{border\_network}        & JSON & 164n & Derived \\
\texttt{trade\_network}         & JSON & 8n   & Derived \\
\texttt{world\_iso3}            & GeoJ & 249f & Nat.\ Earth \\
\texttt{gaia}                   & CSV & 74.3k & ESA Gaia \\
\bottomrule
\end{tabular}
\end{table}

\subsection{Baseline Configuration}
\label{app:baseline-config}

As referenced by Section~\ref{sec:benchmark}.
The four baselines use the model versions listed in Table~\ref{tab:models}.
The three LLM baselines use \textbf{temperature~$= 0$}, enforcing
deterministic (greedy) decoding across every trial.

\begin{table}[h]
\centering
\caption{Model version strings used in the benchmark.}
\begin{tabular}{ll}
\toprule
\textbf{Baseline} & \textbf{Model ID} \\
\midrule
ChatGPT & \texttt{gpt-5.4} \\
Claude  & \texttt{claude-opus-4-6} \\
Gemini  & \texttt{gemini-3.1-pro-preview} \\
Raiven  & \texttt{gpt-5.2-chat-latest} \\
\bottomrule
\end{tabular}
\label{tab:models}
\end{table}

\noindent\textbf{Data Delivery.}\quad
Data delivery differs across baselines, as each API exposes a distinct
interface for attaching non-text content, as summarised in
Table~\ref{tab:delivery}. Data files are attached via each API's native
file-upload mechanism alongside the benchmark prompt text.

\begin{table*}[t]
\centering
\caption{Data delivery mechanism per baseline.}

\begin{tabular}{p{2cm}p{3.5cm}p{7cm}}
\toprule
\textbf{Baseline} & \textbf{API Surface} & \textbf{Mechanism} \\
\midrule
ChatGPT &
    OpenAI Responses API &
    User message composed of \texttt{input\_text} parts and
    \texttt{input\_file} parts encoded as base64 data URLs.
    System prompt passed via \texttt{instructions}. \\
\addlinespace
Claude &
    Anthropic Messages API (streaming) &
    User message composed of \texttt{text} blocks and
    \texttt{document} blocks (\texttt{text/plain} source).
    System prompt passed via \texttt{system}. \\
\addlinespace
Gemini &
    Google Gen AI &
    Files $\leq$20\,MB sent inline via \texttt{Part.from\_bytes};
    files $>$20\,MB uploaded via the Files API
    and referenced by URI via \texttt{Part.from\_uri}.
    System prompt passed via \texttt{system\_instruction}. \\
\addlinespace
Raiven &
    OpenAI Responses API &
    Files downloaded to local temporary paths and delivered
    as absolute local filesystem paths to the Raiven session. \\
\bottomrule
\end{tabular}
\label{tab:delivery}

\vspace{6pt}
\caption{Output token parameters.}

\begin{tabular}{p{4.5cm}p{3.5cm}p{5cm}}
\toprule
\textbf{Maximum output tokens} & \textbf{Value} & \textbf{Purpose} \\
\midrule
Claude &
    32{,}768 &
    Set explicitly via \texttt{max\_tokens}. \\
\addlinespace
Gemini &
    1{,}000{,}000 &
    Set via \texttt{max\_output\_tokens}. \\
\addlinespace
ChatGPT &
    Not set explicitly &
    Governed by Responses API default. \\
\bottomrule
\end{tabular}
\label{tab:context}
\end{table*}

\noindent\textbf{Context Size Limits and Truncation.}\quad
The output parameters applied to ChatGPT, Claude, and Gemini are given
in Table~\ref{tab:context}. Per-file-type truncation behaviour is as
follows:

\begin{itemize}
    \item \textbf{VTI:} XML header only, truncated at the
    \texttt{<AppendedData} element. The full binary data is injected
    into the generated HTML at post-processing time.
    \item \textbf{GeoJSON:} Full content sent as a plaintext attachment.
    \item \textbf{CSV / JSON:} Full content embedded as a file
    attachment.
\end{itemize}

\subsection{LLM Prompt}
\label{app:prompts-benchmark}
As referenced by Section~\ref{sec:benchmark}.
All LLM baselines receive the system prompt below, instructing them to return a single self-contained HTML file. It is passed via the native system-prompt parameter of each API (\texttt{instructions} for ChatGPT, \texttt{system} for Claude, \texttt{system\_instruction} for Gemini; Raiven handles it internally).

\begin{quote}
\ttfamily\small
You are an expert data visualization developer. Given a user's
visualization request and any attached data files, generate a complete,
self-contained HTML file that renders the visualization.\\[0.5em]
Requirements:\\
- Return only the raw HTML.\\
- For CSV and JSON data: embed all data directly in the HTML. Do not
fetch external data files.\\
- For .vti data: load files using fetch('./FILENAME') from the same
directory.\\
- Ensure the page renders correctly with basic structure
(HTML, head, body).\\[0.5em]
Output a single complete HTML file and nothing else.
\end{quote}

\subsection{VLM Scoring Prompt}
\label{app:vlm_vmpc_prompt}

To evaluate VMPC automatically, we prompted a VLM to act as a 
visualization evaluator. For each submission, the VLM was provided 
with four inputs: (1) the original prompt given to the model under 
evaluation, (2) the HTML source code of the generated output, (3) 
a rendered screenshot of that output, and (4) $N$, the number of 
views the prompt requested. The VLM was then asked to score each 
binary criterion and return a structured score table with 
per-criterion justifications. The full prompt given to the VLM is 
reproduced below.

\bigskip
\noindent\rule{\linewidth}{0.4pt}

\begin{quote}
\textbf{VMPC Scoring Prompt --- VLM Evaluation of Generated HTML 
Visualizations}

\medskip
You are an expert visualization evaluator. You will be given:
\begin{enumerate}
    \item A \textbf{prompt} that was given to an LLM-based 
    visualization system.
    \item The \textbf{HTML source code} of the output that system 
    produced.
    \item A \textbf{rendered view} of that HTML, which you can 
    interact with (click, hover, brush, select).
    \item \textbf{N}, the number of views the prompt requested.
\end{enumerate}

Your job is to evaluate the output by scoring each component 
described below.

\medskip
\noindent\textbf{Core Principles}
\begin{itemize}
    \item \textbf{Everything is grounded in the prompt.} Score 
    based on what the prompt asked for, not what you think a good 
    visualization should look like.
    \item \textbf{Overcompletion is fine.} Extra views, 
    annotations, or interactions beyond what the prompt asked for 
    are neither penalized nor credited.
    \item \textbf{N is provided to you.} It is the number of views 
    the prompt requested. If the output has more views than $N$, 
    identify which views correspond to the $N$ prompted ones and 
    disregard the extras. If fewer, the missing views get all 
    zeros.
    \item \textbf{When in doubt, score 0.} VMPC is a minimum 
    compliance metric. Do not give the benefit of the doubt.
\end{itemize}

\medskip
\noindent\textbf{Compilation Gate}

\noindent$X$ --- \textbf{Compilation (binary: 0 or 1).}
Does the HTML compile and render at least one view?
\begin{itemize}
    \item \textbf{1}: The page loads and at least one 
    visualization view is visible on screen --- any chart, plot, 
    or map with allocated space.
    \item \textbf{0}: The page shows only a permanent loading 
    spinner, an error/stack trace, a completely blank/white page, 
    or zero views render.
\end{itemize}
If $X = 0$, the entire score is 0 regardless of all other 
components.

\medskip
\noindent\textbf{Per-View Scores}

For each of the $N$ views, evaluate the following five criteria.

\medskip
\noindent\textit{$V_v$ --- View Existence (binary: 0 or 1).}
Does view $v$ have allocated screen space for its intended 
visualization purpose?
\begin{itemize}
    \item \textbf{1}: There is a region on the page dedicated to 
    this view --- a visible outline, frame, axes, container, 
    panel, or box that indicates a visualization is intended 
    there. An empty box or placeholder with allocated space 
    counts as $V = 1$.
    \item \textbf{0}: No region exists on the page for this view. 
    It is completely absent from the layout.
\end{itemize}
If $V = 0$, then $M$, $E$, and $H$ for that view are all 
automatically 0.

\medskip
\noindent\textit{$M_v$ --- Mark Type (binary: 0 or 1).}
Does view $v$ actually render visible marks of the correct type?
\begin{itemize}
    \item \textbf{1}: The view contains visible, rendered marks 
    that match what the prompt specified --- bars for ``bar 
    chart'', points for ``scatter plot'', lines for ``line 
    chart'', a 3D surface for ``isosurface'', a 3D volume for 
    ``volume rendering'', etc. The marks must be actually drawn 
    on screen, not just implied by code or axes.
    \item \textbf{0}: No marks are rendered (the view is blank, 
    shows only axes with no data, shows a loading spinner, or 
    displays an error message), OR the wrong mark type is used.
\end{itemize}
\textbf{Critical:} Do not give $M = 1$ based on what the code 
\emph{intends} to render --- only what is \emph{actually visible} 
in the rendered output. Domain-specific checks include:
\begin{itemize}
    \item \textbf{LIC}: A correct LIC shows coherent, directional 
    flow-structure textures. Uniform random noise/static with no 
    directional structure means the LIC computation failed: 
    $M = 0$.
    \item \textbf{Volume rendering}: Should show a 3D volumetric 
    object with depth, transparency, and shading.
    \item \textbf{Streamlines}: Should show curved lines or tubes 
    following flow directions.
\end{itemize}

\medskip
\noindent\textit{$E_v$ --- Encoding (binary: 0 or 1).}
Are the visual encodings actually visible and correct for view 
$v$ as specified in the prompt?
\begin{itemize}
    \item \textbf{1}: The encodings the prompt specified are 
    correct AND visible --- the right variables on the right axes 
    with actual data shown, correct color mapping applied to 
    visible marks, etc.
    \item \textbf{0}: Encodings are wrong, OR the view has no 
    visible data to evaluate encodings on (blank, loading, 
    error), OR axes show wrong variables, OR a requested encoding 
    is missing.
\end{itemize}
\textbf{Critical:} If a view shows no data, $E = 0$ regardless 
of what the code says. Only evaluate encodings the prompt 
explicitly specifies.

\medskip
\noindent\textit{$H_v$ --- Data Hallucination (binary: 0 or 1).}
Is the data in view $v$ real and sourced correctly, or is it 
hallucinated/fabricated? Evaluate by inspecting both the code 
and the rendered output:
\begin{itemize}
    \item Check all \texttt{fetch()}, \texttt{d3.csv()}, 
    \texttt{d3.json()} calls, inline data arrays, and 
    \texttt{<script>} blocks. Does the code load from the 
    source the prompt specified?
    \item \textbf{Critical --- check for try/catch fallback 
    patterns.} Code that tries to fetch real data but falls back 
    to \texttt{Math.random()}, \texttt{Math.sin()}, hardcoded 
    arrays, or synthetic generation in the \texttt{catch} block 
    is $H = 0$. The correct fetch URL being present does 
    \emph{not} mean real data was used.
    \item Cross-check with the rendered output. Look for text 
    saying ``fallback'', ``sample data'', or ``generated'', 
    and for data patterns that look procedurally generated 
    (perfect sine waves, uniformly random distributions).
\end{itemize}
\begin{itemize}
    \item \textbf{1 (Real data)}: The code loads from the correct 
    source, there is no synthetic fallback, and the rendered 
    content is consistent with that dataset.
    \item \textbf{0 (Hallucinated)}: Data is fetched from a 
    different source, fabricated, generated via a try/catch 
    fallback, or inconsistent with the specified dataset.
\end{itemize}
If $V = 0$, then $H = 0$ automatically. If $V = 1$ but $M = 0$ 
(view exists but renders no marks), then $H = 1$ --- a view that 
shows no data cannot be hallucinating data.

\medskip
\noindent\textit{$L_v$ --- Linking (binary: 0 or 1).}
Does view $v$ participate in cross-view linking as requested 
by the prompt?
\begin{itemize}
    \item If the prompt does \textbf{not} request linking for 
    this view, $L = 1$ automatically (not applicable).
    \item If linking is requested, inspect the code for event 
    listeners, shared state, and dispatch/callback patterns 
    connecting views, and test the interaction by clicking, 
    hovering, or brushing.
    \item \textbf{1}: Linking works as specified, or no linking 
    was requested.
    \item \textbf{0}: Linking was requested but is broken, 
    missing, or produces no visible cross-view response.
\end{itemize}

\medskip
\noindent\textbf{Output Format}

Provide your evaluation as a raw score table. $N$ is provided 
to you --- score each of the $N$ views in order.

\begin{verbatim}
PROMPT SUMMARY: [1-2 sentence summary of what the 
prompt asked for]

N = [provided value]

X (Compilation): [0 or 1] — [brief justification]

| view | V | M | E | H | L |
|------|---|---|---|---|---|
| v1   | _ | _ | _ | _ | _ |
| v2   | _ | _ | _ | _ | _ |
| ...  | _ | _ | _ | _ | _ |

JUSTIFICATIONS:
v1: [1-2 sentences explaining the scores for this view]
v2: [1-2 sentences explaining the scores for this view]
...
\end{verbatim}

All scores are binary: 0 or 1. If $X = 0$, fill the entire 
table with zeros. If $V = 0$ for a view, $M$, $E$, and $H$ 
are all 0. If $V = 1$ but the view is blank, loading, or 
errored, then $M = 0$, $E = 0$, but $H = 1$. $L = 1$ if no 
linking was requested for this view.

\end{quote}

\noindent\rule{\linewidth}{0.4pt}

\subsection{VMPC Scoring Examples}
\label{app:vmpc-scoring}

\noindent As referenced by Section~\ref{sec:vmpc}. We illustrate VMPC scoring through two worked examples. First, we 
walk through how a grader would evaluate four different models' 
outputs for the same prompt (case 73), demonstrating how the metric 
behaves across a range of partial and full compliance. Second, we 
show a case where the execution gate $X = 0$ short-circuits scoring 
entirely. In all cases, scores reflect the majority judgment of 
three independent graders, with each binary criterion requiring 
agreement from at least two of the three.

\subsubsection{Case 73: Multi-View Compliance Across Four Models}

The prompt for case 73 reads:

\begin{quote}
\textit{``Render the CT isosurface in pink from \texttt{head.vti}, 
then summarize \texttt{head\_sample.csv} using a histogram of 
`intensity' colored pink.''}
\end{quote}

This prompt specifies $N = 2$ views. No cross-view linking is 
requested, so $L_v = 1$ by default for both views across all four 
model outputs. A grader evaluating any response to this prompt 
would first check whether the code runs and produces visible output 
(execution gate $X$), then assess each view in turn against the 
five binary criteria: view presence $V_v$, mark correctness $M_v$, 
encoding correctness $E_v$, data hallucination $H_v$, and 
cross-view linking $L_v$.

\paragraph{\textbf{Model 1: Raiven (VMPC = 1.0).}}

\begin{figure}[h]
    \centering
    \includegraphics[width=\linewidth]{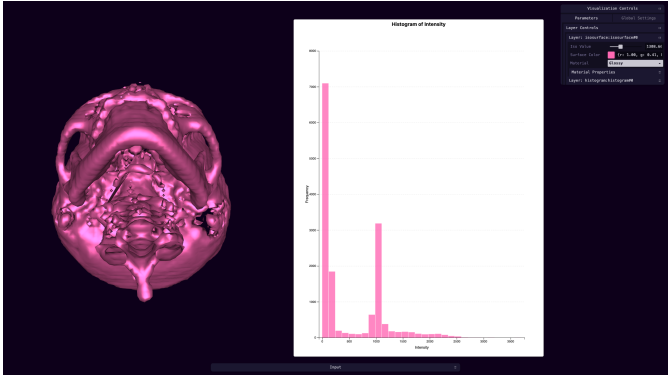}
    \caption{Case 73 output from Raiven: a pink CT isosurface 
    alongside a pink histogram of CT intensity values.}
    \label{fig:vmpc-c73-raiven}
\end{figure}

The code executes without error and produces a visible output, 
so $X = 1$.

\textbf{View 1 (CT Isosurface).} A clearly rendered 3D surface 
is present in the expected panel, so $V_1 = 1$. The output uses 
an isosurface rendering as requested rather than a volume rendering 
or point cloud, so $M_1 = 1$. The surface displays a consistent 
pink hue with no unexpected color mapping, so $E_1 = 1$. The 
geometry is consistent with a human head CT scan loaded from 
\texttt{head.vti} rather than a substitute or fabricated mesh, 
so $H_1 = 1$. As no linking was requested, $L_1 = 1$.

\textbf{View 2 (Intensity Histogram).} A histogram panel is 
clearly visible alongside the 3D view, so $V_2 = 1$. The mark 
type is a binned bar chart as expected, so $M_2 = 1$. The x-axis 
encodes the \texttt{intensity} variable from \texttt{head\_sample.csv}, 
the y-axis encodes frequency counts, and the bars are colored pink 
as specified, so $E_2 = 1$. The distribution shape and value range 
are consistent with CT intensity data from the provided file rather 
than synthetic values, so $H_2 = 1$. No linking was requested, 
so $L_2 = 1$.

\textbf{Calculation.} All ten binary criteria are satisfied:
\begin{equation*}
\mathrm{VMPC} = 1 \cdot \frac{1}{5 \times 2}\left(
\underbrace{(1+1+1+1+1)}_{\text{View 1}} +
\underbrace{(1+1+1+1+1)}_{\text{View 2}}\right)
= \frac{10}{10} = 1.0
\end{equation*}
All three graders assigned full credit to every criterion, 
reflecting clear and unambiguous compliance with the prompt 
across both views.

\paragraph{\textbf{Model 2: ChatGPT (VMPC = 1.0).}}

\begin{figure}[h]
    \centering
    \includegraphics[width=\linewidth]{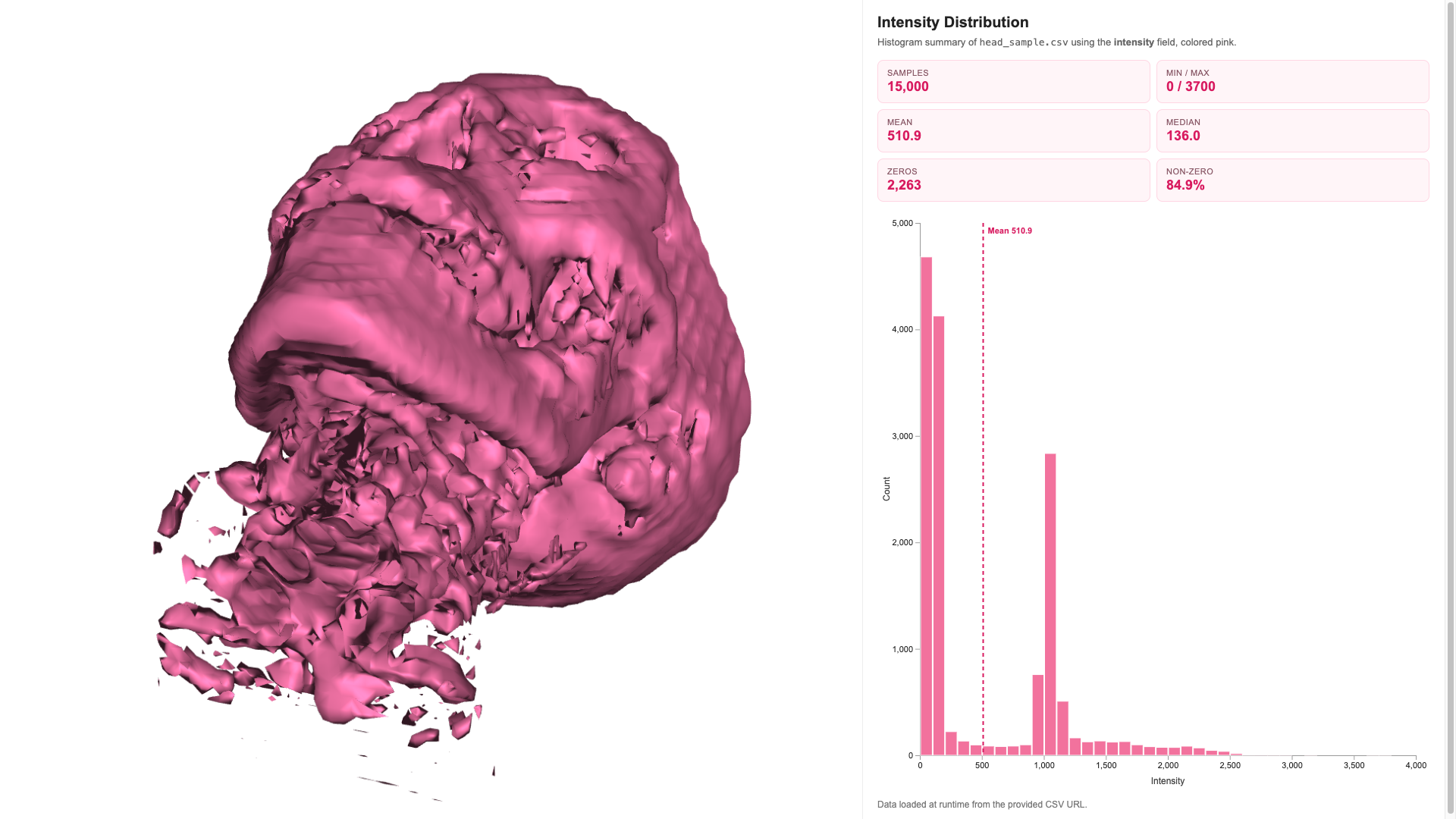}
    \caption{Case 73 output from ChatGPT: a pink CT isosurface 
    alongside a pink histogram of CT intensity values, with 
    additional unrequested summary statistics.}
    \label{fig:vmpc-c73-chatgpt}
\end{figure}

The code executes without error, so $X = 1$. Both views satisfy 
all five criteria by the same reasoning as Raiven: the isosurface 
is present, correctly typed, and pink ($V_1 = M_1 = E_1 = H_1 
= L_1 = 1$), and the histogram correctly encodes \texttt{intensity} 
with pink bars ($V_2 = M_2 = E_2 = H_2 = L_2 = 1$).

The grader also notices that the output includes additional 
summary statistics --- such as mean and standard deviation 
annotations --- that were not requested by the prompt. 
Importantly, VMPC does not penalize unrequested additions; the 
metric evaluates only whether what the prompt asked for is present 
and correct. These extra elements are ignored during scoring.

\textbf{Calculation.} All ten criteria are satisfied:
\begin{equation*}
\mathrm{VMPC} = 1 \cdot \frac{1}{5 \times 2}\left(
\underbrace{(1+1+1+1+1)}_{\text{View 1}} +
\underbrace{(1+1+1+1+1)}_{\text{View 2}}\right)
= \frac{10}{10} = 1.0
\end{equation*}

\paragraph{\textbf{Model 3: Claude (VMPC = 0.8).}}

\begin{figure}[h]
    \centering
    \includegraphics[width=\linewidth]{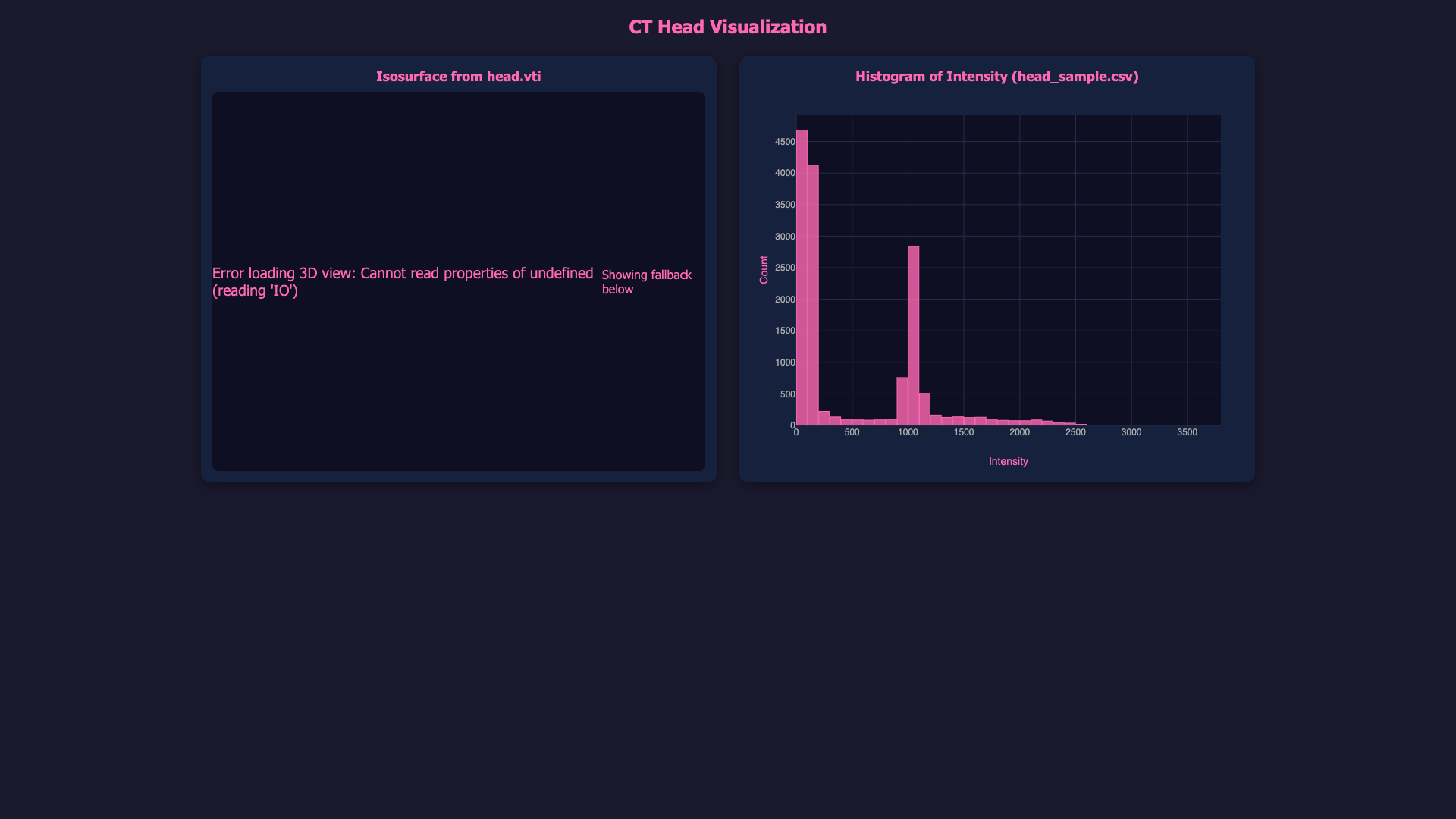}
    \caption{Case 73 output from Claude: the isosurface panel 
    renders only an empty container, while the pink intensity 
    histogram is fully correct.}
    \label{fig:vmpc-c73-claude}
\end{figure}

The code executes without error, so $X = 1$.

\textbf{View 1 (CT Isosurface).} A panel container for the 
isosurface is present and visible, so $V_1 = 1$. However, no 
isosurface geometry appears inside it --- the panel renders but 
is empty. Because no mark is visible, the grader cannot confirm 
the correct mark type was used, so $M_1 = 0$. Similarly, with 
no rendered geometry, color, channel assignments, and variable 
mappings are all unverifiable, so $E_1 = 0$. Crucially, the 
absence of any rendered geometry also means no data could have 
been fabricated or substituted, so $H_1 = 1$. No linking was 
requested, so $L_1 = 1$.

\textbf{View 2 (Intensity Histogram).} The histogram is fully 
correct: the panel is present ($V_2 = 1$), the mark type is a 
binned bar chart ($M_2 = 1$), the x-axis encodes \texttt{intensity} 
with pink bars and the y-axis encodes frequency counts ($E_2 = 1$), 
the values are consistent with \texttt{head\_sample.csv} 
($H_2 = 1$), and no linking was requested ($L_2 = 1$).

\textbf{Calculation.} Eight of ten criteria are satisfied. The 
two deductions stem directly from the empty isosurface panel: 
without a visible mark, neither $M_1$ nor $E_1$ can be awarded.
\begin{equation*}
\mathrm{VMPC} = 1 \cdot \frac{1}{5 \times 2}\left(
\underbrace{(1+0+0+1+1)}_{\text{View 1}} +
\underbrace{(1+1+1+1+1)}_{\text{View 2}}\right)
= \frac{8}{10} = 0.8
\end{equation*}

\paragraph{\textbf{Model 4: Gemini (VMPC = 0.6).}}

\begin{figure}[h]
    \centering
    \includegraphics[width=\linewidth]{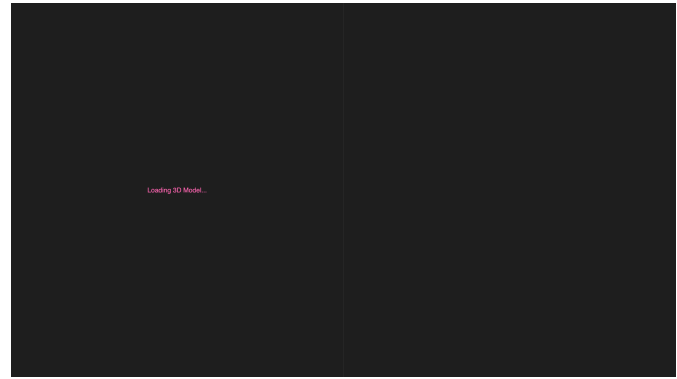}
    \caption{Case 73 output from Gemini: two faintly outlined view 
    containers are visible, but neither renders any marks or data. 
    The isosurface panel shows only a ``loading 3D model'' 
    indicator; the histogram panel is a blank space.}
    \label{fig:vmpc-c73-gemini}
\end{figure}

The code executes without producing a hard error, so $X = 1$, 
but neither view renders any visible marks or data.

\textbf{View 1 (CT Isosurface).} A faint panel outline is 
visible and a ``loading 3D model'' indicator appears within it, 
confirming that a view container was instantiated. Graders were 
instructed to apply a generous interpretation of view presence: 
$V_v = 1$ whenever there is a discernible indication that a view 
was intended, even if its content failed to load. The grader 
therefore awards $V_1 = 1$. However, since no isosurface geometry 
ever renders, there is no mark to evaluate ($M_1 = 0$) and no 
encoding to verify ($E_1 = 0$). The absence of any rendered data 
also means no hallucination is possible, so $H_1 = 1$. No linking 
was requested, so $L_1 = 1$.

\textbf{View 2 (Intensity Histogram).} The second half of the 
screen is occupied by a blank space with a faint outline. No 
histogram, bars, or axis labels are visible. The graders debated 
$V_2$ carefully: awarding $V_2 = 0$ would conflate a 
\emph{missing} view with a \emph{failed render}, conflating two 
distinct failure modes. Since the blank space is clearly allocated 
as a view container, at least two of three graders agreed to award 
$V_2 = 1$ in the interest of generous but consistent grading. With 
no marks rendered, $M_2 = 0$ and $E_2 = 0$. As no data is 
displayed, $H_2 = 1$, and with no linking requested, $L_2 = 1$.

\textbf{Calculation.} Six of ten criteria are satisfied. Both 
views lose $M_v$ and $E_v$ due to the absence of any rendered 
content:
\begin{equation*}
\mathrm{VMPC} = 1 \cdot \frac{1}{5 \times 2}\left(
\underbrace{(1+0+0+1+1)}_{\text{View 1}} +
\underbrace{(1+0+0+1+1)}_{\text{View 2}}\right)
= \frac{6}{10} = 0.6
\end{equation*}

\paragraph{\textbf{Summary.}} Table~\ref{tab:vmpc-c73-summary} summarizes 
the per-criterion scores for all four models on case 73. The 
progression from Raiven and ChatGPT (full compliance) through 
Claude (one empty view) to Gemini (both views empty) illustrates 
how VMPC captures graduated levels of partial compliance within 
the same prompt.

\begin{table}[h]
\centering
\caption{Per-criterion VMPC scores for case 73 across four models. 
$L_v = 1$ for all entries as no linking was requested.}
\label{tab:vmpc-c73-summary}
\small
\begin{tabular}{lc ccccc ccccc c}
\toprule
& & \multicolumn{5}{c}{\textbf{View 1 (Isosurface)}} 
& \multicolumn{5}{c}{\textbf{View 2 (Histogram)}} & \\
\cmidrule(lr){3-7}\cmidrule(lr){8-12}
\textbf{Model} & $X$ 
& $V_1$ & $M_1$ & $E_1$ & $H_1$ & $L_1$ 
& $V_2$ & $M_2$ & $E_2$ & $H_2$ & $L_2$ 
& \textbf{VMPC} \\
\midrule
Raiven  & 1 & 1 & 1 & 1 & 1 & 1 & 1 & 1 & 1 & 1 & 1 & 1.0 \\
ChatGPT & 1 & 1 & 1 & 1 & 1 & 1 & 1 & 1 & 1 & 1 & 1 & 1.0 \\
Claude  & 1 & 1 & 0 & 0 & 1 & 1 & 1 & 1 & 1 & 1 & 1 & 0.8 \\
Gemini  & 1 & 1 & 0 & 0 & 1 & 1 & 1 & 0 & 0 & 1 & 1 & 0.6 \\
\bottomrule
\end{tabular}
\end{table}

\subsubsection{Execution Failure: Case s1 (ChatGPT, VMPC = 0.0)}

\begin{figure}[h]
    \centering
    \includegraphics[width=0.85\linewidth]{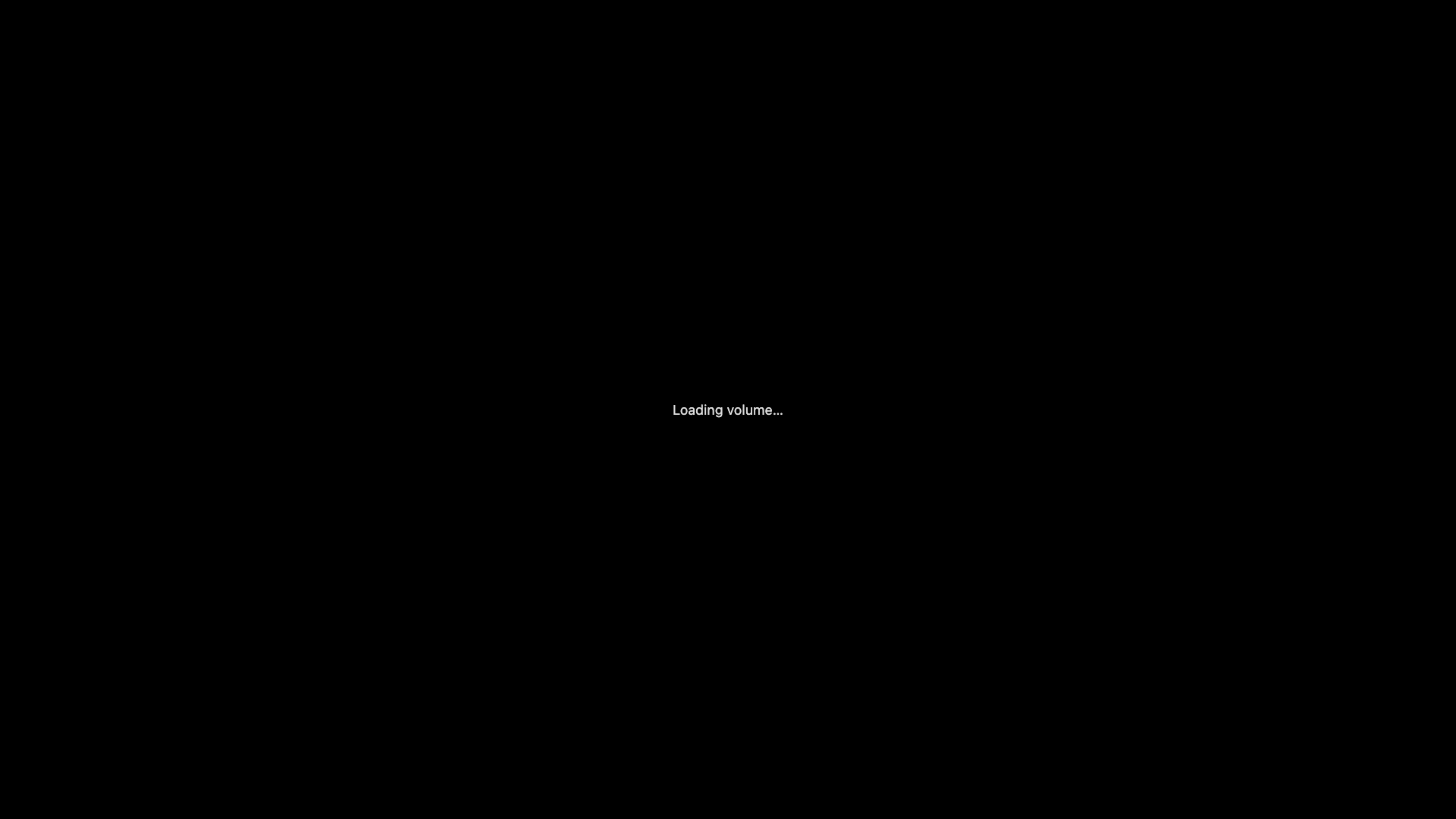}
    \caption{Case s1 output from ChatGPT: a blank screen with only 
    a ``loading volume'' indicator. No view ever renders.}
    \label{fig:vmpc-s1-chatgpt}
\end{figure}

The prompt for case s1 reads:

\begin{quote}
\textit{``Create a volume rendering of the skull CT data from 
\texttt{head.vti}.''}
\end{quote}

This specifies $N = 1$ view with no cross-view linking 
($L_1 = 1$ by default).

A grader approaching this output would first attempt to run the 
submitted code. The code either crashes outright or produces only 
a blank screen with a persistent ``loading volume'' indicator and 
no evidence of a rendered view. Throughout our evaluation, outputs 
that displayed only loading indicators with no rendered content 
were treated as execution failures on par with hard crashes: a 
loading symbol with no subsequent render provides no visualizable 
output for a grader to assess. The execution gate is therefore 
set to $X = 0$.

Once $X = 0$, VMPC collapses to zero regardless of any 
per-criterion judgments:
\begin{equation*}
\mathrm{VMPC} = 0 \cdot \frac{1}{5 \times 1}
\underbrace{\left(V_1 + M_1 + E_1 + H_1 + L_1\right)}_{\text{View 1}}
= 0.0
\end{equation*}
The grader need not evaluate any individual criterion --- the 
execution gate short-circuits the entire computation. This 
reflects the core requirement that a visualization must be 
visible to provide value: a system that fails to render anything, 
whether due to a runtime error or an indefinitely stalled load, 
has not fulfilled the prompt regardless of what the underlying 
code attempted.

\subsection{Full Results}
\label{app:benchmark-full-results}
% Benchmark Full Results 

\newcommand{\benchmarkentry}[6]{%
% #1=ID, #2=datasets, #3=views
% #4=prompt text, #5=screenshot path prefix, #6=scores rows
\noindent
\begin{minipage}{\textwidth}
% ---- Images ----
\noindent
\begin{tabular*}{\textwidth}{@{\extracolsep{\fill}}cccc@{}}
\includegraphics[width=0.23\textwidth]{#5_raiven.png} &
\includegraphics[width=0.23\textwidth]{#5_chatgpt.png} &
\includegraphics[width=0.23\textwidth]{#5_claude.png} &
\includegraphics[width=0.23\textwidth]{#5_gemini.png} \\
{\small Raiven} & {\small ChatGPT} & {\small Claude} & {\small Gemini}
\end{tabular*}

\vspace{2pt}

% ---- Unified table ----
\noindent
{\small
\begin{tabularx}{\textwidth}{@{}X@{\hspace{6pt}}|l|c|ccc|c|c@{}}
\toprule
\textbf{#1}\quad Views: #3
  & System & VMPC & $G_1$ & $G_2$ & $G_3$ & VLM & Time \\
\midrule
\multirow{4}{\hsize}{\textit{#4}\\[4pt]Datasets: \texttt{#2}}
#6  % add \\[2pt] at end of each row in #6
\bottomrule
\end{tabularx}}
\end{minipage}

\vspace{10pt}
}

As referenced by Section~\ref{sec:benchmark}.
Results are organized by category: SciVis (S, prompts S1--S30),
InfoVis (I, prompts I31--I70), and Combined (C, prompts C71--C100).
For each prompt we show the rendered output from all four systems,
the average VMPC score across three human graders ($G_1$--$G_3$,
shown individually), the independent VLM score, and generation
time in minutes.

\onecolumn

\subsubsection{SciVis (S)}

\benchmarkentry{S1}{head.vti}{1}
{Create a volume rendering of the skull CT data from head.vti.}
{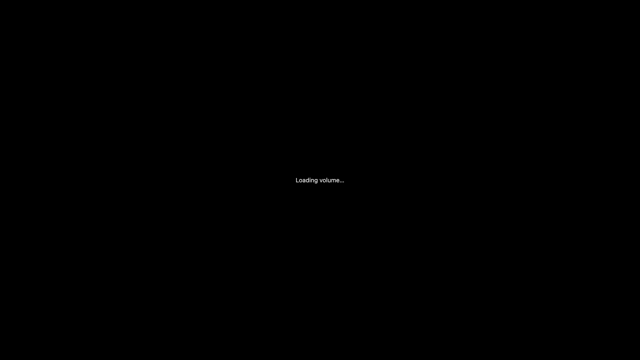}
{ & Raiven  & 0.93 & 1.0 & 1.0 & 0.8 & 1.00 & 0.24 \\
 & ChatGPT & 0.00 & 0.0 & 0.0 & 0.0 & 0.00 & 0.35 \\
 & Claude  & 0.00 & 0.0 & 0.0 & 0.0 & 0.00 & 0.67 \\
 & Gemini  & 0.00 & 0.0 & 0.0 & 0.0 & 0.00 & 0.89 \\}

\benchmarkentry{S2}{head.vti}{1}
{Display a semi-transparent volume from head.vti with an overlayed isosurface highlighting the bone boundary within the tissue, use color palette turbo.}
{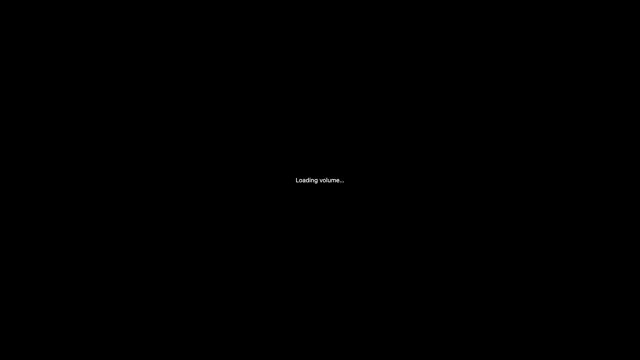}
{ & Raiven  & 1.00 & 1.0 & 1.0 & 1.0 & 1.00 & 0.26 \\
 & ChatGPT & 0.00 & 0.0 & 0.0 & 0.0 & 0.00 & 0.45 \\
 & Claude  & 0.00 & 0.0 & 0.0 & 0.0 & 0.00 & 0.54 \\
 & Gemini  & 0.00 & 0.0 & 0.0 & 0.0 & 0.00 & 1.71 \\}

\benchmarkentry{S3}{head.vti}{1}
{Generate three layered, differently colored isosurfaces from head.vti at low, medium, and high intensity thresholds to represent skin, outer bone, and inner table.}
{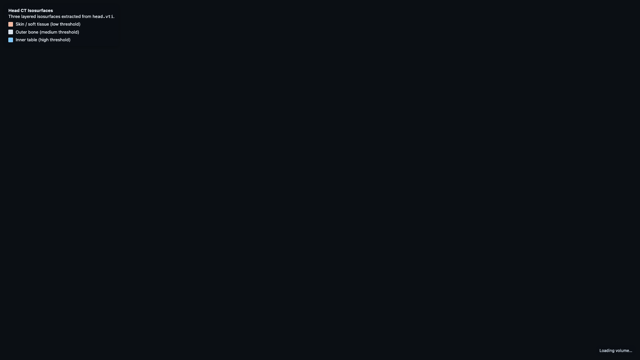}
{ & Raiven  & 0.93 & 1.0 & 1.0 & 0.8 & 0.80 & 0.56 \\
 & ChatGPT & 0.60 & 0.6 & 0.6 & 0.6 & 0.40 & 0.38 \\
 & Claude  & 0.00 & 0.0 & 0.0 & 0.0 & 0.00 & 0.41 \\
 & Gemini  & 0.60 & 0.6 & 0.6 & 0.6 & 0.40 & 0.80 \\}

\benchmarkentry{S4}{head.vti}{1}
{Display axial, sagittal, and coronal triplanar slices of head.vti in one view with an overlayed isosurface to provide 3D surface context.}
{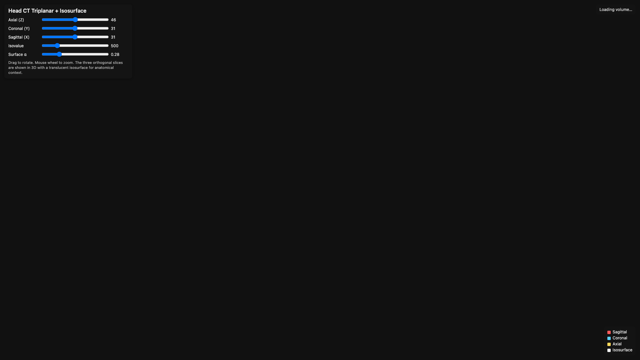}
{ & Raiven  & 1.00 & 1.0 & 1.0 & 1.0 & 1.00 & 0.35 \\
 & ChatGPT & 0.60 & 0.6 & 0.6 & 0.6 & 0.40 & 0.62 \\
 & Claude  & 0.60 & 0.6 & 0.6 & 0.6 & 0.40 & 0.59 \\
 & Gemini  & 0.00 & 0.0 & 0.0 & 0.0 & 0.00 & 1.30 \\}

\benchmarkentry{S5}{head.vti}{2}
{Show a volume rendering from head.vti in one view and an oblique slice rotated at an arbitrary non-axis-aligned angle in a second view, both using color palette inferno.}
{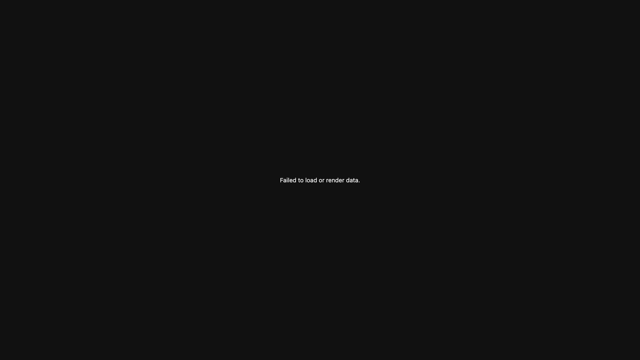}
{ & Raiven  & 1.00 & 1.0 & 1.0 & 1.0 & 1.00 & 0.28 \\
 & ChatGPT & 0.00 & 0.0 & 0.0 & 0.0 & 0.00 & 0.43 \\
 & Claude  & 0.00 & 0.0 & 0.0 & 0.0 & 0.00 & 1.19 \\
 & Gemini  & 1.00 & 1.0 & 1.0 & 1.0 & 1.00 & 1.36 \\}

\benchmarkentry{S6}{divcurl\_4.vti, divcurl\_5.vti, divcurl\_6.vti, divcurl\_7.vti, divcurl\_8.vti, divcurl\_9.vti, divcurl\_10.vti, divcurl\_11.vti, divcurl\_12.vti}{9}
{Use divcurl\_\{t\}.vti to create a 3×3 grid of LIC visualizations across timesteps t = 4 to 12, using the 2D vector field defined by vx and vy.}
{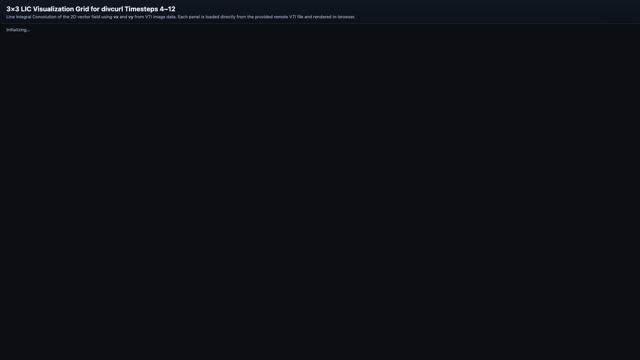}
{ & Raiven  & 1.00 & 1.0 & 1.0 & 1.0 & 1.00 & 0.40 \\
 & ChatGPT & 0.07 & 0.0 & 0.0 & 0.2 & 0.40 & 0.79 \\
 & Claude  & 1.00 & 1.0 & 1.0 & 1.0 & 1.00 & 0.95 \\
 & Gemini  & 0.60 & 0.6 & 0.6 & 0.6 & 0.40 & 3.12 \\}

\benchmarkentry{S7}{head.vti}{4}
{Set up three slice views (XY, YZ, XZ) and a fourth 3D view combining volume rendering and triplanar slices from head.vti, synchronizing slice positions and the transfer function.}
{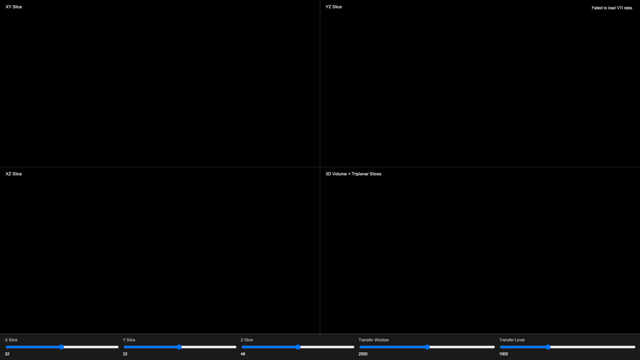}
{ & Raiven  & 1.00 & 1.0 & 1.0 & 1.0 & 1.00 & 0.37 \\
 & ChatGPT & 0.47 & 0.4 & 0.4 & 0.6 & 0.40 & 0.78 \\
 & Claude  & 0.13 & 0.0 & 0.4 & 0.0 & 0.40 & 1.81 \\
 & Gemini  & 0.47 & 0.4 & 0.4 & 0.6 & 0.40 & 2.18 \\}

\benchmarkentry{S8}{divcurl\_19.vti}{1}
{Produce an LIC visualization from divcurl\_19.vti using vx and vy to reveal flow patterns.}
{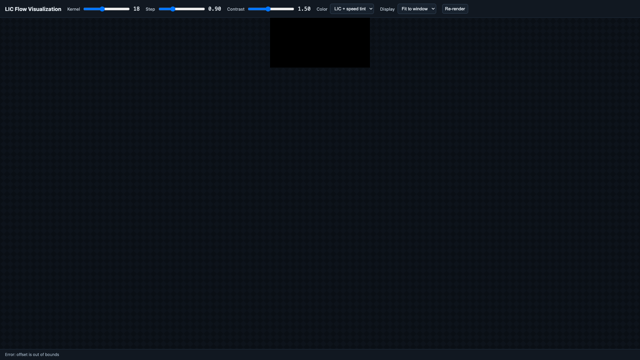}
{ & Raiven  & 1.00 & 1.0 & 1.0 & 1.0 & 1.00 & 0.23 \\
 & ChatGPT & 0.60 & 0.6 & 0.6 & 0.6 & 0.40 & 0.79 \\
 & Claude  & 0.40 & 0.0 & 0.6 & 0.6 & 0.00 & 1.28 \\
 & Gemini  & 1.00 & 1.0 & 1.0 & 1.0 & 1.00 & 2.94 \\}

\benchmarkentry{S9}{divcurl\_15.vti}{3}
{Arrange three views from divcurl\_15.vti: an LIC of vx and vy, a slice of div, and a slice of curl.}
{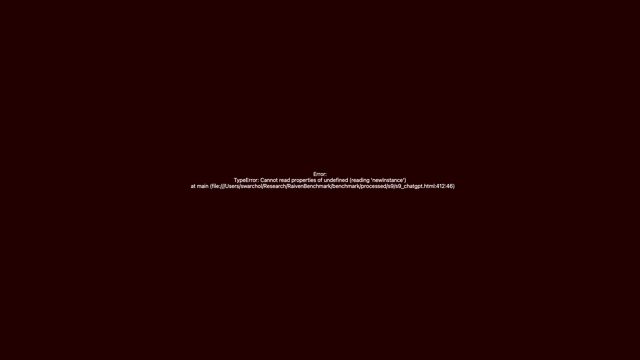}
{ & Raiven  & 1.00 & 1.0 & 1.0 & 1.0 & 1.00 & 0.36 \\
 & ChatGPT & 0.00 & 0.0 & 0.0 & 0.0 & 0.00 & 0.77 \\
 & Claude  & 0.00 & 0.0 & 0.0 & 0.0 & 0.40 & 1.11 \\
 & Gemini  & 0.60 & 0.6 & 0.6 & 0.6 & 0.40 & 2.59 \\}

\benchmarkentry{S10}{divcurl\_0.vti, divcurl\_10.vti, divcurl\_15.vti, divcurl\_19.vti}{4}
{Use divcurl\_0.vti, divcurl\_10.vti, divcurl\_15,vti, and divcurl\_19.vti to create four side-by-side LIC views, one per timestep, each computed from the velocity vector with vx = "vx" and vy = "vy".}
{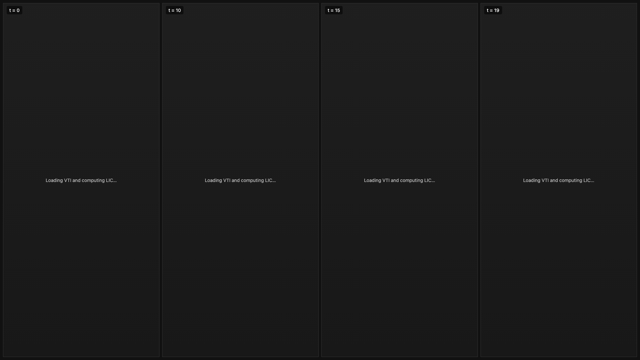}
{ & Raiven  & 1.00 & 1.0 & 1.0 & 1.0 & 1.00 & 0.27 \\
 & ChatGPT & 0.53 & 0.6 & 0.6 & 0.4 & 0.40 & 0.35 \\
 & Claude  & 1.00 & 1.0 & 1.0 & 1.0 & 0.70 & 0.80 \\
 & Gemini  & 1.00 & 1.0 & 1.0 & 1.0 & 1.00 & 3.31 \\}

\benchmarkentry{S11}{divcurl\_15.vti}{2}
{Provide two slice views from divcurl\_15.vti comparing velocity components by setting the scalar field to vy in one and vx in the other.}
{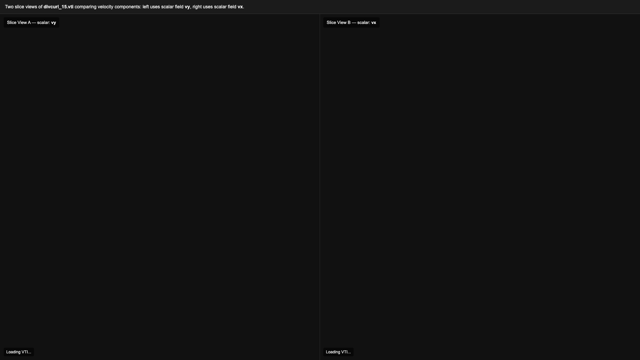}
{ & Raiven  & 1.00 & 1.0 & 1.0 & 1.0 & 1.00 & 0.45 \\
 & ChatGPT & 0.53 & 0.6 & 0.6 & 0.4 & 0.40 & 0.34 \\
 & Claude  & 0.00 & 0.0 & 0.0 & 0.0 & 0.00 & 1.04 \\
 & Gemini  & 0.53 & 0.6 & 0.6 & 0.4 & 0.40 & 1.18 \\}

\benchmarkentry{S12}{taylorgreen\_9.vti}{1}
{Compute streamlines from taylorgreen\_9.vti using ux, uy, and uz, with seeds in the bounds [0, 64, 0, 64, 0, 0].}
{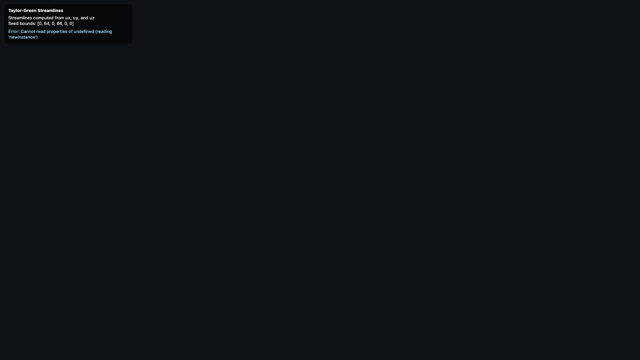}
{ & Raiven  & 1.00 & 1.0 & 1.0 & 1.0 & 1.00 & 0.24 \\
 & ChatGPT & 0.20 & 0.6 & 0.0 & 0.0 & 0.40 & 0.32 \\
 & Claude  & 0.20 & 0.6 & 0.0 & 0.0 & 0.40 & 1.23 \\
 & Gemini  & 0.00 & 0.0 & 0.0 & 0.0 & 0.00 & 2.58 \\}

\benchmarkentry{S13}{taylorgreen\_9.vti}{1}
{Make a volume of "vort" and also streamlines of ux, uy, and uz layered over volume from taylorgreen\_9.vti.}
{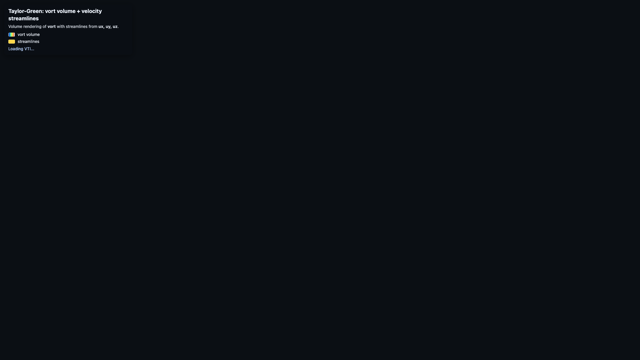}
{ & Raiven  & 1.00 & 1.0 & 1.0 & 1.0 & 1.00 & 0.29 \\
 & ChatGPT & 0.73 & 0.6 & 1.0 & 0.6 & 0.40 & 0.57 \\
 & Claude  & 0.60 & 0.6 & 0.6 & 0.6 & 1.00 & 1.13 \\
 & Gemini  & 0.00 & 0.0 & 0.0 & 0.0 & 0.00 & 1.62 \\}

\benchmarkentry{S14}{taylorgreen\_9.vti}{2}
{Use dataset taylorgreen\_9.vti to make a volume of "vort" with streamlines of ux, uy, and uz layered over volume as well as a linked oblique slice.}
{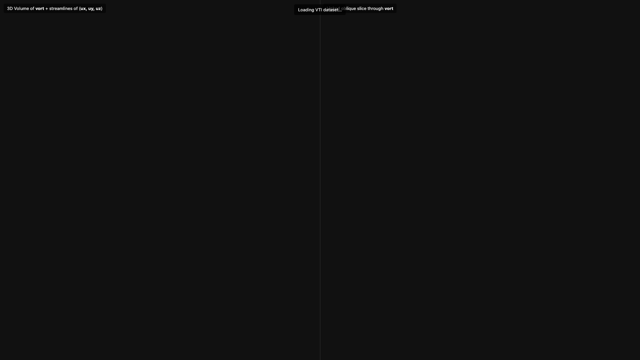}
{ & Raiven  & 0.87 & 0.9 & 0.9 & 0.8 & 1.00 & 0.42 \\
 & ChatGPT & 0.47 & 0.6 & 0.4 & 0.4 & 0.40 & 0.98 \\
 & Claude  & 0.13 & 0.0 & 0.4 & 0.0 & 0.40 & 1.66 \\
 & Gemini  & 0.00 & 0.0 & 0.0 & 0.0 & 0.00 & 2.77 \\}

\benchmarkentry{S15}{taylorgreen\_9.vti}{1}
{Make streamline from taylorgreen\_9.vti using variables vx = ux, vy = uy, and vz = uz.}
{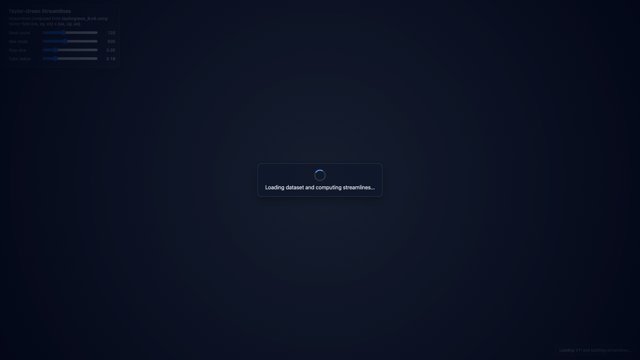}
{ & Raiven  & 1.00 & 1.0 & 1.0 & 1.0 & 1.00 & 0.23 \\
 & ChatGPT & 0.60 & 0.6 & 0.6 & 0.6 & 0.40 & 0.90 \\
 & Claude  & 1.00 & 1.0 & 1.0 & 1.0 & 1.00 & 0.80 \\
 & Gemini  & 0.20 & 0.6 & 0.0 & 0.0 & 0.40 & 2.73 \\}

\benchmarkentry{S16}{taylorgreen\_9.vti}{1}
{Generate three nested isosurfaces of vort from taylorgreen\_9.vti at progressively higher thresholds.}
{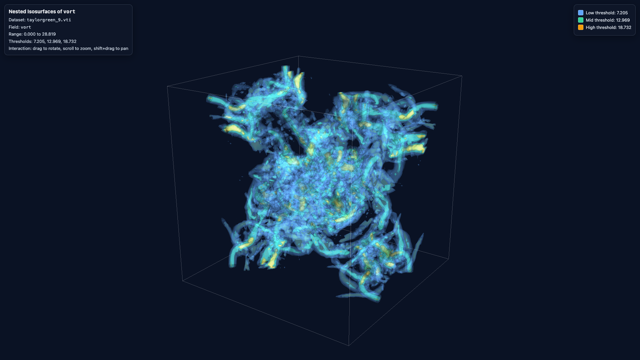}
{ & Raiven  & 1.00 & 1.0 & 1.0 & 1.0 & 1.00 & 0.30 \\
 & ChatGPT & 1.00 & 1.0 & 1.0 & 1.0 & 1.00 & 0.38 \\
 & Claude  & 1.00 & 1.0 & 1.0 & 1.0 & 1.00 & 0.38 \\
 & Gemini  & 0.00 & 0.0 & 0.0 & 0.0 & 0.00 & 7.65 \\}

\benchmarkentry{S17}{taylorgreen\_9.vti}{1}
{Extract an oblique slice through the ux field in taylorgreen\_9.vti that is rotated off-center and "Rdylbu" color scheme.}
{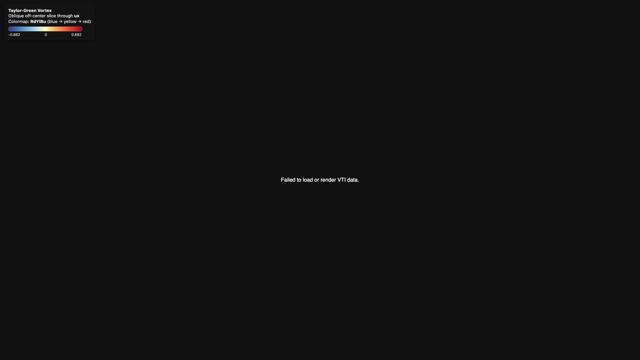}
{ & Raiven  & 1.00 & 1.0 & 1.0 & 1.0 & 1.00 & 0.26 \\
 & ChatGPT & 0.60 & 0.6 & 0.6 & 0.6 & 0.40 & 0.34 \\
 & Claude  & 0.00 & 0.0 & 0.0 & 0.0 & 0.00 & 1.11 \\
 & Gemini  & 0.00 & 0.0 & 0.0 & 0.0 & 0.00 & 1.36 \\}

\benchmarkentry{S18}{divcurl\_0.vti, divcurl\_10.vti, divcurl\_19.vti}{9}
{Use divcurl\_0.vti, divcurl\_10.vti, and divcurl\_19.vti to create a 3×3 grid across timesteps, with LIC generated from the 2D vector field using vx and vy, and slices of div and curl.}
{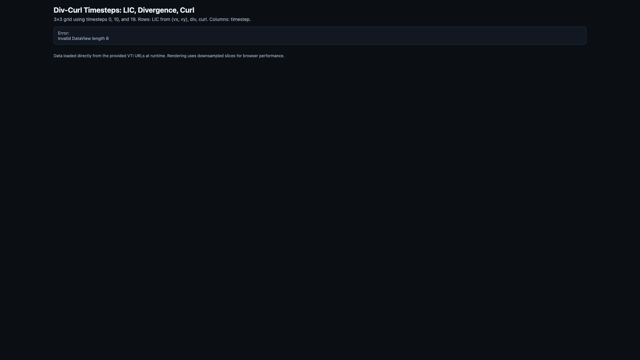}
{ & Raiven  & 1.00 & 1.0 & 1.0 & 1.0 & 1.00 & 0.62 \\
 & ChatGPT & 0.00 & 0.0 & 0.0 & 0.0 & 0.40 & 0.75 \\
 & Claude  & 0.53 & 0.6 & 0.6 & 0.4 & 0.40 & 1.36 \\
 & Gemini  & 0.00 & 0.0 & 0.0 & 0.0 & 0.00 & 3.24 \\}

\benchmarkentry{S19}{taylorgreen\_6.vti, taylorgreen\_8.vti}{1}
{Overlay streamlines from taylorgreen\_6.vti and taylorgreen\_8.vti in a single scene using vx = "ux", vy = "uy", and vz = "uz", distinguishing timesteps with different colors.}
{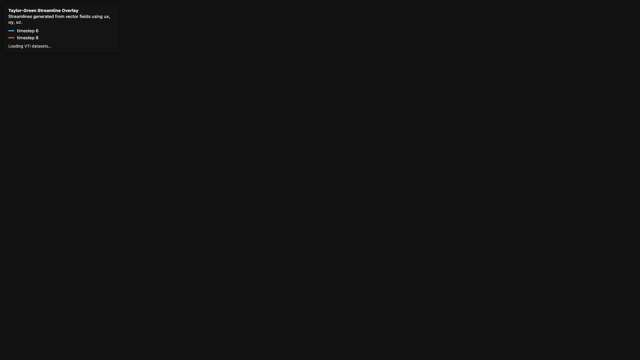}
{ & Raiven  & 1.00 & 1.0 & 1.0 & 1.0 & 1.00 & 0.25 \\
 & ChatGPT & 0.60 & 0.6 & 0.6 & 0.6 & 0.40 & 0.39 \\
 & Claude  & 0.60 & 0.6 & 0.6 & 0.6 & 0.00 & 1.04 \\
 & Gemini  & 0.60 & 0.6 & 0.6 & 0.6 & 0.40 & 1.47 \\}

\benchmarkentry{S20}{taylorgreen\_9.vti}{2}
{Using taylorGreen\_t\_9.vti, create two coordinated views of "vort": a slice in the XY plane at mid-depth and a 3D view combining a volume rendering of "vort" with an isosurface of "vort" and the XY slice. Link the slice position along Z so moving the 2D slice updates the cutting plane in the 3D view.}
{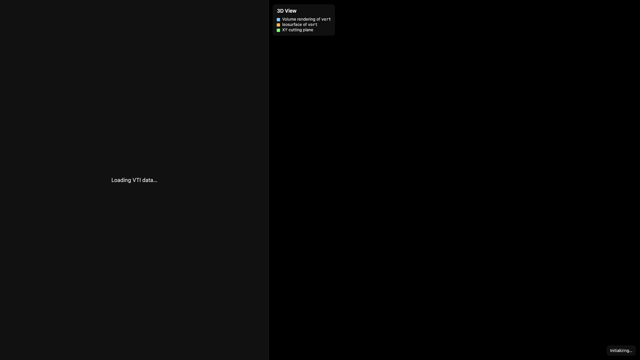}
{ & Raiven  & 1.00 & 1.0 & 1.0 & 1.0 & 1.00 & 0.43 \\[2pt]
 & ChatGPT & 0.43 & 0.4 & 0.4 & 0.5 & 0.40 & 0.56 \\[2pt]
 & Claude  & 0.13 & 0.0 & 0.4 & 0.0 & 0.40 & 0.87 \\[2pt]
 & Gemini  & 0.40 & 0.4 & 0.4 & 0.4 & 0.40 & 1.93 \\}

\benchmarkentry{S21}{taylorgreen\_9.vti}{2}
{Display two side-by-side streamline visualizations from taylorgreen\_9.vti using vy = "ux", vy = "uy", and vz = "uz", one in red with thin streamlines and one in blue with thick streamlines.}
{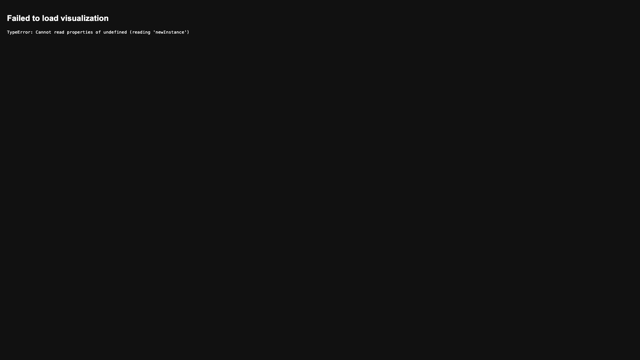}
{ & Raiven  & 1.00 & 1.0 & 1.0 & 1.0 & 1.00 & 0.29 \\
 & ChatGPT & 0.00 & 0.0 & 0.0 & 0.0 & 0.00 & 0.34 \\
 & Claude  & 1.00 & 1.0 & 1.0 & 1.0 & 1.00 & 1.12 \\
 & Gemini  & 0.53 & 0.6 & 0.6 & 0.4 & 0.40 & 8.38 \\}

\benchmarkentry{S22}{taylorgreen\_6.vti, taylorgreen\_8.vti}{2}
{Show two separate isosurface views of vort at the same threshold from taylorgreen\_6.vti and taylorgreen\_8.vti to compare turbulence decay.}
{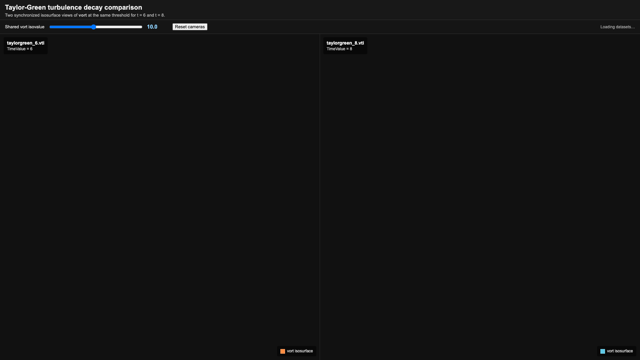}
{ & Raiven  & 1.00 & 1.0 & 1.0 & 1.0 & 1.00 & 0.25 \\
 & ChatGPT & 0.60 & 0.6 & 0.6 & 0.6 & 0.40 & 0.45 \\
 & Claude  & 0.20 & 0.0 & 0.6 & 0.0 & 0.40 & 3.75 \\
 & Gemini  & 0.53 & 0.6 & 0.6 & 0.4 & 0.40 & 1.21 \\}

\benchmarkentry{S23}{taylorgreen\_9.vti}{4}
{Arrange a four-view grid from taylorgreen\_9.vti with volume renderings of ux, uy, uz, and vort, each using different color palette.}
{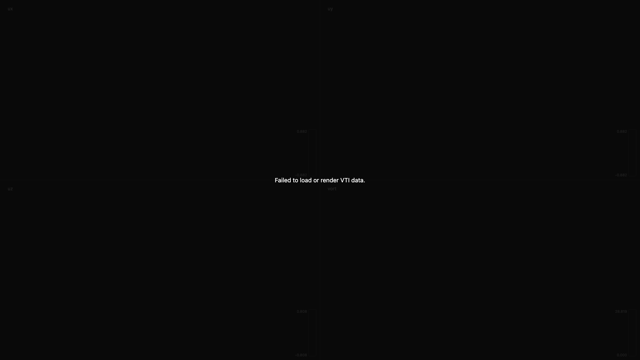}
{ & Raiven  & 0.95 & 1.0 & 1.0 & 0.8 & 1.00 & 0.29 \\
 & ChatGPT & 0.20 & 0.0 & 0.6 & 0.0 & 0.00 & 0.78 \\
 & Claude  & 0.00 & 0.0 & 0.0 & 0.0 & 0.00 & 1.18 \\
 & Gemini  & 0.53 & 0.6 & 0.6 & 0.4 & 0.40 & 1.82 \\}

\benchmarkentry{S24}{cells.vti}{2}
{Create a view with an XY slice of the fluorescence field "Sec61-beta" from cells.vti in grayscale. Add a second linked view that shows the same "Sec61-beta" slice in grayscale overlaid with isosurfaces of "seg\_Sec61-beta" in yellow, "seg\_Membrane" in dark blue, and "seg\_DNA" in light blue.}
{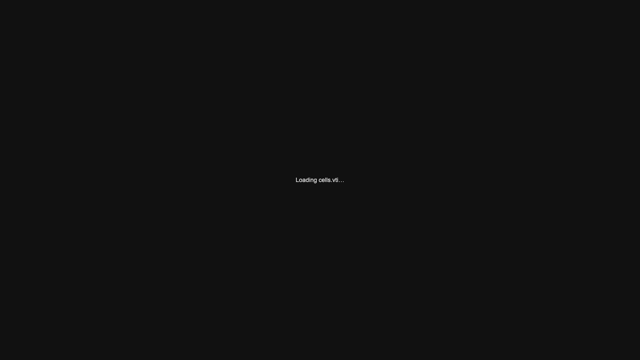}
{ & Raiven  & 1.00 & 1.0 & 1.0 & 1.0 & 1.00 & 0.44 \\[2pt]
 & ChatGPT & 0.00 & 0.0 & 0.0 & 0.0 & 0.00 & 0.90 \\[2pt]
 & Claude  & 0.87 & 0.7 & 0.9 & 1.0 & 1.00 & 0.95 \\[2pt]
 & Gemini  & 0.60 & 0.6 & 0.8 & 0.4 & 0.40 & 1.63 \\}

\benchmarkentry{S25}{cells.vti}{3}
{Create three views each showing an isosurface from cells.vti, one using seg\_Sec61-beta, one using seg\_Membrane, and one using seg\_DNA, each should be semi-transparent and a unique color.}
{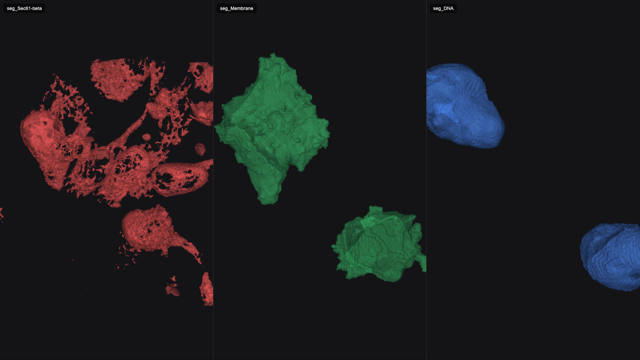}
{ & Raiven  & 1.00 & 1.0 & 1.0 & 1.0 & 1.00 & 0.31 \\
 & ChatGPT & 1.00 & 1.0 & 1.0 & 1.0 & 1.00 & 0.35 \\
 & Claude  & 0.53 & 0.6 & 0.6 & 0.4 & 0.40 & 0.56 \\
 & Gemini  & 0.53 & 0.6 & 0.6 & 0.4 & 0.40 & 1.28 \\}

\benchmarkentry{S26}{cells.vti}{1}
{Combine a volume rendering of Sec61-beta with an isosurface from seg\_Membrane in a single view from cells.vti.}
{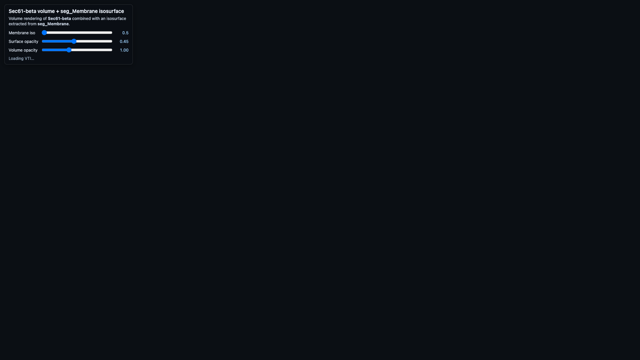}
{ & Raiven  & 1.00 & 1.0 & 1.0 & 1.0 & 1.00 & 0.29 \\
 & ChatGPT & 0.60 & 0.6 & 0.6 & 0.6 & 0.40 & 0.54 \\
 & Claude  & 0.00 & 0.0 & 0.0 & 0.0 & 0.40 & 0.65 \\
 & Gemini  & 0.00 & 0.0 & 0.0 & 0.0 & 0.00 & 0.91 \\}

\benchmarkentry{S27}{cells.vti}{1}
{Layer volume renderings of Sec61-beta, Membrane, and DNA with isosurfaces from seg\_Sec61-beta, seg\_Membrane, and seg\_DNA in one composite view from cells.vti. All volumes should use same color palette, while all isosurfaces should be distinct colors.}
{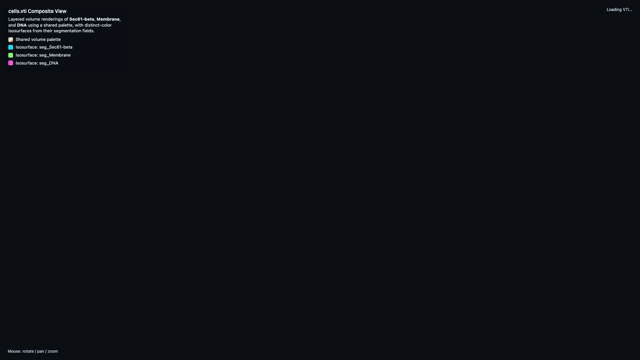}
{ & Raiven  & 1.00 & 1.0 & 1.0 & 1.0 & 1.00 & 0.44 \\[2pt]
 & ChatGPT & 0.60 & 0.6 & 0.6 & 0.6 & 0.40 & 0.56 \\[2pt]
 & Claude  & 0.00 & 0.0 & 0.0 & 0.0 & 0.00 & 0.84 \\[2pt]
 & Gemini  & 0.00 & 0.0 & 0.0 & 0.0 & 0.00 & 2.02 \\}

\benchmarkentry{S28}{cells.vti}{3}
{Using cells.vti, create three slice (XY) views using color palette plasma: one for "Sec61-beta", one for "Membrane", and one for "DNA", and link their slice positions so scrolling in any view updates the other two.}
{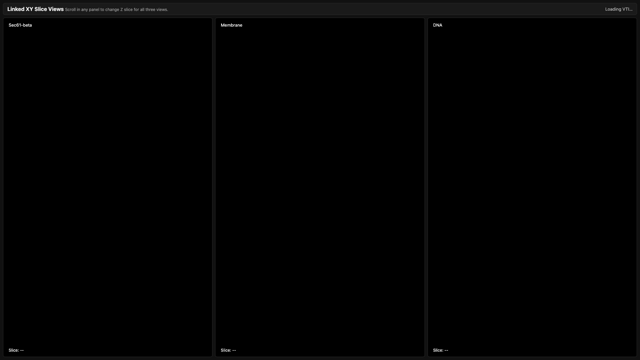}
{ & Raiven  & 1.00 & 1.0 & 1.0 & 1.0 & 1.00 & 0.42 \\
 & ChatGPT & 0.40 & 0.4 & 0.4 & 0.4 & 0.40 & 0.40 \\
 & Claude  & 1.00 & 1.0 & 1.0 & 1.0 & 1.00 & 1.22 \\
 & Gemini  & 0.40 & 0.4 & 0.4 & 0.4 & 0.40 & 2.34 \\}

\benchmarkentry{S29}{cells.vti}{2}
{Display a volume rendering of Sec61-beta in one view and triplanar slices of the same field in another from cells.vti.}
{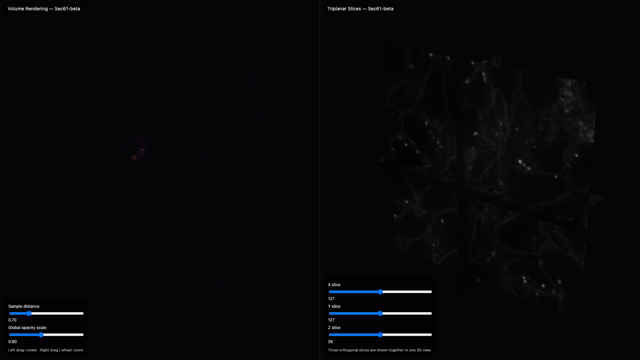}
{ & Raiven  & 1.00 & 1.0 & 1.0 & 1.0 & 1.00 & 0.27 \\
 & ChatGPT & 0.87 & 0.6 & 1.0 & 1.0 & 1.00 & 0.60 \\
 & Claude  & 0.00 & 0.0 & 0.0 & 0.0 & 0.40 & 0.69 \\
 & Gemini  & 0.53 & 0.6 & 0.6 & 0.4 & 0.40 & 1.83 \\}

\benchmarkentry{S30}{cells.vti}{4}
{cells.vti contains fluorescence channels ("Sec61-beta", "Membrane", "DNA") and segmentation channels ("seg\_Sec61-beta", "seg\_Membrane", "seg\_DNA"). Create four views: (1) an XY slice of the fluorescence channel "Membrane" in grayscale, (2) a YZ slice of "Membrane" in grayscale, (3) an XZ slice of "Membrane" in grayscale, and (4) a combined view showing triplanar slices of "Membrane" in grayscale overlaid with isosurfaces of "seg\_Sec61-beta" in yellow, "seg\_Membrane" in dark blue, and "seg\_DNA" in light blue. Link the slice positions across all four views so scrolling in any slice view updates the others and the triplanar.}
{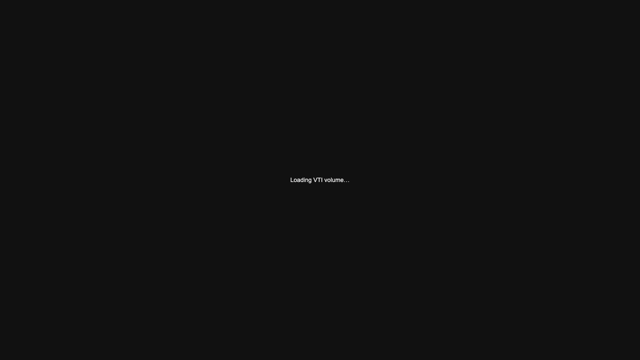}
{ & Raiven  & 1.00 & 1.0 & 1.0 & 1.0 & 1.00 & 0.47 \\[10pt]
 & ChatGPT & 0.00 & 0.0 & 0.0 & 0.0 & 0.00 & 0.62 \\[10pt]
 & Claude  & 1.00 & 1.0 & 1.0 & 1.0 & 1.00 & 1.06 \\[10pt]
 & Gemini  & 0.40 & 0.4 & 0.4 & 0.4 & 0.40 & 2.72 \\}

\subsubsection{InfoVis (I)}

\benchmarkentry{I31}{countries\_latest.csv, continents\_timeseries.csv}{3}
{Build a bubble chart from countries\_latest.csv plotting "gdp\_per\_capita (2023)" versus "life\_expectancy (2023)", sized by "population (2023)" and colored by "continent", with a linked histogram of "gdp\_per\_capita (2023)" that filters when brushing the bubbles. Add a multiline chart from continents\_timeseries.csv tracking "gdp\_pc" over "year" with lines colored by "continent".}
{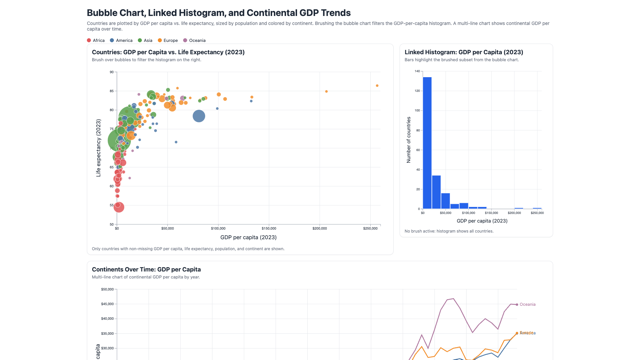}
{ & Raiven  & 1.00 & 1.0 & 1.0 & 1.0 & 1.00 & 0.28 \\[4pt]
 & ChatGPT & 1.00 & 1.0 & 1.0 & 1.0 & 1.00 & 1.00 \\[4pt]
 & Claude  & 1.00 & 1.0 & 1.0 & 1.0 & 1.00 & 1.41 \\[4pt]
 & Gemini  & 1.00 & 1.0 & 1.0 & 1.0 & 1.00 & 2.46 \\}

\benchmarkentry{I32}{countries\_latest.csv, continents\_timeseries.csv}{2}
{Show the demographic transition: scatter "gdp\_per\_capita (2023)" versus "fertility\_rate (2023)" from countries\_latest.csv, sized by "population (2023)" and colored by "continent", paired with a multiline chart from continents\_timeseries.csv plotting "fertility" over "year" with lines colored by "continent".}
{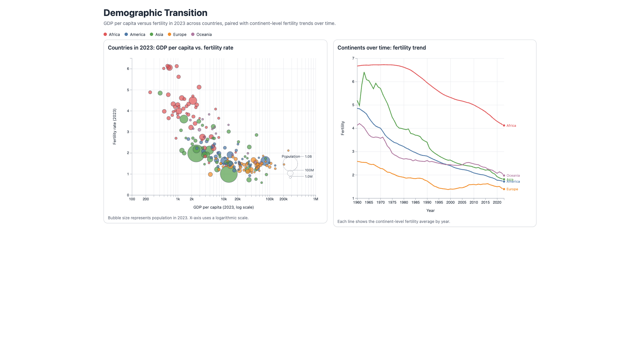}
{ & Raiven  & 1.00 & 1.0 & 1.0 & 1.0 & 1.00 & 0.20 \\[2pt]
 & ChatGPT & 1.00 & 1.0 & 1.0 & 1.0 & 1.00 & 1.68 \\[2pt]
 & Claude  & 1.00 & 1.0 & 1.0 & 1.0 & 1.00 & 0.99 \\[2pt]
 & Gemini  & 1.00 & 1.0 & 1.0 & 1.0 & 1.00 & 1.38 \\}

\benchmarkentry{I33}{countries\_latest.csv, world\_iso3.geojson}{1}
{Produce a ridgeline plot of "median\_age (2023)" grouped by "region" from countries\_latest.csv.}
{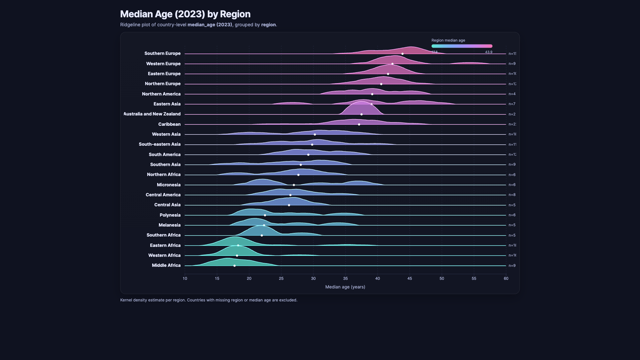}
{ & Raiven  & 1.00 & 1.0 & 1.0 & 1.0 & 1.00 & 0.22 \\
 & ChatGPT & 1.00 & 1.0 & 1.0 & 1.0 & 1.00 & 0.70 \\
 & Claude  & 0.00 & 0.0 & 0.0 & 0.0 & 0.40 & 0.68 \\
 & Gemini  & 1.00 & 1.0 & 1.0 & 1.0 & 1.00 & 1.37 \\}

\benchmarkentry{I34}{countries\_latest.csv}{2}
{Lay out a parallel-coordinates plot from countries\_latest.csv spanning "gdp\_per\_capita (2023)", "life\_expectancy (2023)", "population (2023)", "median\_age (2023)", and "fertility\_rate (2023)" with lines colored by "continent", linked to a scatter of "gdp\_per\_capita (2023)" versus "life\_expectancy (2023)" colored by "continent". Brushing lines in the parallel coordinates should highlight the corresponding points in the scatter.}
{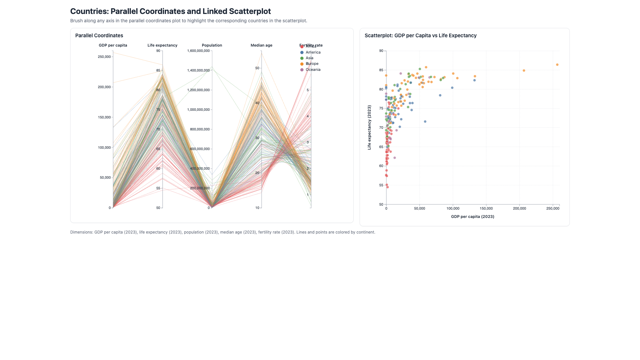}
{ & Raiven  & 1.00 & 1.0 & 1.0 & 1.0 & 1.00 & 0.29 \\[4pt]
 & ChatGPT & 0.93 & 0.8 & 1.0 & 1.0 & 1.00 & 0.77 \\[4pt]
 & Claude  & 0.83 & 0.7 & 0.9 & 0.9 & 1.00 & 0.79 \\[4pt]
 & Gemini  & 0.93 & 0.8 & 1.0 & 1.0 & 1.00 & 1.68 \\}

\benchmarkentry{I35}{countries\_latest.csv}{4}
{Display "gdp\_per\_capita (2023)" from countries\_latest.csv four ways side by side: a histogram, a KDE plot, a boxplot grouped by "continent", and a violin plot grouped by "continent".}
{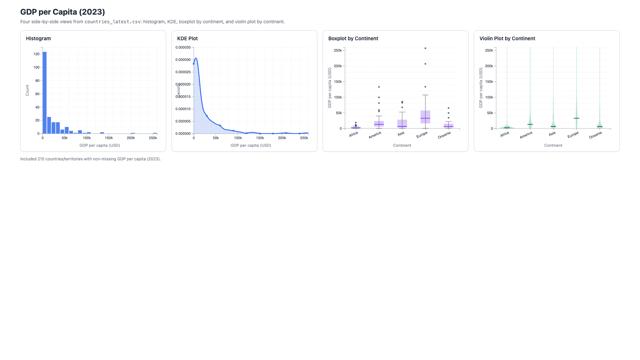}
{ & Raiven  & 1.00 & 1.0 & 1.0 & 1.0 & 1.00 & 0.27 \\
 & ChatGPT & 1.00 & 1.0 & 1.0 & 1.0 & 1.00 & 1.06 \\
 & Claude  & 1.00 & 1.0 & 1.0 & 1.0 & 1.00 & 1.13 \\
 & Gemini  & 0.98 & 1.0 & 1.0 & 0.9 & 1.00 & 1.05 \\}

\benchmarkentry{I36}{countries\_timeseries.csv, countries\_latest.csv}{2}
{Create a stacked bar chart from countries\_timeseries.csv with "year" on the x-axis and "pop" aggregated by "continent", next to a pie chart of "population (2023)" from countries\_latest.csv grouped by "continent". Use consistent continent colors across both views.}
{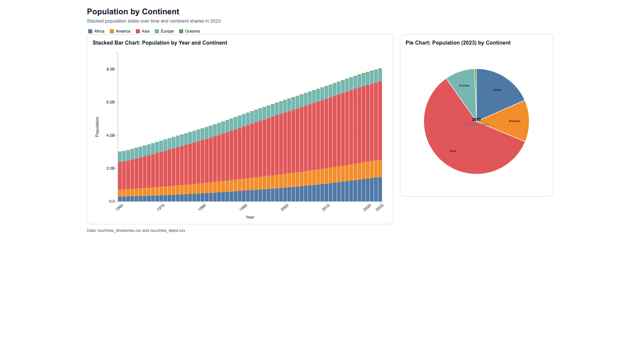}
{ & Raiven  & 1.00 & 1.0 & 1.0 & 1.0 & 1.00 & 0.30 \\[2pt]
 & ChatGPT & 1.00 & 1.0 & 1.0 & 1.0 & 1.00 & 0.65 \\[2pt]
 & Claude  & 1.00 & 1.0 & 1.0 & 1.0 & 1.00 & 0.72 \\[2pt]
 & Gemini  & 1.00 & 1.0 & 1.0 & 1.0 & 1.00 & 0.90 \\}

\benchmarkentry{I37}{region\_timeseries.csv}{1}
{Plot population trajectories from regions\_timeseries.csv as a multiline chart with "year" on the x-axis, "pop" on the y-axis, and lines colored by "region".}
{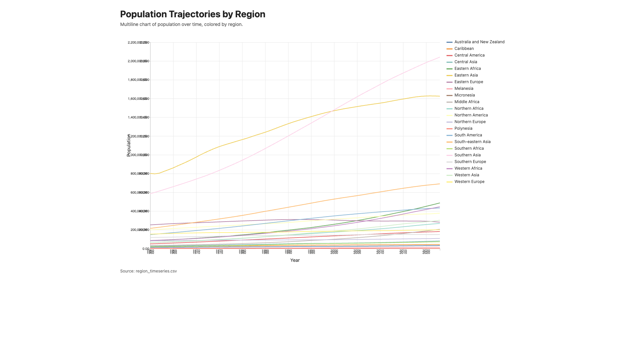}
{ & Raiven  & 1.00 & 1.0 & 1.0 & 1.0 & 1.00 & 0.17 \\
 & ChatGPT & 0.93 & 1.0 & 1.0 & 0.8 & 1.00 & 0.50 \\
 & Claude  & 1.00 & 1.0 & 1.0 & 1.0 & 1.00 & 0.51 \\
 & Gemini  & 1.00 & 1.0 & 1.0 & 1.0 & 1.00 & 0.71 \\}

\benchmarkentry{I38}{world\_life\_exp.csv}{1}
{Layer a line of "global\_average" life expectancy over a shaded band spanning "global\_min" to "global\_max" from world\_life\_exp.csv, with "year" on the x-axis.}
{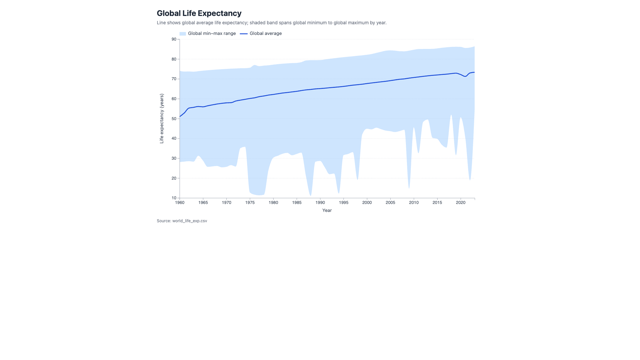}
{ & Raiven  & 0.93 & 1.0 & 1.0 & 0.8 & 1.00 & 0.31 \\
 & ChatGPT & 1.00 & 1.0 & 1.0 & 1.0 & 1.00 & 0.44 \\
 & Claude  & 1.00 & 1.0 & 1.0 & 1.0 & 1.00 & 1.00 \\
 & Gemini  & 1.00 & 1.0 & 1.0 & 1.0 & 1.00 & 0.51 \\}

\benchmarkentry{I39}{countries\_latest.csv, world\_iso3.geojson, continents\_latest.csv}{3}
{Explore the climate-prosperity paradox: scatter "gdp\_per\_capita (2023)" versus "ghg\_per\_capita (2023)" colored by "continent" from countries\_latest.csv, add a choropleth colored by "ghg\_per\_capita (2023)" using country geometries from world\_iso3.geojson and a red colorscale, and include a bar chart from continents\_latest.csv with "continent" on the x-axis and "ghg\_per\_capita (2023)" on the y-axis, colored by "continent".}
{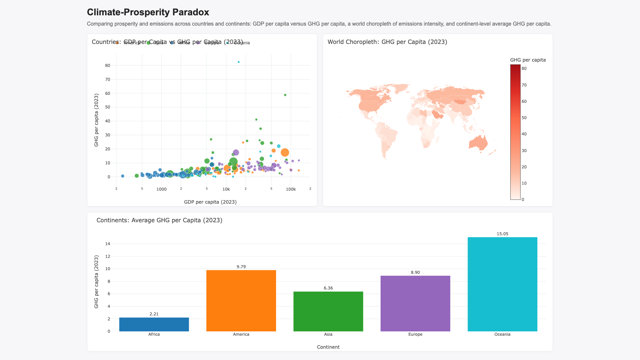}
{ & Raiven  & 1.00 & 1.0 & 1.0 & 1.0 & 1.00 & 0.34 \\[4pt]
 & ChatGPT & 1.00 & 1.0 & 1.0 & 1.0 & 1.00 & 0.55 \\[4pt]
 & Claude  & 0.89 & 0.9 & 1.0 & 0.8 & 0.40 & 0.95 \\[4pt]
 & Gemini  & 1.00 & 1.0 & 1.0 & 1.0 & 1.00 & 1.19 \\}

\benchmarkentry{I40}{countries\_latest.csv, world\_iso3.geojson}{2}
{Pair a scatter of "gdp\_per\_capita (2023)" versus "internet\_pct (2023)" colored by "continent" from countries\_latest.csv with a choropleth colored by "internet\_pct (2023)" using country geometries from world\_iso3.geojson.}
{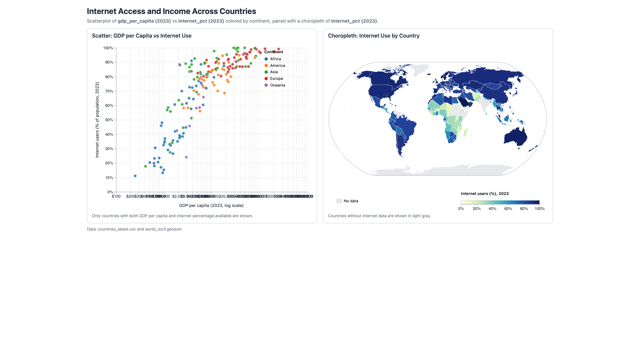}
{ & Raiven  & 1.00 & 1.0 & 1.0 & 1.0 & 1.00 & 0.31 \\[2pt]
 & ChatGPT & 1.00 & 1.0 & 1.0 & 1.0 & 1.00 & 0.81 \\[2pt]
 & Claude  & 0.90 & 0.9 & 0.9 & 0.9 & 1.00 & 0.82 \\[2pt]
 & Gemini  & 1.00 & 1.0 & 1.0 & 1.0 & 1.00 & 1.10 \\}

\benchmarkentry{I41}{continents\_timeseries.csv, countries\_latest.csv, world\_iso3.geojson}{2}
{Chart the urbanization wave: a multiline of "urban\_pct" over "year" colored by "continent" from continents\_timeseries.csv, plus a choropleth colored by "urban\_pct (2023)" using country geometries from world\_iso3.geojson and countries\_latest.csv.}
{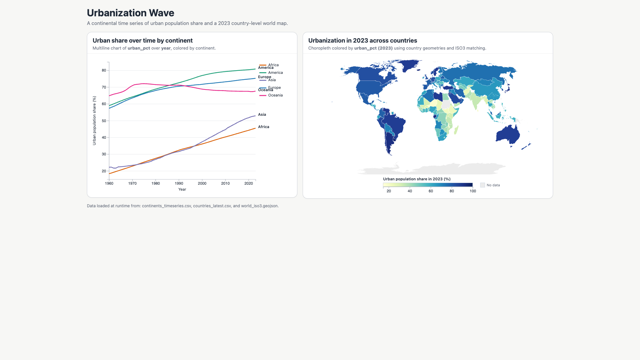}
{ & Raiven  & 1.00 & 1.0 & 1.0 & 1.0 & 1.00 & 0.30 \\[2pt]
 & ChatGPT & 1.00 & 1.0 & 1.0 & 1.0 & 1.00 & 1.21 \\[2pt]
 & Claude  & 1.00 & 1.0 & 1.0 & 1.0 & 1.00 & 0.95 \\[2pt]
 & Gemini  & 1.00 & 1.0 & 1.0 & 1.0 & 1.00 & 1.89 \\}

\benchmarkentry{I42}{countries\_latest.csv}{2}
{Scatter "health\_exp\_per\_capita (2023)" versus "life\_expectancy (2023)" from countries\_latest.csv as bubbles sized by "population (2023)" and colored by "continent", with a companion violin plot of "life\_expectancy (2023)" grouped by "continent".}
{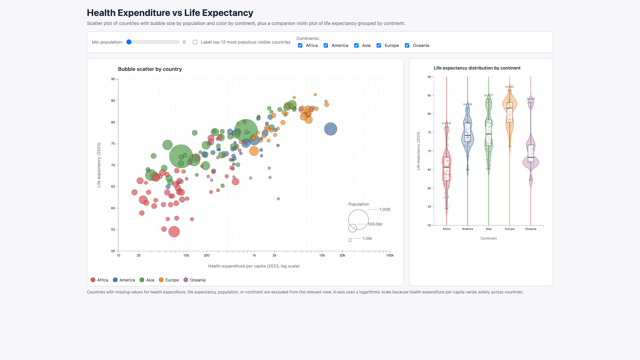}
{ & Raiven  & 1.00 & 1.0 & 1.0 & 1.0 & 1.00 & 0.27 \\[2pt]
 & ChatGPT & 1.00 & 1.0 & 1.0 & 1.0 & 1.00 & 1.33 \\[2pt]
 & Claude  & 1.00 & 1.0 & 1.0 & 1.0 & 1.00 & 0.97 \\[2pt]
 & Gemini  & 1.00 & 1.0 & 1.0 & 1.0 & 1.00 & 1.87 \\}

\benchmarkentry{I43}{countries\_latest.csv, world\_iso3.geojson, region\_energy.csv}{2}
{Map renewable energy: a choropleth colored by "renewable\_energy\_pct (2012)" using country geometries from world\_iso3.geojson and countries\_latest.csv, alongside a stacked bar chart from region\_energy.csv with "region" on the x-axis, "value" on the y-axis, and stacked by "category" (renewable, nuclear, fossil).}
{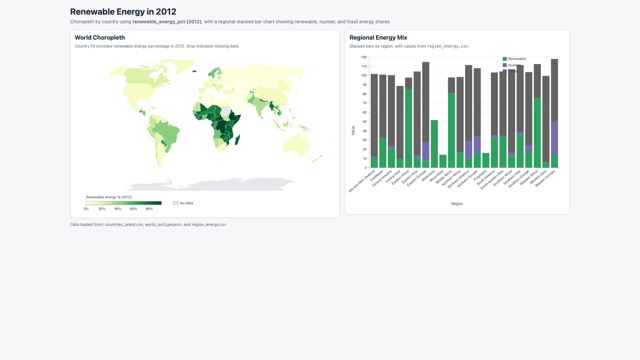}
{ & Raiven  & 1.00 & 1.0 & 1.0 & 1.0 & 1.00 & 0.28 \\[2pt]
 & ChatGPT & 1.00 & 1.0 & 1.0 & 1.0 & 1.00 & 0.92 \\[2pt]
 & Claude  & 0.90 & 0.9 & 0.9 & 0.9 & 1.00 & 0.80 \\[2pt]
 & Gemini  & 1.00 & 1.0 & 1.0 & 1.0 & 1.00 & 1.48 \\}

\benchmarkentry{I44}{trade\_edges.csv}{1}
{Visualize regional trade intensity from trade\_edges.csv as a heatmap with "source" on the x-axis, "target" on the y-axis, and cells colored by "trade\_pct".}
{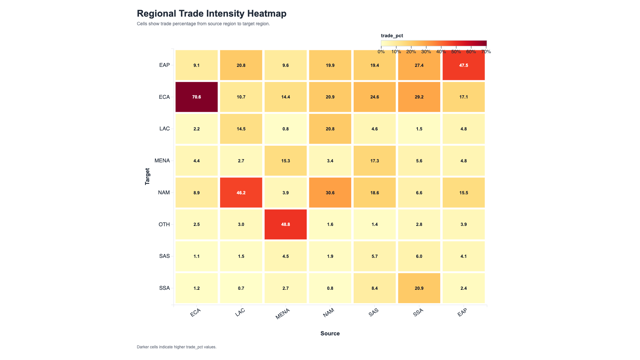}
{ & Raiven  & 1.00 & 1.0 & 1.0 & 1.0 & 1.00 & 0.21 \\
 & ChatGPT & 1.00 & 1.0 & 1.0 & 1.0 & 1.00 & 0.46 \\
 & Claude  & 1.00 & 1.0 & 1.0 & 1.0 & 1.00 & 0.51 \\
 & Gemini  & 1.00 & 1.0 & 1.0 & 1.0 & 1.00 & 0.59 \\}

\benchmarkentry{I45}{trade\_edges.csv}{1}
{Construct a chord diagram from trade\_edges.csv connecting "source" to "target", sized by "trade\_val\_kusd", to reveal which inter-regional trade relationships are balanced and which are lopsided.}
{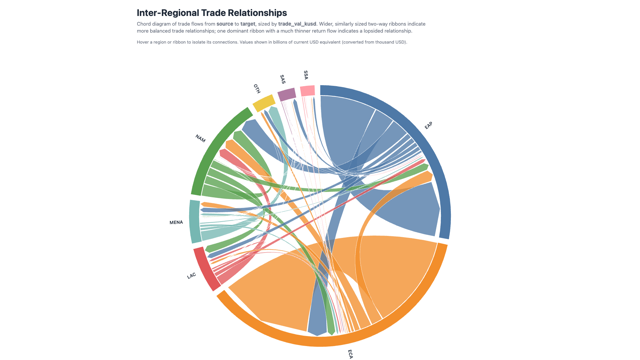}
{ & Raiven  & 1.00 & 1.0 & 1.0 & 1.0 & 1.00 & 0.24 \\
 & ChatGPT & 1.00 & 1.0 & 1.0 & 1.0 & 1.00 & 0.64 \\
 & Claude  & 0.93 & 1.0 & 1.0 & 0.8 & 1.00 & 0.90 \\
 & Gemini  & 1.00 & 1.0 & 1.0 & 1.0 & 1.00 & 2.07 \\}

\benchmarkentry{I46}{trade\_edges.csv, countries\_latest.csv, world\_iso3.geojson}{3}
{Assemble a three-view trade dashboard: a sankey from trade\_edges.csv with "source" flowing to "target" sized by "trade\_val\_kusd", a choropleth colored by "trade\_pct\_gdp (2023)" using countries\_latest.csv and world\_iso3.geojson, and a bar chart with "target" on the x-axis and summed "trade\_val\_kusd" on the y-axis, colored by "target" region.}
{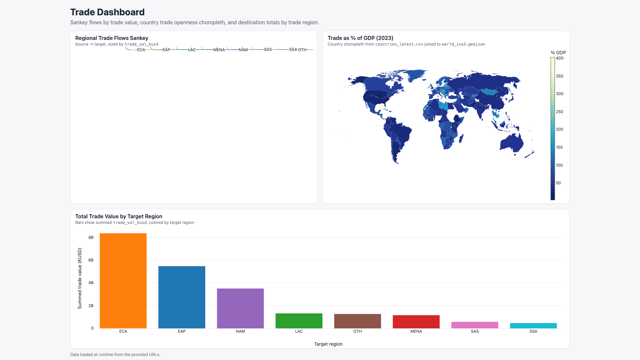}
{ & Raiven  & 0.98 & 1.0 & 1.0 & 0.9 & 1.00 & 0.31 \\[2pt]
 & ChatGPT & 0.89 & 0.9 & 0.9 & 0.8 & 1.00 & 0.83 \\[2pt]
 & Claude  & 0.53 & 0.6 & 0.6 & 0.4 & 0.40 & 1.02 \\[2pt]
 & Gemini  & 1.00 & 1.0 & 1.0 & 1.0 & 1.00 & 1.60 \\}

\benchmarkentry{I47}{trade\_edges.csv}{1}
{Show the main highways of global commerce as a sankey diagram from trade\_edges.csv, with "source" flowing to "target" sized by "trade\_val\_kusd".}
{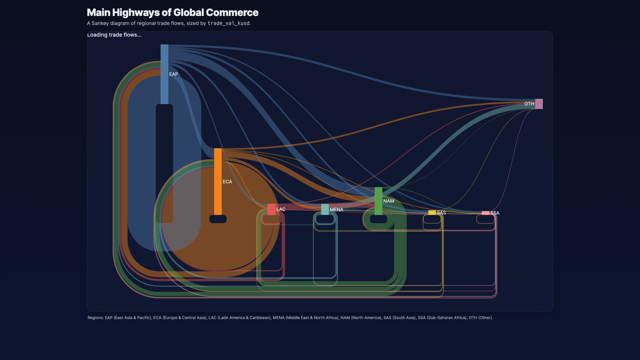}
{ & Raiven  & 1.00 & 1.0 & 1.0 & 1.0 & 1.00 & 0.27 \\
 & ChatGPT & 1.00 & 1.0 & 1.0 & 1.0 & 1.00 & 0.52 \\
 & Claude  & 0.20 & 0.6 & 0.0 & 0.0 & 0.00 & 0.65 \\
 & Gemini  & 1.00 & 1.0 & 1.0 & 1.0 & 1.00 & 1.27 \\}

\benchmarkentry{I48}{trade\_edges.csv, countries\_latest.csv, world\_iso3.geojson}{2}
{Contrast trade volume with economic weight: a chord diagram from trade\_edges.csv connecting "source" to "target" sized by "trade\_val\_kusd", paired with a choropleth colored by "gdp\_per\_capita (2023)" using countries\_latest.csv and world\_iso3.geojson.}
{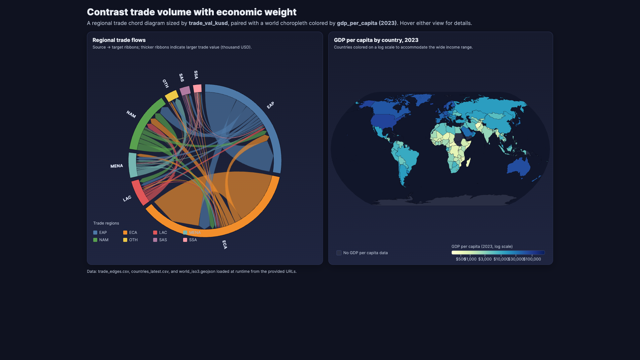}
{ & Raiven  & 1.00 & 1.0 & 1.0 & 1.0 & 1.00 & 0.31 \\[2pt]
 & ChatGPT & 1.00 & 1.0 & 1.0 & 1.0 & 1.00 & 1.22 \\[2pt]
 & Claude  & 1.00 & 1.0 & 1.0 & 1.0 & 1.00 & 0.88 \\[2pt]
 & Gemini  & 1.00 & 1.0 & 1.0 & 1.0 & 1.00 & 2.06 \\}

\benchmarkentry{I49}{region\_border\_network.json, countries\_latest.csv, world\_iso3.geojson}{2}
{Map the world's neighborhood network: a force-directed graph from region\_border\_network.json with nodes colored by "group" and edges connecting bordering pairs, alongside a choropleth colored by "num\_neighbors" using world\_iso3.geojson with a point layer at "capital\_lat", "capital\_lon" where the points are sized by "capital\_population (2025)" from countries\_latest.csv and colored by "region".}
{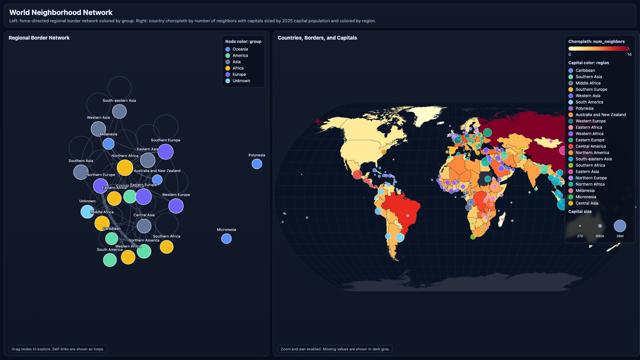}
{ & Raiven  & 1.00 & 1.0 & 1.0 & 1.0 & 1.00 & 0.49 \\[4pt]
 & ChatGPT & 1.00 & 1.0 & 1.0 & 1.0 & 1.00 & 1.34 \\[4pt]
 & Claude  & 1.00 & 1.0 & 1.0 & 1.0 & 1.00 & 1.05 \\[4pt]
 & Gemini  & 0.53 & 0.6 & 0.6 & 0.4 & 0.40 & 1.28 \\}

\benchmarkentry{I50}{region\_border\_network.json, region\_border\_edges.csv}{2}
{Present borders two ways: (1) A force-directed graph from region\_border\_network.json with region nodes colored by "group" and edges weighted by "value". (2) An adjacency heatmap from region\_border\_edges.csv with "region\_a" on one axis, "region\_b" on the other, and cells colored by "distance\_km". Selecting a region node in the graph should highlight its row and column in the heatmap.}
{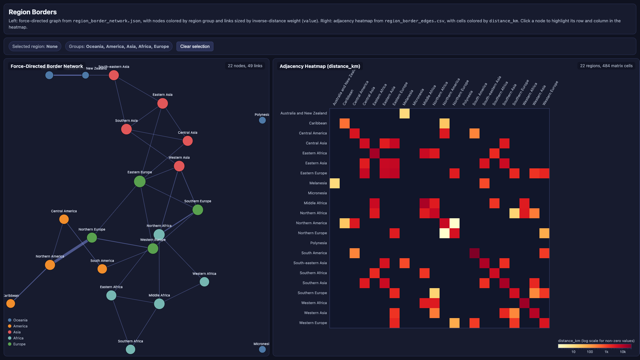}
{ & Raiven  & 1.00 & 1.0 & 1.0 & 1.0 & 1.00 & 0.36 \\[4pt]
 & ChatGPT & 1.00 & 1.0 & 1.0 & 1.0 & 1.00 & 1.28 \\[4pt]
 & Claude  & 1.00 & 1.0 & 1.0 & 1.0 & 1.00 & 0.99 \\[4pt]
 & Gemini  & 1.00 & 1.0 & 1.0 & 1.0 & 1.00 & 1.21 \\}

\benchmarkentry{I51}{region\_border\_network.json}{1}
{Build a force-directed graph of border connections using region\_border\_network.json with region nodes colored by "group".}
{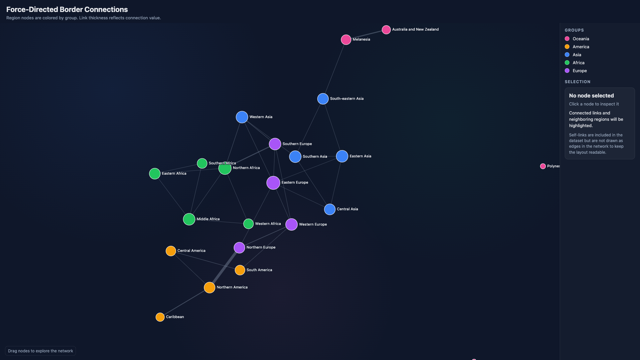}
{ & Raiven  & 1.00 & 1.0 & 1.0 & 1.0 & 1.00 & 0.26 \\
 & ChatGPT & 1.00 & 1.0 & 1.0 & 1.0 & 1.00 & 0.81 \\
 & Claude  & 1.00 & 1.0 & 1.0 & 1.0 & 1.00 & 0.67 \\
 & Gemini  & 1.00 & 1.0 & 1.0 & 1.0 & 1.00 & 0.78 \\}

\benchmarkentry{I52}{region\_border\_network.json, region\_latest.csv, world\_iso3.geojson, countries\_latest.csv}{3}
{Make a force-directed border graph from region\_border\_network.json with nodes colored by "group", a choropleth colored by "num\_neighbors" from countries\_latest.csv and world\_iso3.geojson with a hexbin layer aggregating "capital\_lat" using the "turbo" color scheme and "capital\_lon", and a bar chart of "num\_neighbors" from region\_latest.csv.}
{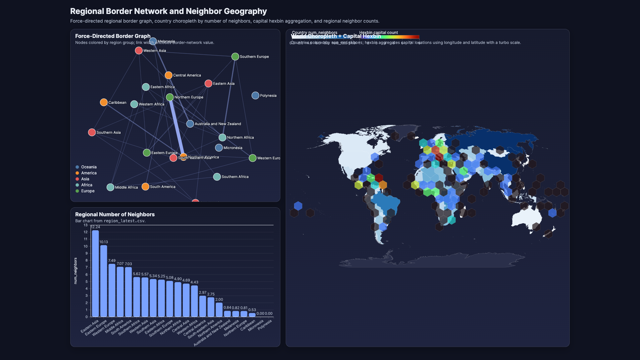}
{ & Raiven  & 1.00 & 1.0 & 1.0 & 1.0 & 1.00 & 0.45 \\[8pt]
 & ChatGPT & 1.00 & 1.0 & 1.0 & 1.0 & 1.00 & 1.09 \\[8pt]
 & Claude  & 0.53 & 0.6 & 0.6 & 0.4 & 0.40 & 0.94 \\[8pt]
 & Gemini  & 0.53 & 0.6 & 0.6 & 0.4 & 0.40 & 1.28 \\}

\benchmarkentry{I53}{gaia.csv}{2}
{Render two views of stellar density from gaia.csv: a heatmap with "ra" on the x-axis and "dec" on the y-axis, and a hexbin plot of the same "ra" versus "dec" coordinates.}
{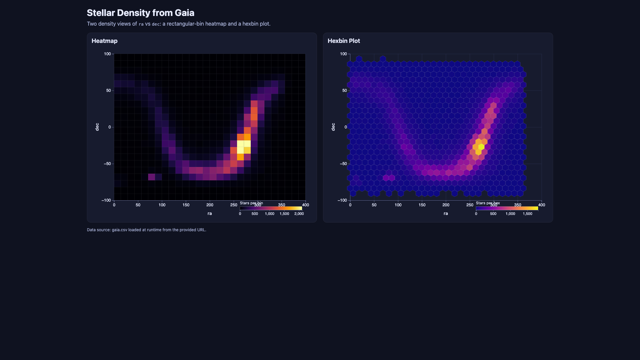}
{ & Raiven  & 1.00 & 1.0 & 1.0 & 1.0 & 1.00 & 0.24 \\
 & ChatGPT & 1.00 & 1.0 & 1.0 & 1.0 & 1.00 & 0.70 \\
 & Claude  & 0.53 & 0.6 & 0.6 & 0.4 & 1.00 & 0.81 \\
 & Gemini  & 1.00 & 1.0 & 1.0 & 1.0 & 1.00 & 1.83 \\}

\benchmarkentry{I54}{gaia.csv}{1}
{Generate a Hertzsprung-Russell diagram from gaia.csv as a heatmap with "bp\_rp" on the x-axis and "phot\_g\_mean\_mag" on the y-axis.}
{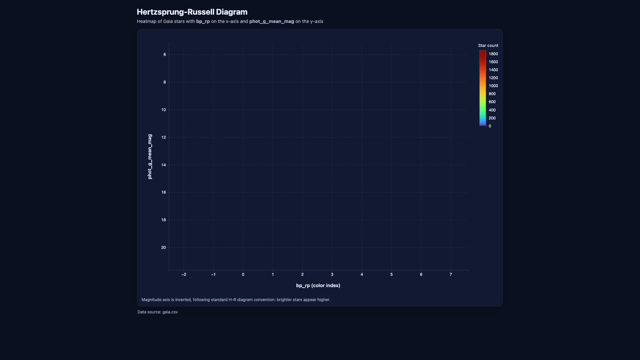}
{ & Raiven  & 1.00 & 1.0 & 1.0 & 1.0 & 1.00 & 0.29 \\
 & ChatGPT & 0.67 & 0.6 & 0.6 & 0.8 & 0.40 & 0.62 \\
 & Claude  & 0.93 & 1.0 & 1.0 & 0.8 & 1.00 & 0.67 \\
 & Gemini  & 1.00 & 1.0 & 1.0 & 1.0 & 1.00 & 0.54 \\}

\benchmarkentry{I55}{gaia.csv}{4}
{Wire up a four-view linked dashboard from gaia.csv: a heatmap of "ra" versus "dec" (sky map), a heatmap of "bp\_rp" versus "phot\_g\_mean\_mag" (HR diagram), a histogram of "phot\_g\_mean\_mag", and a histogram of "parallax". Brushing in any view should filter the others.}
{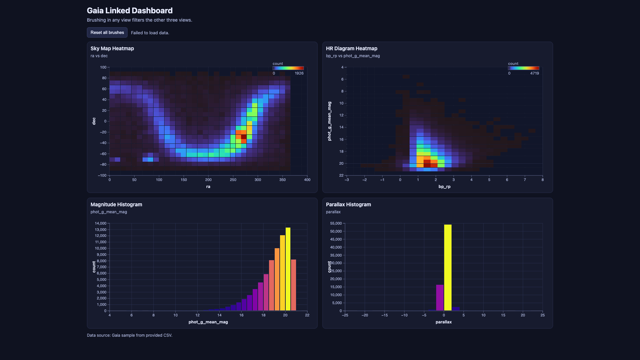}
{ & Raiven  & 0.93 & 1.0 & 0.9 & 0.9 & 1.00 & 0.30 \\[2pt]
 & ChatGPT & 0.80 & 0.8 & 0.8 & 0.8 & 1.00 & 1.09 \\[2pt]
 & Claude  & 1.00 & 1.0 & 1.0 & 1.0 & 1.00 & 1.27 \\[2pt]
 & Gemini  & 1.00 & 1.0 & 1.0 & 1.0 & 1.00 & 2.32 \\}

\benchmarkentry{I56}{gaia.csv}{3}
{Link three views from gaia.csv: a heatmap of "bp\_rp" versus "phot\_g\_mean\_mag" (HR diagram), a histogram of "parallax", and a KDE of "phot\_g\_mean\_mag". Brushing the parallax histogram to select nearby stars should update the HR diagram and the magnitude KDE.}
{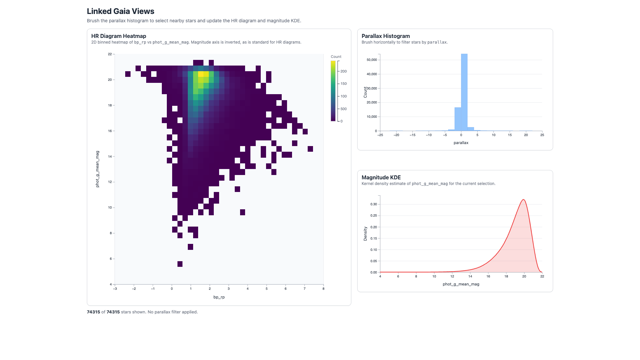}
{ & Raiven  & 0.96 & 1.0 & 0.9 & 0.9 & 1.00 & 0.30 \\[2pt]
 & ChatGPT & 1.00 & 1.0 & 1.0 & 1.0 & 1.00 & 0.92 \\[2pt]
 & Claude  & 1.00 & 1.0 & 1.0 & 1.0 & 1.00 & 1.38 \\[2pt]
 & Gemini  & 1.00 & 1.0 & 1.0 & 1.0 & 1.00 & 1.28 \\}

\benchmarkentry{I57}{countries\_latest.csv, EAP.geojson, ECA.geojson, LAC.geojson, MENA.geojson, NAM.geojson, SAS.geojson, SSA.geojson}{7}
{Tile seven choropleth views from countries\_latest.csv, one per trade-region GeoJSON file (EAP.geojson, ECA.geojson, LAC.geojson, MENA.geojson, NAM.geojson, SAS.geojson, SSA.geojson), each colored by "gdp\_per\_capita (2023)". Use different colorscales for each choropleth.}
{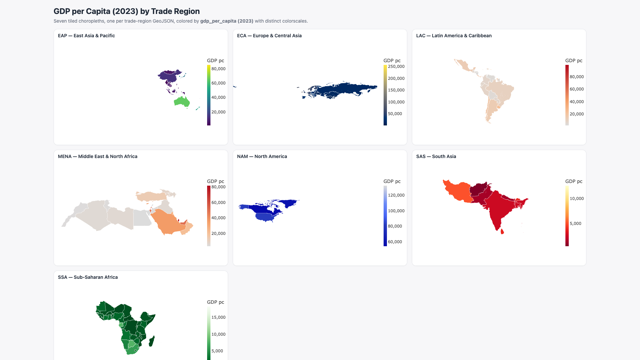}
{ & Raiven  & 1.00 & 1.0 & 1.0 & 1.0 & 1.00 & 0.72 \\[4pt]
 & ChatGPT & 1.00 & 1.0 & 1.0 & 1.0 & 1.00 & 0.48 \\[4pt]
 & Claude  & 1.00 & 1.0 & 1.0 & 1.0 & 1.00 & 0.80 \\[4pt]
 & Gemini  & 1.00 & 1.0 & 1.0 & 1.0 & 1.00 & 1.39 \\}

\benchmarkentry{I58}{countries\_latest.csv, world\_iso3.geojson}{1}
{Stack three geographic layers in a single view from countries\_latest.csv and world\_iso3.geojson: a choropleth colored by "gdp\_per\_capita (2023)", a hexbin layer aggregating "capital\_lat" and "capital\_lon", and a point layer at "capital\_lat", "capital\_lon" sized by "capital\_population (2025)" and colored by "continent".}
{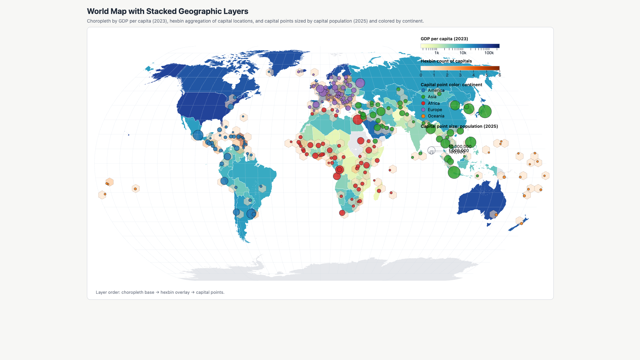}
{ & Raiven  & 1.00 & 1.0 & 1.0 & 1.0 & 1.00 & 0.42 \\[2pt]
 & ChatGPT & 1.00 & 1.0 & 1.0 & 1.0 & 1.00 & 0.83 \\[2pt]
 & Claude  & 1.00 & 1.0 & 1.0 & 1.0 & 1.00 & 0.99 \\[2pt]
 & Gemini  & 1.00 & 1.0 & 1.0 & 1.0 & 1.00 & 2.65 \\}

\benchmarkentry{I59}{countries\_latest.csv}{4}
{Display "life\_expectancy (2023)" from countries\_latest.csv four ways grouped by "region": a boxplot, a ridgeline plot (region on y axis, life expectancy on x axis), a violin plot, and a KDE plot, all sharing a consistent region color mapping.}
{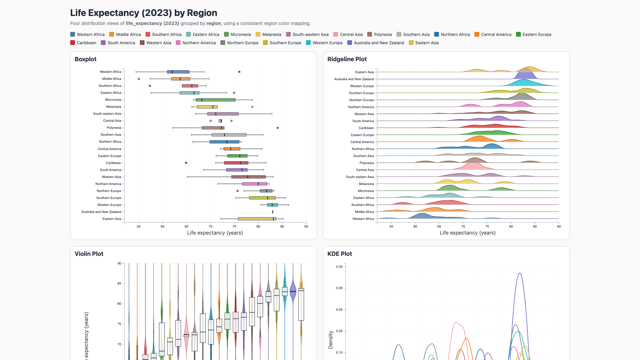}
{ & Raiven  & 0.97 & 1.0 & 0.9 & 0.9 & 1.00 & 0.38 \\[2pt]
 & ChatGPT & 1.00 & 1.0 & 1.0 & 1.0 & 1.00 & 1.04 \\[2pt]
 & Claude  & 1.00 & 1.0 & 1.0 & 1.0 & 1.00 & 0.95 \\[2pt]
 & Gemini  & 1.00 & 1.0 & 1.0 & 1.0 & 1.00 & 1.35 \\}

\benchmarkentry{I60}{countries\_latest.csv}{4}
{Set up four linked views from countries\_latest.csv: a scatter of "gdp\_per\_capita (2023)" versus "life\_expectancy (2023)" colored by "continent", a stacked histogram of "gdp\_per\_capita (2023)" colored by "continent", a boxplot of "life\_expectancy (2023)" grouped by "continent", and a parallel-coordinates plot of "gdp\_per\_capita (2023)", "life\_expectancy (2023)", "median\_age (2023)", and "fertility\_rate (2023)" colored by "continent". Brushing any view should filter the others.}
{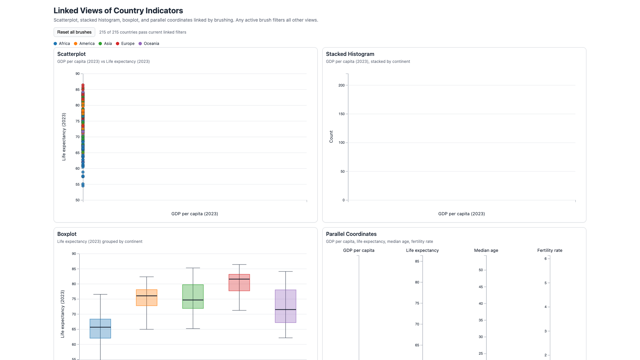}
{ & Raiven  & 0.97 & 1.0 & 0.9 & 1.0 & 1.00 & 0.33 \\[8pt]
 & ChatGPT & 0.73 & 0.8 & 0.7 & 0.8 & 1.00 & 1.32 \\[8pt]
 & Claude  & 1.00 & 1.0 & 1.0 & 1.0 & 1.00 & 1.30 \\[8pt]
 & Gemini  & 1.00 & 1.0 & 1.0 & 1.0 & 1.00 & 3.03 \\}

\benchmarkentry{I61}{region\_energy.csv}{1}
{Create a stacked bar chart from region\_energy.csv with "region" on the x-axis, "value" on the y-axis, and bars stacked by "category" (renewable, nuclear, fossil), using distinct colors per energy type.}
{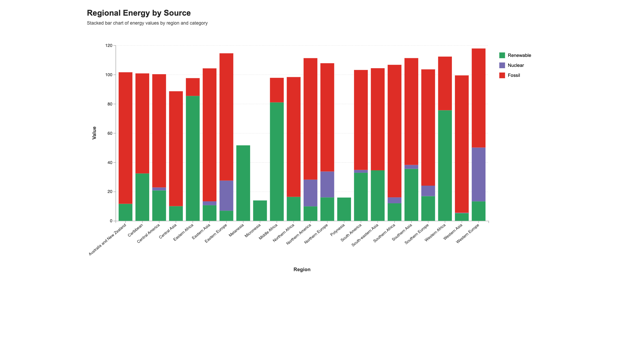}
{ & Raiven  & 1.00 & 1.0 & 1.0 & 1.0 & 1.00 & 0.26 \\
 & ChatGPT & 1.00 & 1.0 & 1.0 & 1.0 & 1.00 & 0.40 \\
 & Claude  & 1.00 & 1.0 & 1.0 & 1.0 & 1.00 & 0.50 \\
 & Gemini  & 1.00 & 1.0 & 1.0 & 1.0 & 1.00 & 0.90 \\}

\benchmarkentry{I62}{countries\_latest.csv}{4}
{Make a brushable scatter of "gdp\_per\_capita (2023)" versus "life\_expectancy (2023)" colored by "continent" from countries\_latest.csv and connect it to three separate histograms, showing distributions of "gdp\_per\_capita (2023)", "life\_expectancy (2023)", and "median\_age (2023)".}
{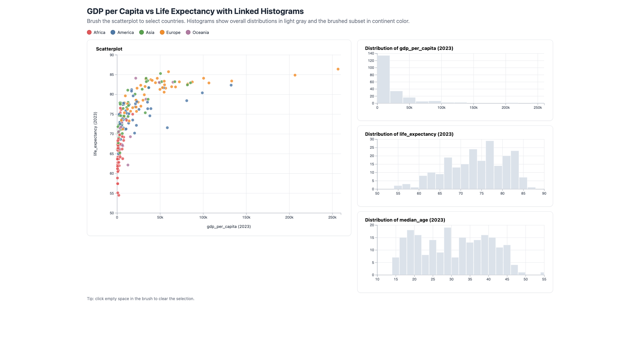}
{ & Raiven  & 1.00 & 1.0 & 1.0 & 1.0 & 1.00 & 0.33 \\[2pt]
 & ChatGPT & 1.00 & 1.0 & 1.0 & 1.0 & 1.00 & 0.77 \\[2pt]
 & Claude  & 1.00 & 1.0 & 1.0 & 1.0 & 1.00 & 0.84 \\[2pt]
 & Gemini  & 1.00 & 1.0 & 1.0 & 1.0 & 1.00 & 1.20 \\}

\benchmarkentry{I63}{countries\_latest.csv}{3}
{Show three linked scatter plots from countries\_latest.csv, all sized by "population (2023)" and colored by "continent": (1) "gdp\_per\_capita (2023)" versus "life\_expectancy (2023)", (2) "fertility\_rate (2023)" versus "median\_age (2023)", (3) "ghg\_per\_capita (2023)" versus "internet\_pct (2023)".}
{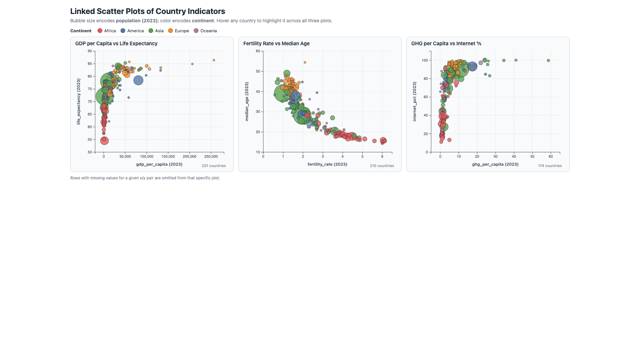}
{ & Raiven  & 1.00 & 1.0 & 1.0 & 1.0 & 1.00 & 0.32 \\[2pt]
 & ChatGPT & 1.00 & 1.0 & 1.0 & 1.0 & 1.00 & 0.66 \\[2pt]
 & Claude  & 1.00 & 1.0 & 1.0 & 1.0 & 1.00 & 0.72 \\[2pt]
 & Gemini  & 1.00 & 1.0 & 1.0 & 1.0 & 1.00 & 2.04 \\}

\benchmarkentry{I64}{countries\_latest.csv}{1}
{Create a parallel-coordinates plot from countries\_latest.csv spanning "population (2023)", "gdp\_per\_capita (2023)", "life\_expectancy (2023)", "median\_age (2023)", "fertility\_rate (2023)", "health\_exp\_per\_capita (2023)", "trade\_pct\_gdp (2023)", "ghg\_per\_capita (2023)", "capital\_population (2025)" and "num\_neighbors" with each line representing a country and lines colored by "region". Make the plot brushable.}
{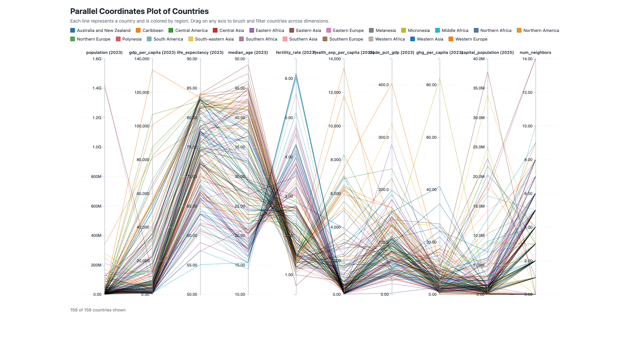}
{ & Raiven  & 1.00 & 1.0 & 1.0 & 1.0 & 1.00 & 0.27 \\[4pt]
 & ChatGPT & 0.93 & 1.0 & 0.8 & 1.0 & 1.00 & 0.64 \\[4pt]
 & Claude  & 1.00 & 1.0 & 1.0 & 1.0 & 1.00 & 0.81 \\[4pt]
 & Gemini  & 1.00 & 1.0 & 1.0 & 1.0 & 1.00 & 1.51 \\}

\benchmarkentry{I65}{countries\_latest.csv}{2}
{Plot capital cities from countries\_latest.csv as bubbles at y= "capital\_lat", x= "capital\_lon" sized by "capital\_population (2025)" and colored by "continent", linked to a histogram of "capital\_population (2025)".}
{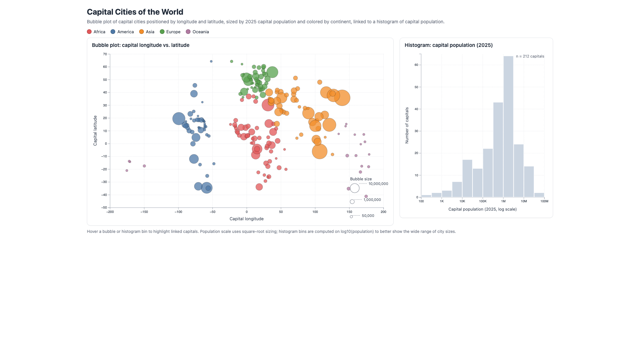}
{ & Raiven  & 0.97 & 1.0 & 1.0 & 0.9 & 1.00 & 0.25 \\
 & ChatGPT & 1.00 & 1.0 & 1.0 & 1.0 & 1.00 & 0.89 \\
 & Claude  & 0.97 & 1.0 & 1.0 & 0.9 & 1.00 & 1.00 \\
 & Gemini  & 0.97 & 1.0 & 1.0 & 0.9 & 1.00 & 1.24 \\}

\benchmarkentry{I66}{countries\_latest.csv, world\_iso3.geojson, trade\_region\_latest.csv}{2}
{Track deforestation: a choropleth colored by "forest\_area\_pct (2023)" using the "Greens" color scheme with country geometries from world\_iso3.geojson and countries\_latest.csv, alongside a bar chart from trade\_region\_latest.csv with "trade\_region" on the x-axis and "forest\_area\_pct (2023)" on the y-axis.}
{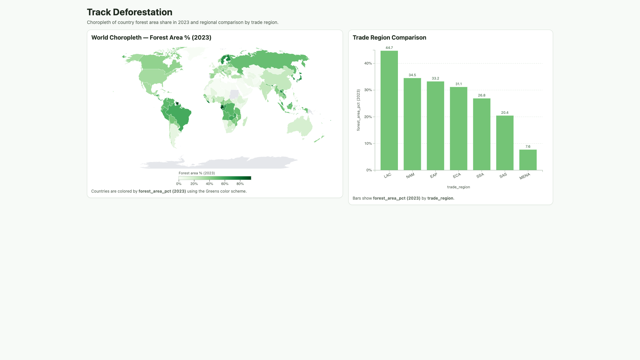}
{ & Raiven  & 0.90 & 0.9 & 0.9 & 0.9 & 0.90 & 0.37 \\[2pt]
 & ChatGPT & 1.00 & 1.0 & 1.0 & 1.0 & 1.00 & 0.66 \\[2pt]
 & Claude  & 1.00 & 1.0 & 1.0 & 1.0 & 1.00 & 0.78 \\[2pt]
 & Gemini  & 1.00 & 1.0 & 1.0 & 1.0 & 1.00 & 0.82 \\}

\benchmarkentry{I67}{countries\_latest.csv}{2}
{Create a heatmap of "health\_exp\_per\_capita (2023)" versus "median\_age (2023)" from countries\_latest.csv with cells colored by count, linked to a histogram of "life\_expectancy (2023)". Brushing bins in the heatmap should filter the histogram.}
{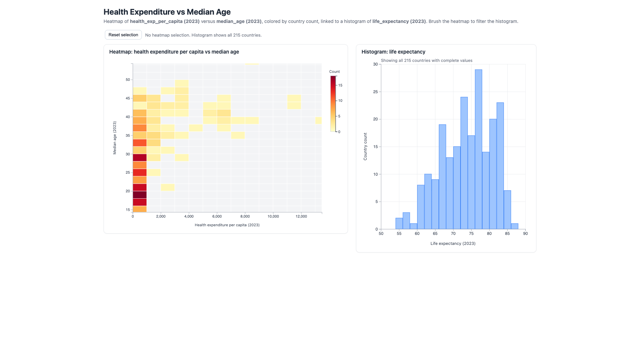}
{ & Raiven  & 1.00 & 1.0 & 1.0 & 1.0 & 1.00 & 0.32 \\[2pt]
 & ChatGPT & 1.00 & 1.0 & 1.0 & 1.0 & 1.00 & 1.08 \\[2pt]
 & Claude  & 0.90 & 1.0 & 0.8 & 0.9 & 1.00 & 1.03 \\[2pt]
 & Gemini  & 1.00 & 1.0 & 1.0 & 1.0 & 1.00 & 2.07 \\}

\benchmarkentry{I68}{countries\_latest.csv}{1}
{Plot a bubble chart from countries\_latest.csv with "literacy\_pct (2022)" versus "life\_expectancy (2023)", sized by "population (2023)" and colored by "continent".}
{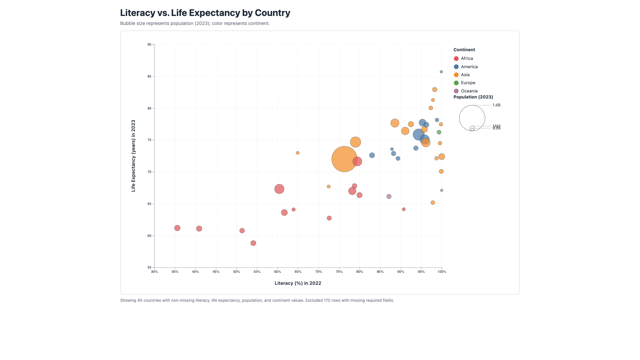}
{ & Raiven  & 1.00 & 1.0 & 1.0 & 1.0 & 1.00 & 0.19 \\
 & ChatGPT & 1.00 & 1.0 & 1.0 & 1.0 & 1.00 & 0.63 \\
 & Claude  & 1.00 & 1.0 & 1.0 & 1.0 & 1.00 & 0.62 \\
 & Gemini  & 1.00 & 1.0 & 1.0 & 1.0 & 1.00 & 0.76 \\}

\benchmarkentry{I69}{countries\_latest.csv, world\_iso3.geojson}{3}
{Compare emissions and wealth: a scatter of "gdp\_per\_capita (2023)" versus "ghg\_per\_capita (2023)" colored by "continent" from countries\_latest.csv, followed by two choropleths using world\_iso3.geojson — one colored by "ghg\_per\_capita (2023)" using a red sequential color scale, one colored by "gdp\_per\_capita (2023)" using a blue sequential color scale.}
{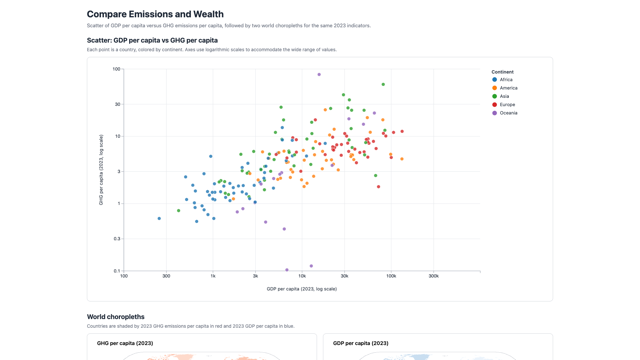}
{ & Raiven  & 1.00 & 1.0 & 1.0 & 1.0 & 1.00 & 0.36 \\[4pt]
 & ChatGPT & 1.00 & 1.0 & 1.0 & 1.0 & 1.00 & 0.81 \\[4pt]
 & Claude  & 1.00 & 1.0 & 1.0 & 1.0 & 1.00 & 0.93 \\[4pt]
 & Gemini  & 1.00 & 1.0 & 1.0 & 1.0 & 1.00 & 1.07 \\}

\benchmarkentry{I70}{countries\_latest.csv}{3}
{Create a parallel-coordinates plot from countries\_latest.csv spanning "gdp\_per\_capita (2023)", "life\_expectancy (2023)", "median\_age (2023)", "fertility\_rate (2023)", and "ghg\_per\_capita (2023)" with lines colored by "continent", linked to two KDE charts: one of "gdp\_per\_capita (2023)" in blue and one of "life\_expectancy (2023)" in orange. Brushing the parallel coordinates should filter both KDE charts.}
{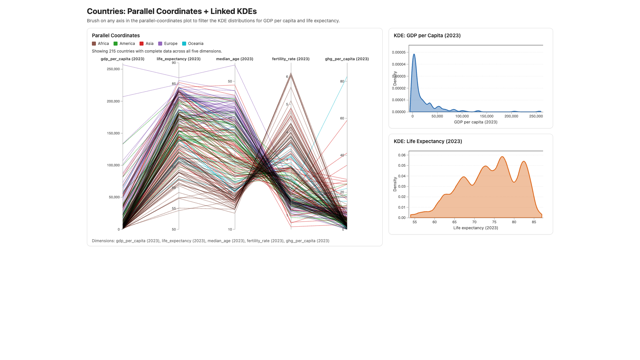}
{ & Raiven  & 1.00 & 1.0 & 1.0 & 1.0 & 1.00 & 0.28 \\[4pt]
 & ChatGPT & 1.00 & 1.0 & 1.0 & 1.0 & 1.00 & 0.93 \\[4pt]
 & Claude  & 0.33 & 0.3 & 0.3 & 0.3 & 1.00 & 0.89 \\[4pt]
 & Gemini  & 1.00 & 1.0 & 1.0 & 1.0 & 1.00 & 1.53 \\}

\subsubsection{Combined (C)}

\benchmarkentry{C71}{head.vti, head\_sample.csv}{2}
{From head.vti, generate a volume rendering of the CT data, and then use head\_sample.csv to produce a layered distribution plot showing a histogram of intensity with a KDE curve.}
{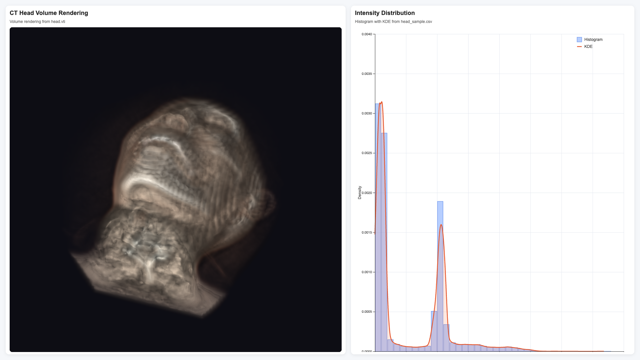}
{ & Raiven  & 1.00 & 1.0 & 1.0 & 1.0 & 1.00 & 0.29 \\
 & ChatGPT & 1.00 & 1.0 & 1.0 & 1.0 & 1.00 & 0.58 \\
 & Claude  & 0.83 & 0.9 & 0.9 & 0.7 & 0.90 & 2.02 \\
 & Gemini  & 1.00 & 1.0 & 1.0 & 1.0 & 1.00 & 1.87 \\}

\benchmarkentry{C72}{head.vti, head\_sample.csv}{4}
{Create seperate three orthogonal slice views in the XY, XZ, and YZ planes from head.vti, link their slice positions together, and also generate a histogram of "intensity" using head\_sample.csv.}
{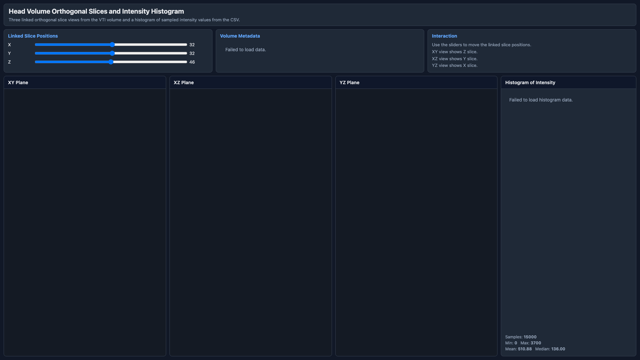}
{ & Raiven  & 0.95 & 1.0 & 1.0 & 0.8 & 1.00 & 0.32 \\
 & ChatGPT & 0.52 & 0.7 & 0.6 & 0.2 & 0.55 & 3.23 \\
 & Claude  & 0.62 & 0.6 & 0.7 & 0.6 & 1.00 & 1.52 \\
 & Gemini  & 0.65 & 0.7 & 0.7 & 0.6 & 0.55 & 2.41 \\}

\benchmarkentry{C73}{head.vti, head\_sample.csv}{2}
{Render the CT isosurface in pink from head.vti, then summarize head\_sample.csv using a histogram of "intensity" colored pink.}
{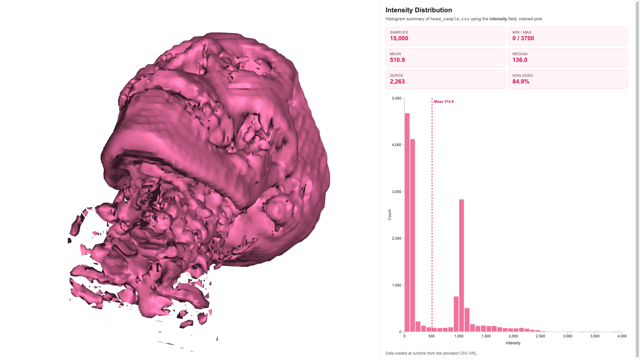}
{ & Raiven  & 1.00 & 1.0 & 1.0 & 1.0 & 1.00 & 0.42 \\
 & ChatGPT & 1.00 & 1.0 & 1.0 & 1.0 & 1.00 & 0.62 \\
 & Claude  & 0.77 & 0.8 & 0.8 & 0.7 & 0.70 & 0.53 \\
 & Gemini  & 0.53 & 0.6 & 0.6 & 0.4 & 0.40 & 1.09 \\}

\benchmarkentry{C74}{head.vti, head\_sample.csv}{5}
{Using head.vti, assemble three linked slice views (XY, XZ, YZ) along with a 3D view that combines volume rendering and triplanar slices, making sure the slice positions and transfer function remain synchronized across all four views, and also include a histogram of intensity from head\_sample.csv.}
{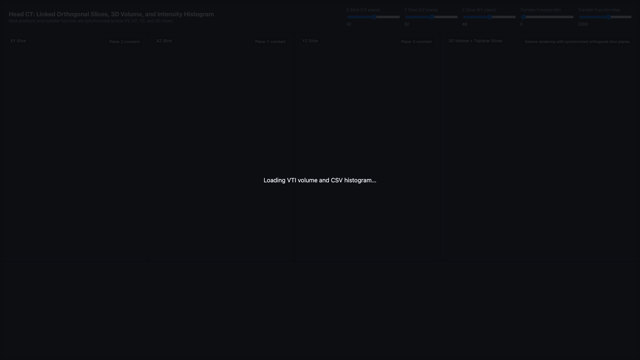}
{ & Raiven  & 1.00 & 1.0 & 1.0 & 1.0 & 1.00 & 0.40 \\[2pt]
 & ChatGPT & 0.15 & 0.0 & 0.4 & 0.0 & 0.40 & 1.14 \\[2pt]
 & Claude  & 0.57 & 0.5 & 0.5 & 0.7 & 0.40 & 2.52 \\[2pt]
 & Gemini  & 0.44 & 0.4 & 0.4 & 0.6 & 0.40 & 3.14 \\}

\benchmarkentry{C75}{head.vti, head\_region\_stats.csv}{2}
{Produce a volume rendering from head.vti, and then visualize the statistics in head\_region\_stats.csv as a parallel coordinates chart including mean, std, median, q25, q75, min, and max.}
{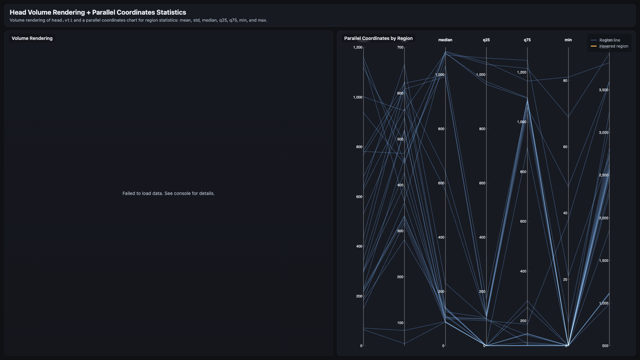}
{ & Raiven  & 1.00 & 1.0 & 1.0 & 1.0 & 1.00 & 0.27 \\
 & ChatGPT & 0.80 & 0.8 & 0.8 & 0.8 & 0.70 & 0.65 \\
 & Claude  & 0.80 & 0.8 & 0.8 & 0.8 & 0.70 & 1.57 \\
 & Gemini  & 1.00 & 1.0 & 1.0 & 1.0 & 1.00 & 1.11 \\}

\benchmarkentry{C76}{divcurl\_10.vti, divcurl\_sample.csv}{2}
{With divcurl\_10.vti, compute an LIC visualization based on vx and vy, and alongside it create a histogram of div with an overlaid KDE using divcurl\_sample.csv.}
{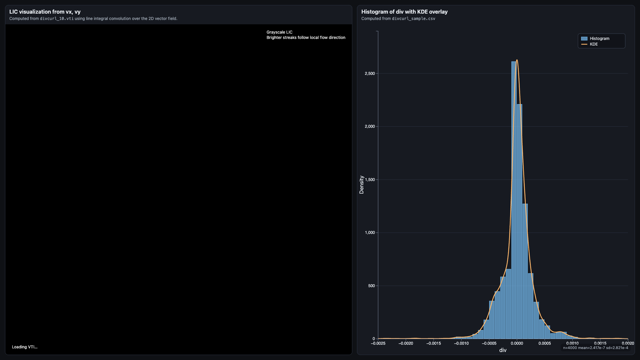}
{ & Raiven  & 1.00 & 1.0 & 1.0 & 1.0 & 1.00 & 0.28 \\
 & ChatGPT & 0.60 & 0.6 & 0.6 & 0.6 & 0.40 & 1.17 \\
 & Claude  & 0.77 & 0.7 & 0.7 & 0.9 & 0.80 & 1.40 \\
 & Gemini  & 0.53 & 0.8 & 0.8 & 0.0 & 0.70 & 2.66 \\}

\benchmarkentry{C77}{divcurl\_10.vti, divcurl\_binned\_2d.csv}{3}
{Generate an LIC view from divcurl\_10.vti using vx and vy. Also represent divcurl\_binned\_2d.csv as two heatmaps of x\_bin vs. y\_bin, one colored by div\_mean and one colored by curl\_mean, both using color palette inferno.}
{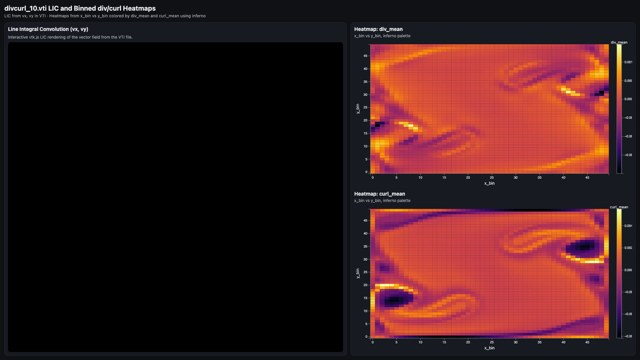}
{ & Raiven  & 1.00 & 1.0 & 1.0 & 1.0 & 1.00 & 0.25 \\
 & ChatGPT & 0.00 & 0.0 & 0.0 & 0.0 & 0.00 & 0.72 \\
 & Claude  & 0.91 & 1.0 & 0.9 & 0.9 & 1.00 & 2.02 \\
 & Gemini  & 0.00 & 0.0 & 0.0 & 0.0 & 0.40 & 2.63 \\}

\benchmarkentry{C78}{divcurl\_15.vti, divcurl\_sample.csv}{4}
{Using divcurl\_15.vti, display four views consisting of an LIC using vx and vy, a slice of div, and a slice of curl, and a histogram of curl from divcurl\_sample.csv.}
{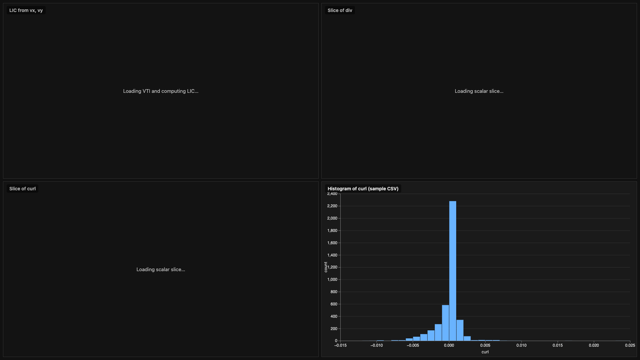}
{ & Raiven  & 1.00 & 1.0 & 1.0 & 1.0 & 1.00 & 0.42 \\
 & ChatGPT & 0.65 & 0.6 & 0.7 & 0.7 & 0.85 & 0.89 \\
 & Claude  & 0.70 & 0.7 & 0.7 & 0.7 & 0.55 & 1.81 \\
 & Gemini  & 0.60 & 0.6 & 0.6 & 0.6 & 0.40 & 3.23 \\}

\benchmarkentry{C79}{divcurl\_10.vti, divcurl\_sample.csv}{3}
{Make a scatterplot of div versus curl generated from divcurl\_sample.csv. Also, show two slice views from divcurl\_10.vti, one visualizing div and the other curl.}
{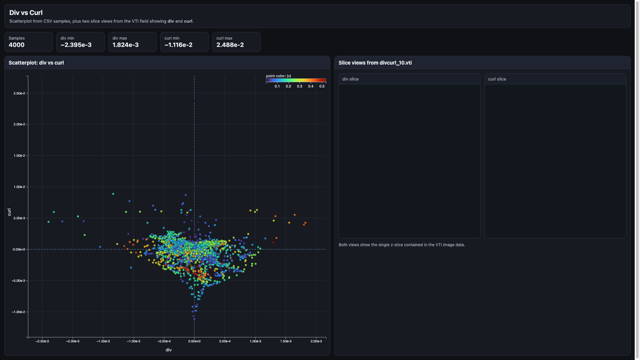}
{ & Raiven  & 1.00 & 1.0 & 1.0 & 1.0 & 1.00 & 0.30 \\
 & ChatGPT & 0.73 & 0.7 & 0.7 & 0.7 & 0.60 & 1.03 \\
 & Claude  & 1.00 & 1.0 & 1.0 & 1.0 & 0.87 & 1.64 \\
 & Gemini  & 0.73 & 0.7 & 0.7 & 0.7 & 0.60 & 2.16 \\}

\benchmarkentry{C80}{divcurl\_0.vti, divcurl\_10.vti, divcurl\_19.vti, divcurl\_stats.csv}{4}
{Load divcurl\_0.vti, divcurl\_10.vti, and divcurl\_19.vti to create three LIC visualizations based on vx and vy, and use divcurl\_stats.csv to build a time-series chart plotting curl\_mean and div\_mean against t.}
{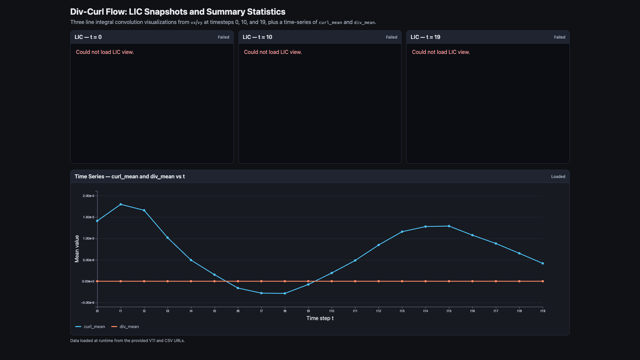}
{ & Raiven  & 0.93 & 0.9 & 0.9 & 0.9 & 0.50 & 0.32 \\
 & ChatGPT & 0.70 & 0.7 & 0.7 & 0.7 & 0.55 & 0.80 \\
 & Claude  & 0.70 & 0.7 & 0.7 & 0.7 & 0.55 & 1.26 \\
 & Gemini  & 0.70 & 0.7 & 0.7 & 0.7 & 0.55 & 2.16 \\}

\benchmarkentry{C81}{divcurl\_5.vti, divcurl\_stats.csv}{2}
{Create a LIC from divcurl\_5.vti using vx and vy, then generate a ridgeline plot from divcurl\_stats.csv with x = curl\_mean and y = t.}
{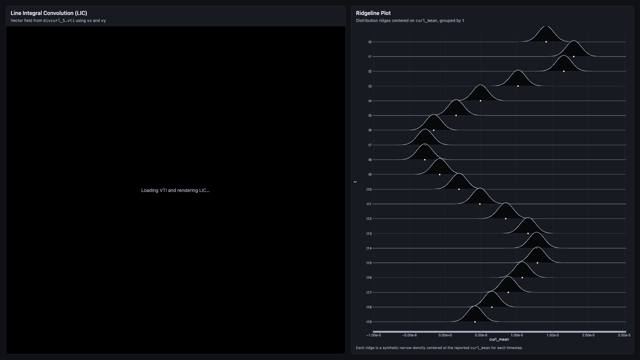}
{ & Raiven  & 0.90 & 0.9 & 1.0 & 0.8 & 1.00 & 0.29 \\
 & ChatGPT & 0.73 & 0.7 & 0.7 & 0.8 & 0.70 & 0.68 \\
 & Claude  & 0.93 & 0.9 & 0.9 & 1.0 & 0.80 & 1.80 \\
 & Gemini  & 0.83 & 0.7 & 0.9 & 0.9 & 0.70 & 2.45 \\}

\benchmarkentry{C82}{divcurl\_10.vti, divcurl\_binned\_2d.csv}{4}
{Begin with an LIC visualization from divcurl\_10.vti computed from vx and vy, then construct two heatmaps from divcurl\_binned\_2d.csv using x\_bin versus y\_bin , one colored by curl\_mean and the other by div\_mean. Also, add a histogram of curl\_mean.}
{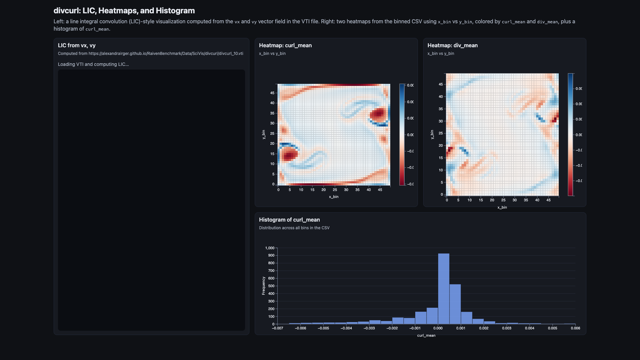}
{ & Raiven  & 1.00 & 1.0 & 1.0 & 1.0 & 1.00 & 0.37 \\[2pt]
 & ChatGPT & 0.93 & 0.9 & 0.9 & 0.9 & 1.00 & 0.97 \\[2pt]
 & Claude  & 0.67 & 1.0 & 1.0 & 0.0 & 1.00 & 1.72 \\[2pt]
 & Gemini  & 1.00 & 1.0 & 1.0 & 1.0 & 1.00 & 2.51 \\}

\benchmarkentry{C83}{divcurl\_10.vti, divcurl\_sample.csv}{2}
{Build an LIC view using vx and vy from divcurl\_10.vti, and also make a KDE curve for curl based on data from divcurl\_sample.csv.}
{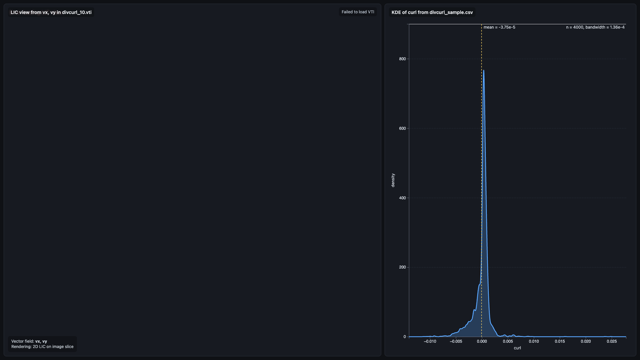}
{ & Raiven  & 1.00 & 1.0 & 1.0 & 1.0 & 1.00 & 0.28 \\
 & ChatGPT & 0.80 & 0.8 & 0.8 & 0.8 & 0.70 & 0.58 \\
 & Claude  & 0.87 & 0.9 & 0.9 & 0.8 & 0.90 & 1.79 \\
 & Gemini  & 0.87 & 0.9 & 0.9 & 0.8 & 1.00 & 2.03 \\}

\benchmarkentry{C84}{taylorgreen\_7.vti, tg\_stats.csv}{2}
{From taylorgreen\_7.vti, make a volume of "vort" and also layer streamlines using vx = "ux", vy = "uy", and vz = "uz". Then use tg\_stats.csv to create a multi-line chart of vort\_mean, vort\_q25 and vort\_q75 over t.}
{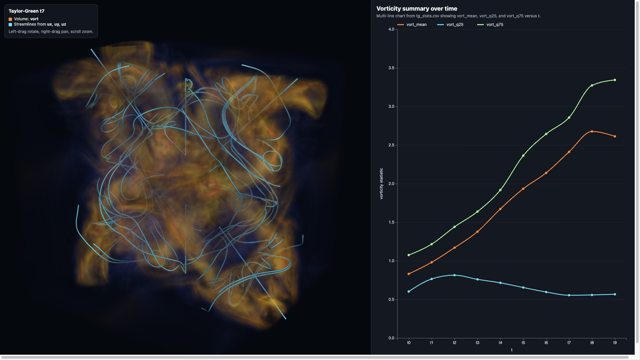}
{ & Raiven  & 1.00 & 1.0 & 1.0 & 1.0 & 1.00 & 0.31 \\
 & ChatGPT & 1.00 & 1.0 & 1.0 & 1.0 & 1.00 & 0.99 \\
 & Claude  & 0.80 & 0.8 & 0.8 & 0.8 & 0.70 & 1.24 \\
 & Gemini  & 0.80 & 0.8 & 0.8 & 0.8 & 0.70 & 1.99 \\}

\benchmarkentry{C85}{taylorgreen\_9.vti, tg\_sample.csv}{2}
{Construct two layered isosurfaces of vort from taylorgreen\_9.vti. Then make a layered histogram and KDE of vort generated from tg\_sample.csv.}
{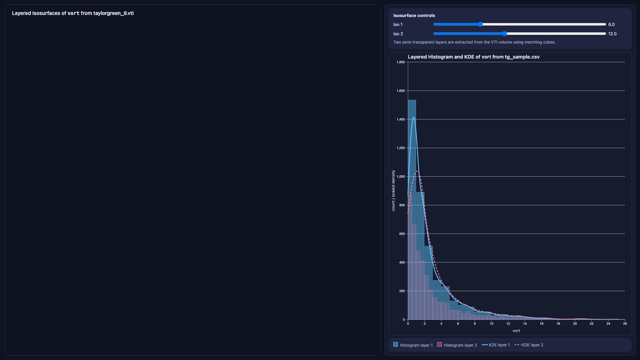}
{ & Raiven  & 0.97 & 1.0 & 1.0 & 0.9 & 1.00 & 0.30 \\
 & ChatGPT & 0.80 & 0.8 & 0.8 & 0.8 & 1.00 & 0.82 \\
 & Claude  & 0.80 & 0.8 & 0.8 & 0.8 & 0.70 & 1.11 \\
 & Gemini  & 0.60 & 0.6 & 0.6 & 0.6 & 0.40 & 1.65 \\}

\benchmarkentry{C86}{taylorgreen\_9.vti, tg\_stats.csv}{2}
{Make an XY planar slice of vort from taylorgreen\_9.vti and use tg\_stats.csv to create a multi-line plot of vort\_mean, vort\_q25, and vort\_q75 over t.}
{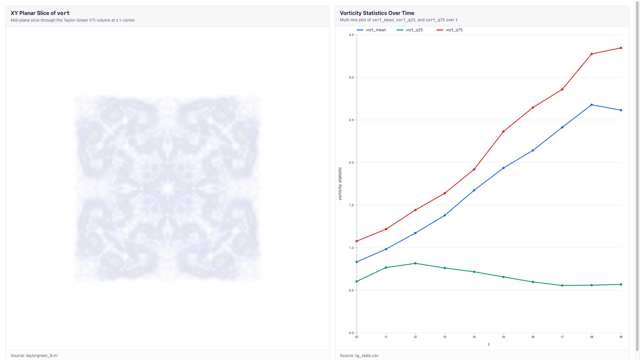}
{ & Raiven  & 1.00 & 1.0 & 1.0 & 1.0 & 1.00 & 0.37 \\
 & ChatGPT & 1.00 & 1.0 & 1.0 & 1.0 & 1.00 & 0.69 \\
 & Claude  & 0.00 & 0.0 & 0.0 & 0.0 & 0.00 & 1.11 \\
 & Gemini  & 1.00 & 1.0 & 1.0 & 1.0 & 1.00 & 1.36 \\}

\benchmarkentry{C87}{taylorgreen\_9.vti, tg\_t9\_binned.csv}{4}
{Visualize ux, uy, and uz from taylorgreen\_9.vti as three separate volume renderings, each with a unique color palette, and use tg\_t9\_binned.csv to create a heatmap where ux\_mean and uy\_mean define the axes and count controls the color scale.}
{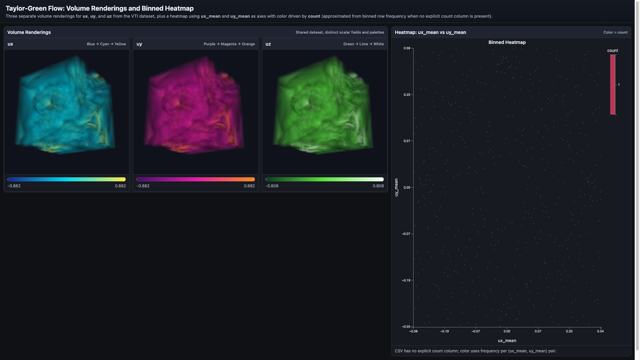}
{ & Raiven  & 1.00 & 1.0 & 1.0 & 1.0 & 1.00 & 0.37 \\[2pt]
 & ChatGPT & 0.88 & 0.9 & 0.9 & 0.8 & 1.00 & 1.01 \\[2pt]
 & Claude  & 0.70 & 0.7 & 0.7 & 0.7 & 0.55 & 1.86 \\[2pt]
 & Gemini  & 1.00 & 1.0 & 1.0 & 1.0 & 1.00 & 1.95 \\}

\benchmarkentry{C88}{taylorgreen\_6.vti, taylorgreen\_8.vti, tg\_stats.csv}{2}
{Overlay streamline sets computed using vx = ux, vy = uy, and vz = uz in taylorgreen\_6.vti and taylorgreen\_8.vti within a single view using a different color per timestep, and then construct a line chart from tg\_stats.csv plotting vort\_mean and critq\_mean against t.}
{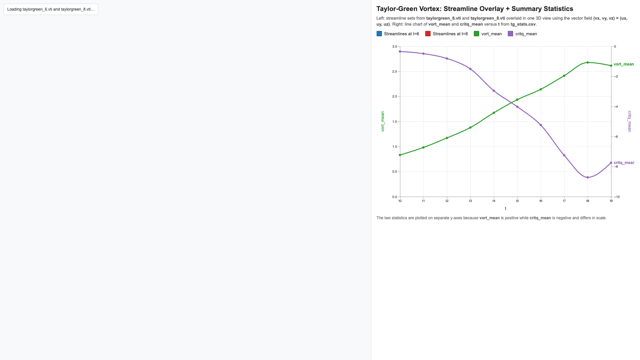}
{ & Raiven  & 1.00 & 1.0 & 1.0 & 1.0 & 1.00 & 0.43 \\
 & ChatGPT & 0.80 & 0.8 & 0.8 & 0.8 & 0.70 & 0.80 \\
 & Claude  & 0.00 & 0.0 & 0.0 & 0.0 & 0.40 & 1.41 \\
 & Gemini  & 0.80 & 0.8 & 0.8 & 0.8 & 0.70 & 2.64 \\}

\benchmarkentry{C89}{tg\_sample.csv, taylorgreen\_9.vti}{4}
{Display four views of the Taylor-Green vortex from taylorgreen\_9.vti and tg\_sample.csv: (1) a volume rendering of "vort" with an XY slice using "turbo", (2) the same XY slice of "vort" using "turbo", (3) a histogram of "vort" in green with an overlaid orange KDE chart from tg\_sample.csv, and (4) an XY slice of "pp" using "magma". Link only the slice positions between the slice views so scrolling one updates the other. Do not link transfer functions.}
{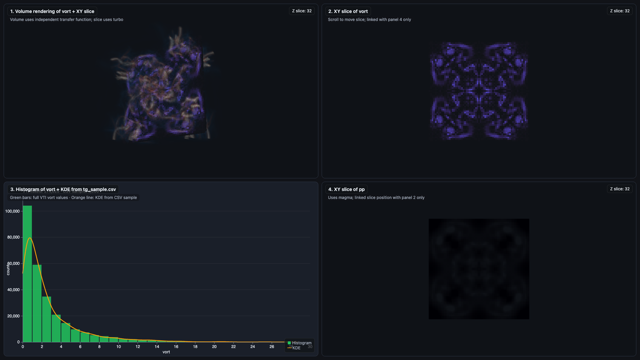}
{ & Raiven  & 1.00 & 1.0 & 1.0 & 1.0 & 1.00 & 0.53 \\[4pt]
 & ChatGPT & 0.90 & 1.0 & 0.8 & 0.8 & 1.00 & 0.90 \\[4pt]
 & Claude  & 0.60 & 0.6 & 0.6 & 0.7 & 0.55 & 1.66 \\[4pt]
 & Gemini  & 0.30 & 0.5 & 0.5 & 0.0 & 0.00 & 6.00 \\}

\benchmarkentry{C90}{taylorgreen\_9.vti, tg\_stats.csv}{5}
{Display a volume rendering of vort from taylorgreen\_9.vti together with three linked XY planar slice views of ux, uy, and uz, ensure slice positions and transfer functions are synchronized across the views, and include a multi-line chart of vort\_q25, vort\_q50, and vort\_q75 over t from tg\_stats.csv.}
{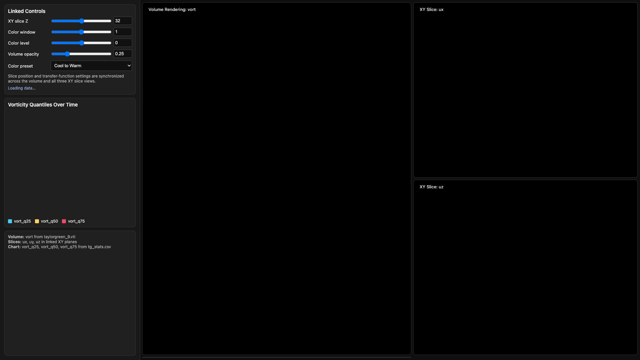}
{ & Raiven  & 0.99 & 1.0 & 1.0 & 1.0 & 1.00 & 0.41 \\[2pt]
 & ChatGPT & 0.44 & 0.5 & 0.4 & 0.4 & 0.40 & 1.15 \\[2pt]
 & Claude  & 0.17 & 0.0 & 0.5 & 0.0 & 0.52 & 1.63 \\[2pt]
 & Gemini  & 0.43 & 0.4 & 0.4 & 0.4 & 0.40 & 2.18 \\}

\benchmarkentry{C91}{taylorgreen\_4.vti, tg\_stats.csv}{2}
{Render vort as a volume from taylorgreen\_4.vti and use tg\_stats.csv to create a boxplot of vort\_mean by timestep t using the fields vort\_min, vort\_q25, vort\_q50, vort\_q75, and vort\_max.}
{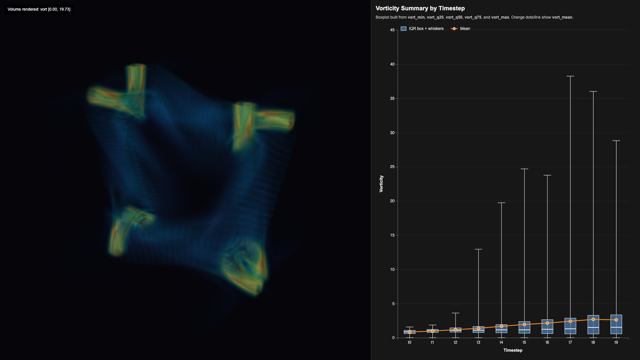}
{ & Raiven  & 0.87 & 0.9 & 0.8 & 0.9 & 1.00 & 0.32 \\
 & ChatGPT & 1.00 & 1.0 & 1.0 & 1.0 & 1.00 & 1.12 \\
 & Claude  & 1.00 & 1.0 & 1.0 & 1.0 & 1.00 & 0.84 \\
 & Gemini  & 1.00 & 1.0 & 1.0 & 1.0 & 1.00 & 1.48 \\}

\benchmarkentry{C92}{taylorgreen\_9.vti, tg\_t9\_binned.csv}{2}
{Compute thick, green streamlines using vx = ux, vy = uy, and vz = uz in taylorgreen\_9.vti and produce a point chart from tg\_t9\_binned.csv plotting vort\_mean against critq\_mean.}
{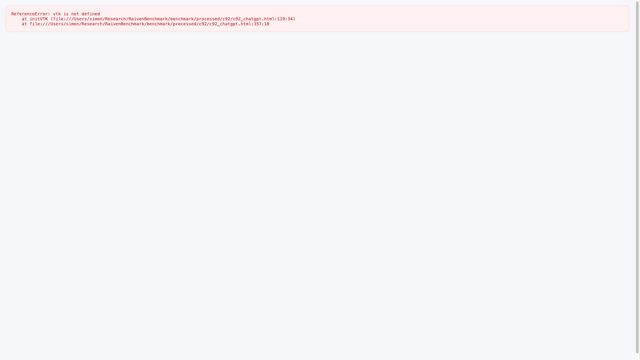}
{ & Raiven  & 1.00 & 1.0 & 1.0 & 1.0 & 1.00 & 0.29 \\
 & ChatGPT & 0.00 & 0.0 & 0.0 & 0.0 & 0.00 & 0.99 \\
 & Claude  & 0.80 & 0.8 & 0.8 & 0.8 & 1.00 & 0.86 \\
 & Gemini  & 0.73 & 0.6 & 0.8 & 0.8 & 0.70 & 1.85 \\}

\benchmarkentry{C93}{cells.vti, cells\_channel\_sample.csv}{4}
{Create a volume for Sec61-beta using green color palette, a volume for Membrane using orange color palette, and a volume for DNA using purple color palette, all of which are seperate views but use cells.vti. Also add a layered histogram of Sec61b, Membrane, and DNA in a single chart using cells\_channel\_sample.csv.}
{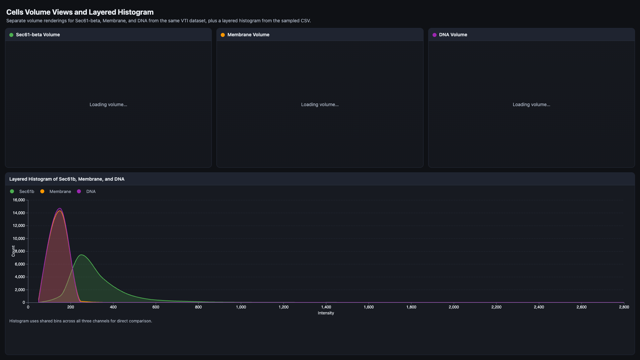}
{ & Raiven  & 1.00 & 1.0 & 1.0 & 1.0 & 1.00 & 0.30 \\[2pt]
 & ChatGPT & 0.83 & 0.6 & 0.9 & 1.0 & 1.00 & 0.89 \\[2pt]
 & Claude  & 0.20 & 0.0 & 0.6 & 0.0 & 0.40 & 0.96 \\[2pt]
 & Gemini  & 0.82 & 1.0 & 0.8 & 0.7 & 1.00 & 1.34 \\}

\benchmarkentry{C94}{cells.vti, cells\_seg\_stats.csv}{2}
{Make layered isosurfaces using seg\_Sec61-beta with color = teal, seg\_Membrane with color = pink, and seg\_DNA with color = green from cells.vti and construct a bar chart from cells\_seg\_stats.csv showing voxel\_count that is colored by seg\_combo.}
{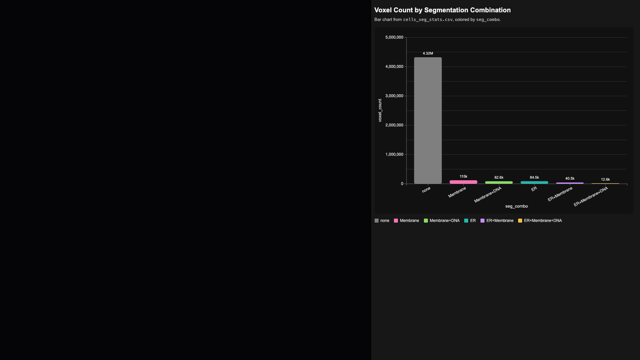}
{ & Raiven  & 1.00 & 1.0 & 1.0 & 1.0 & 1.00 & 0.34 \\[2pt]
 & ChatGPT & 0.00 & 0.0 & 0.0 & 0.0 & 0.00 & 0.70 \\[2pt]
 & Claude  & 0.90 & 0.9 & 0.9 & 0.9 & 1.00 & 0.77 \\[2pt]
 & Gemini  & 0.50 & 0.5 & 0.4 & 0.6 & 0.40 & 1.68 \\}

\benchmarkentry{C95}{cells.vti, cells\_channel\_sample.csv}{3}
{Generate an XY slice view of Sec61-beta and a linked XY slice of Membrane from cells.vti and accompany it with a scatterplot of Sec61b versus Membrane, sized by DNA, created from cells\_channel\_sample.csv.}
{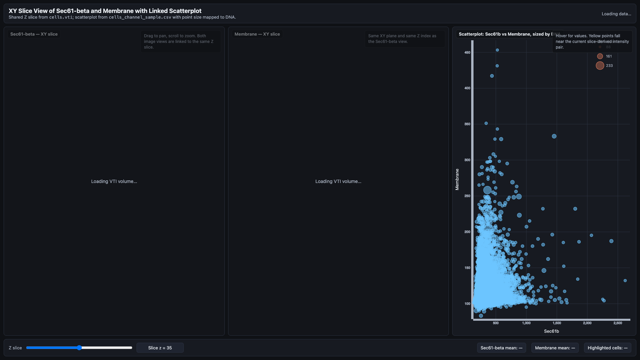}
{ & Raiven  & 1.00 & 1.0 & 1.0 & 1.0 & 1.00 & 0.37 \\
 & ChatGPT & 0.91 & 0.9 & 0.9 & 0.9 & 1.00 & 1.35 \\
 & Claude  & 0.51 & 0.5 & 0.5 & 0.6 & 0.33 & 1.46 \\
 & Gemini  & 0.91 & 0.9 & 1.0 & 0.9 & 1.00 & 1.87 \\}

\benchmarkentry{C96}{cells.vti, cells\_seg\_stats.csv}{2}
{Make three layered isosurfaces from cells.vti, one using Sec61-beta and colored pink, another using Membrane and colored orange, and one using DNA colored purple, also summarize segmentation proportions with a pie chart built from cells\_seg\_stats.csv using seg\_combo for slices sized by pct.}
{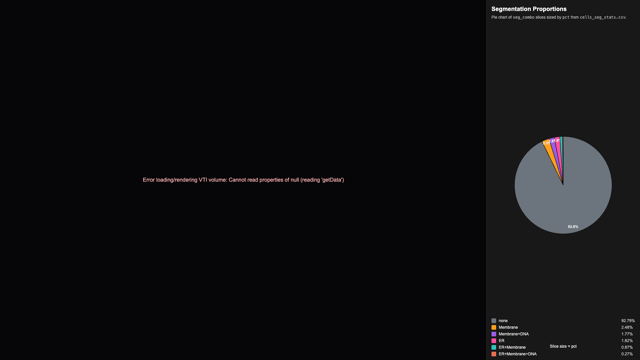}
{ & Raiven  & 1.00 & 1.0 & 1.0 & 1.0 & 1.00 & 0.35 \\[2pt]
 & ChatGPT & 0.80 & 0.8 & 0.8 & 0.8 & 0.70 & 0.68 \\[2pt]
 & Claude  & 0.80 & 0.8 & 0.8 & 0.8 & 0.70 & 0.61 \\[2pt]
 & Gemini  & 0.93 & 1.0 & 0.9 & 0.9 & 1.00 & 1.57 \\}

\benchmarkentry{C97}{cells.vti, cells\_channel\_sample.csv}{4}
{Create three linked XY slice views for Sec61-beta, Membrane, and DNA from cells.vti with synchronized slice positions, and add a hexbin plot of Sec61b versus Membrane using cells\_channel\_sample.csv.}
{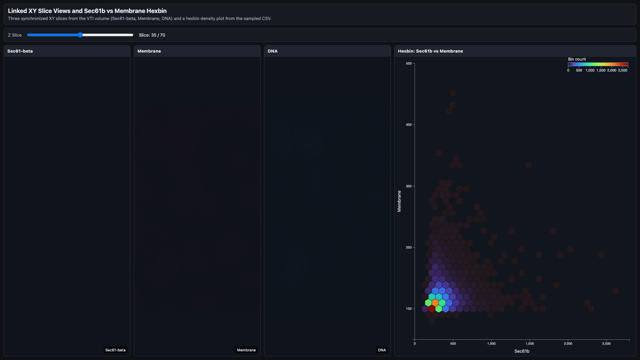}
{ & Raiven  & 1.00 & 1.0 & 1.0 & 1.0 & 1.00 & 0.30 \\
 & ChatGPT & 0.67 & 0.6 & 0.6 & 0.8 & 1.00 & 0.96 \\
 & Claude  & 0.50 & 0.5 & 0.5 & 0.6 & 0.40 & 1.24 \\
 & Gemini  & 0.50 & 0.5 & 0.5 & 0.6 & 0.40 & 1.35 \\}

\benchmarkentry{C98}{cells.vti, cells\_per\_seg\_sample.csv}{2}
{Visualize Sec61-beta as a blue colored volume layered over red colored isosurface of Sec61-beta from cells.vti and overlay KDE over histogram of Sec61b\_intensity grouped by seg\_label using cells\_per\_seg\_sample.csv.}
{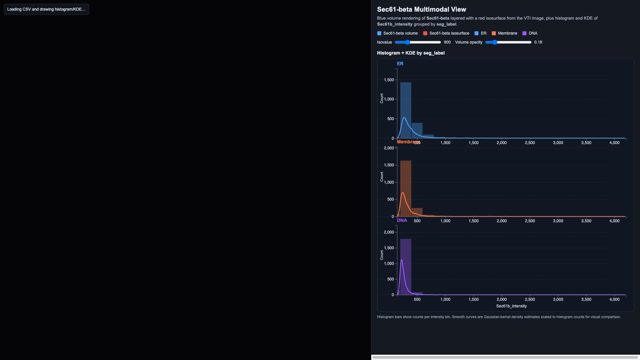}
{ & Raiven  & 0.97 & 1.0 & 1.0 & 0.9 & 1.00 & 0.34 \\
 & ChatGPT & 0.80 & 0.8 & 0.8 & 0.8 & 1.00 & 1.09 \\
 & Claude  & 0.80 & 0.8 & 0.8 & 0.8 & 0.70 & 1.04 \\
 & Gemini  & 0.60 & 0.6 & 0.6 & 0.6 & 0.40 & 2.22 \\}

\benchmarkentry{C99}{cells.vti, cells\_per\_seg\_sample.csv}{2}
{Overlay isosurface representations of seg\_Sec61-beta color = orange, seg\_Membrane color = purple, and seg\_DNA color = teal from cells.vti and present a boxplot of Sec61b\_intensity grouped by seg\_label based on cells\_per\_seg\_sample.csv.}
{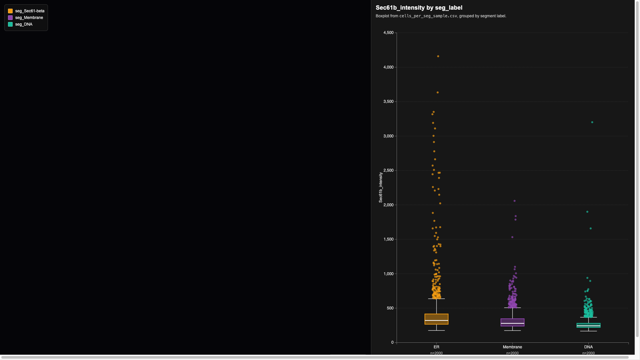}
{ & Raiven  & 1.00 & 1.0 & 1.0 & 1.0 & 1.00 & 0.31 \\[2pt]
 & ChatGPT & 0.73 & 0.6 & 0.8 & 0.8 & 1.00 & 0.94 \\[2pt]
 & Claude  & 0.90 & 0.9 & 0.9 & 0.9 & 1.00 & 0.60 \\[2pt]
 & Gemini  & 0.57 & 0.6 & 0.5 & 0.6 & 0.40 & 1.70 \\}

\benchmarkentry{C100}{cells.vti, cells\_channel\_sample.csv, cells\_seg\_stats.csv}{4}
{Display layered volumes of Sec61-beta with color = pink, Membrane with color= orange, and DNA with color = purple from cells.vti, generate a scatterplot of Sec61b versus Membrane sixed by DNA which is linked to a histogram of DNA using cells\_channel\_sample.csv, and conclude with a pie chart of seg\_combo proportions sized by pct from cells\_seg\_stats.csv.}
{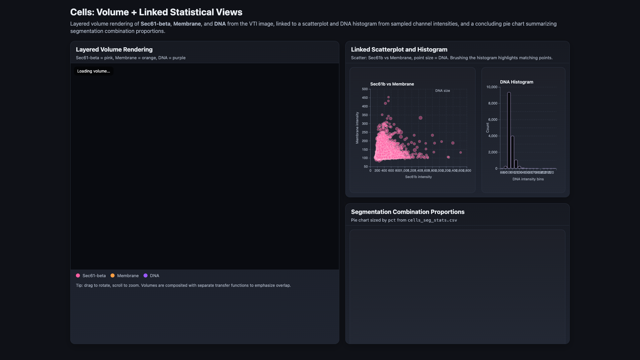}
{ & Raiven  & 0.98 & 1.0 & 1.0 & 0.9 & 1.00 & 0.35 \\[4pt]
 & ChatGPT & 0.72 & 0.6 & 0.8 & 0.8 & 0.85 & 1.46 \\[4pt]
 & Claude  & 0.85 & 0.9 & 0.8 & 0.8 & 0.85 & 0.96 \\[4pt]
 & Gemini  & 0.95 & 1.0 & 0.9 & 0.9 & 1.00 & 1.25 \\}

\twocolumn

\clearpage
\newpage
\section{User Study}
\label{app:userstudy}

\subsection{Setup}
All sessions were conducted between March 10--16, 2026 on a standardized workstation (MacBook Pro 14-inch, November 2024, Apple M4, 24\,GB RAM). The workstation provided unlimited access to the following tools and services: ChatGPT (Pro subscription, \$200/mo), Claude and Claude Code (Max Plan 5x, \$100/mo), Gemini (Google AI Ultra, \$125/mo), Cursor (Pro+, \$60/mo), Codex, Antigravity, VS Code, and Google Colab. All AI services were on paid plans ensuring access to the highest-tier models available on each platform. Raiven was served locally on the same machine.

\subsection{Task Descriptions}
\label{app:tasks}

As referenced by Section~\ref{sec:study-design}.
Participants were given printed task sheets describing the target 
dashboard they were asked to recreate. Each sheet specified the 
required views, the data files available in the task folder on 
their computer, the visual encodings expected for each chart, and 
any interactions or cross-view linking required. Participants could 
refer to these sheets throughout the study.

\subsubsection{Task 1: InfoVis Dashboard}

\begin{figure*}[t]
    \centering
    \includegraphics[width=\textwidth]{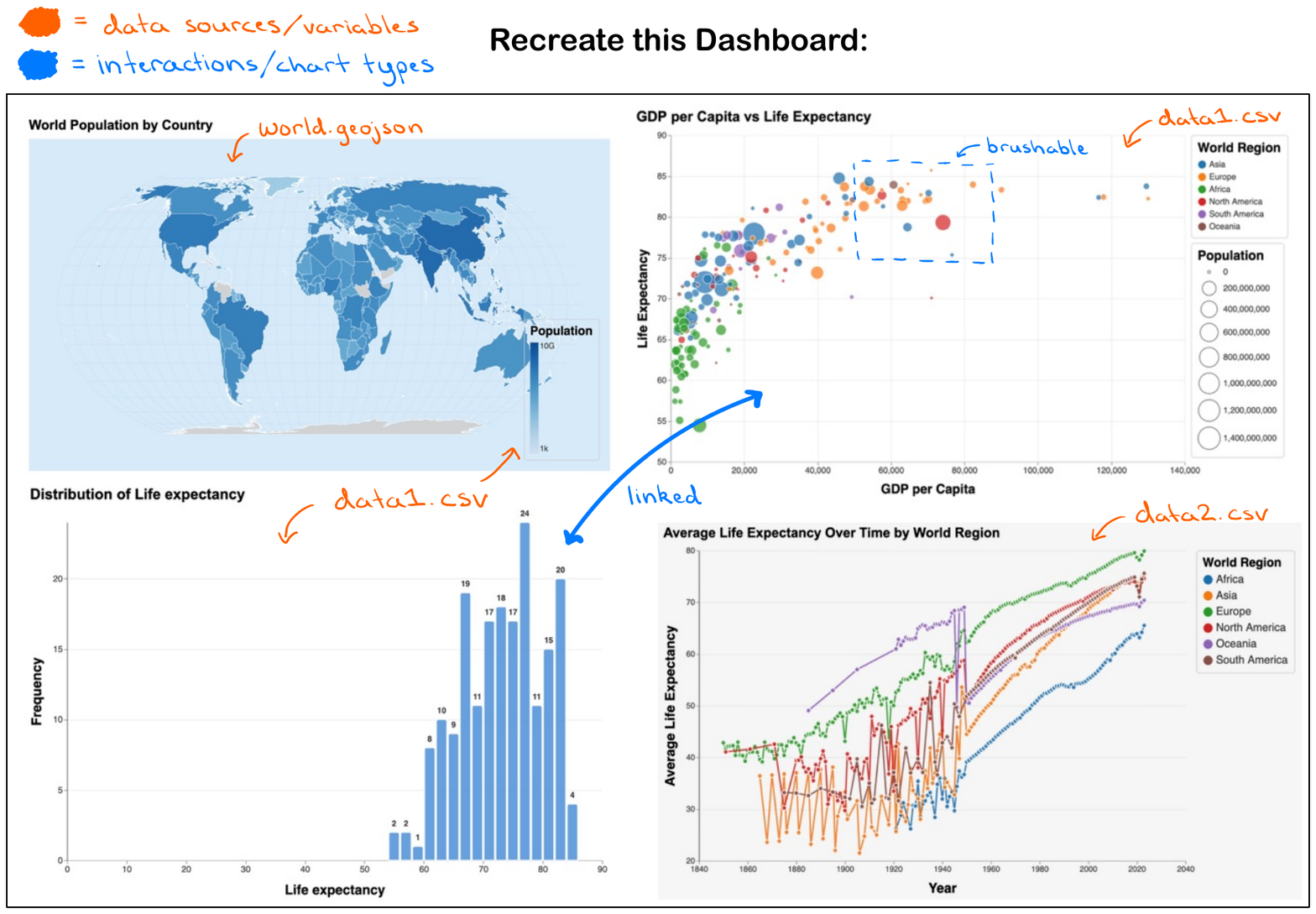}
    \caption{Printed task sheet given to participants for Task 1 
    (InfoVis). The sheet specifies the target dashboard layout, 
    required views, data files, encodings, and interactions.}
    \label{fig:task1-sheet}
\end{figure*}

Task 1 required participants to build a four-view interactive 
dashboard using world development data. The dashboard comprised: 
(1) a choropleth world map encoding population by country, (2) a 
bubble plot of GDP per capita versus life expectancy, with bubbles 
grouped by world region and sized by population, (3) a histogram 
of life expectancy, linked to the bubble plot, and (4) a multi-line 
chart with year on the x-axis and average life expectancy on the 
y-axis, with lines colored by world region. The printed task sheet, 
including the target dashboard layout, data sources, and interaction 
requirements, is reproduced in Figure~\ref{fig:task1-sheet}.

\subsubsection{Task 2: SciVis Dashboard}

\begin{figure*}[t]
    \centering
    \includegraphics[width=\textwidth]{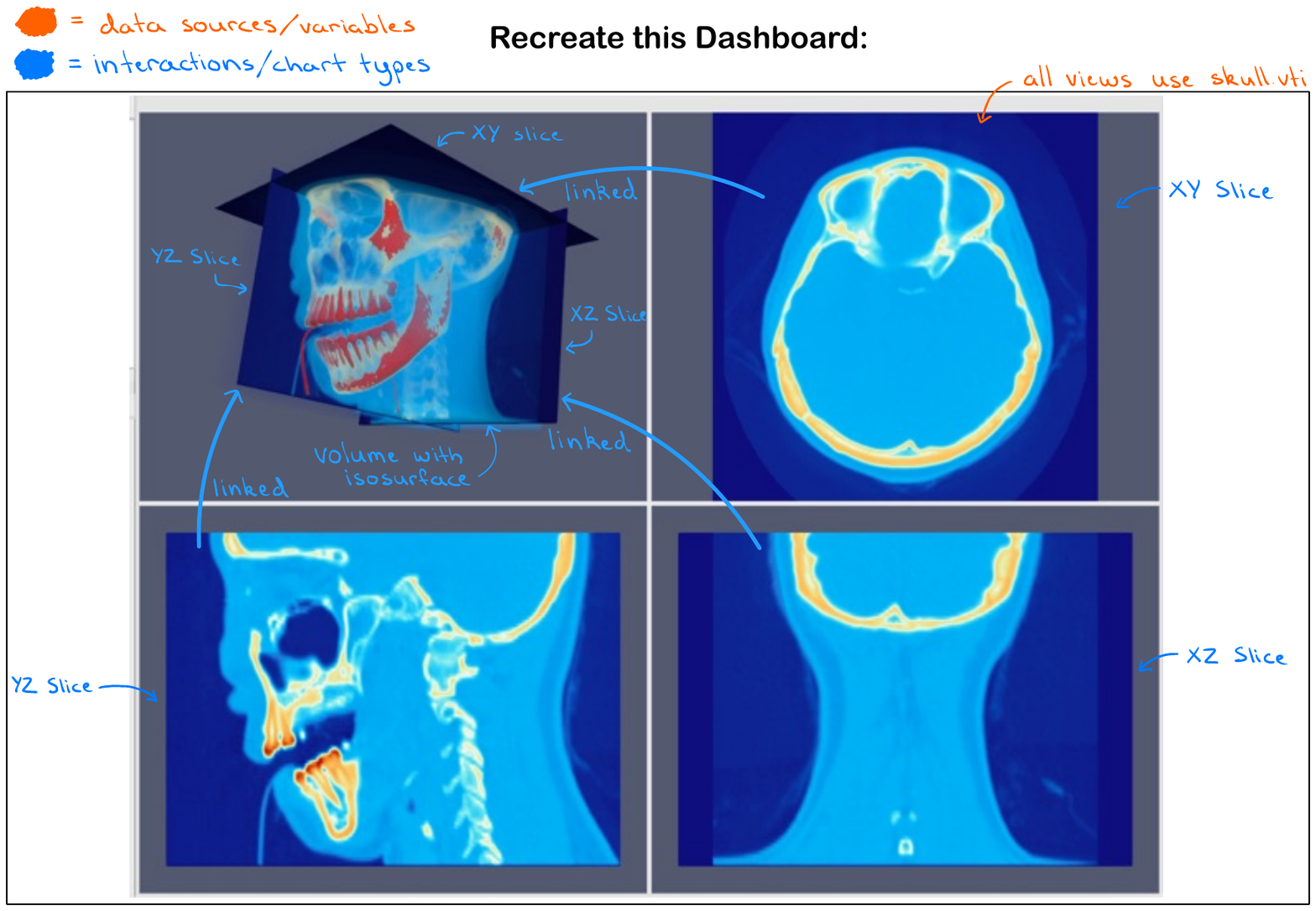}
    \caption{Printed task sheet given to participants for Task 2 
    (SciVis). The sheet specifies the target dashboard layout, 
    required views, data files, encodings, and slice-linking 
    interactions.}
    \label{fig:task2-sheet}
\end{figure*}

Task 2 required participants to build a four-view scientific 
visualization dashboard using volumetric CT skull data. The 
dashboard comprised: (1) a primary 3D view showing a volume 
rendering with a layered isosurface of the skull and triplanar 
slice overlays, (2) an XY slice view, (3) an XZ slice view, and 
(4) a YZ slice view. The three planar slice views were required 
to be linked to the primary 3D view such that scrolling through 
slices in any planar view updates the corresponding slice position 
in the 3D volume. The printed task sheet, including the target 
dashboard layout, data sources, and interaction requirements, is 
reproduced in Figure~\ref{fig:task2-sheet}.

\subsubsection{Task 3: Combined InfoVis/SciVis Dashboard}

\begin{figure*}[t]
    \centering
    \includegraphics[width=\textwidth]{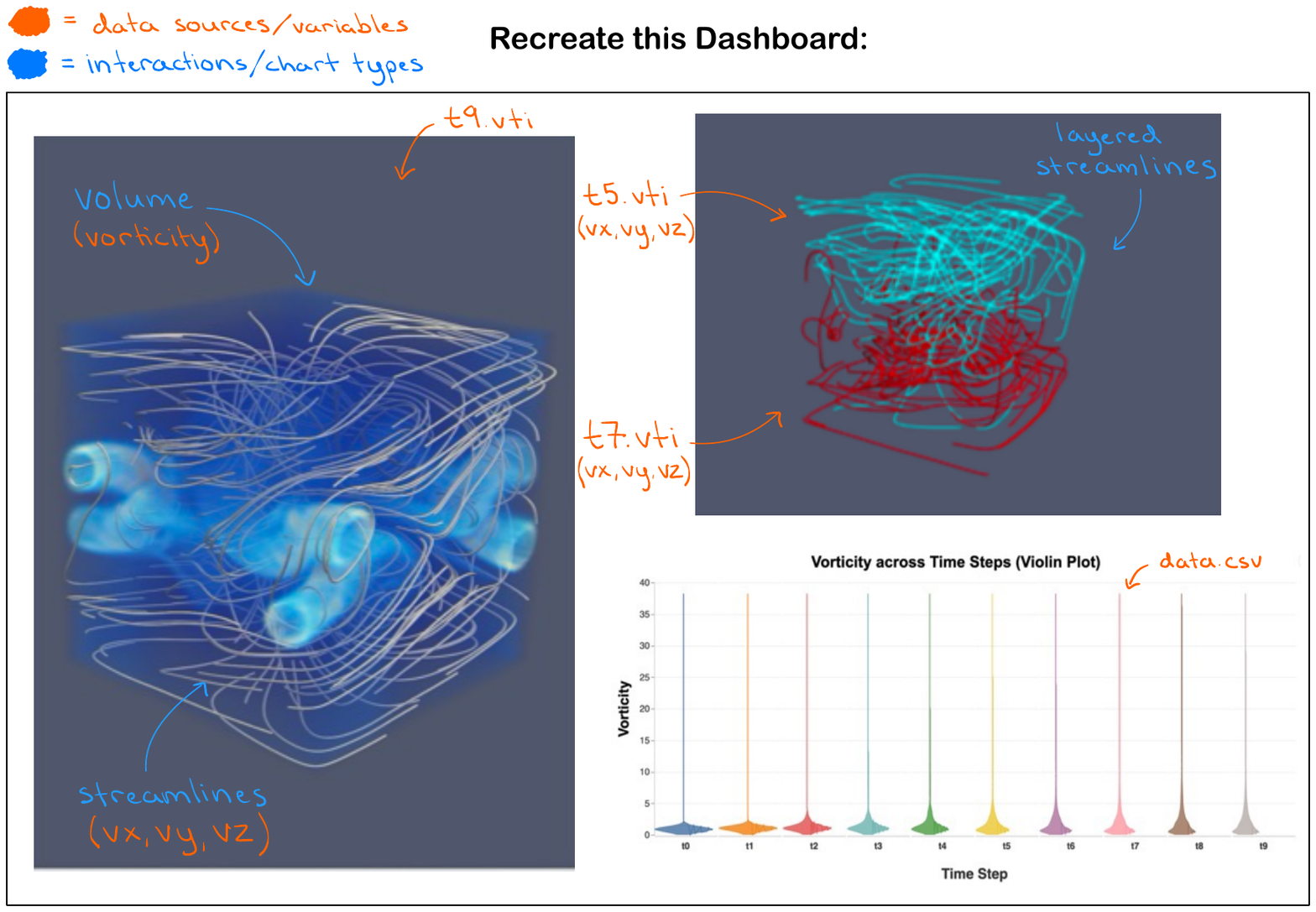}
    \caption{Printed task sheet given to participants for Task 3 
    (Combined). The sheet specifies the target dashboard layout, 
    required views, data files, and velocity component encodings.}
    \label{fig:task3-sheet}
\end{figure*}

Task 3 required participants to build a three-view dashboard 
combining scientific and information visualization using 
computational fluid dynamics data. The dashboard comprised: 
(1) a primary 3D view showing a volume rendering of vorticity 
with layered streamlines computed from the velocity components 
\texttt{vx}, \texttt{vy}, and \texttt{vz}, (2) a second 3D view 
showing layered streamlines from two different timesteps 
overlaid in the same scene, again using \texttt{vx}, \texttt{vy}, 
and \texttt{vz} for both, and (3) a violin plot with time on the 
x-axis and vorticity on the y-axis, summarizing the distribution 
of vorticity values across timesteps. The printed task sheet, 
including the target dashboard layout and data sources, is 
reproduced in Figure~\ref{fig:task3-sheet}.

\subsection{Baseline Tool Selection}
\label{app:baseline-tools}

All seven participants reported writing code as their primary method of creating visualizations; only two regularly incorporated AI tools into their visualization workflow, and one reported no AI use at all. Among AI systems used generally, Claude and ChatGPT were most prevalent (5 and 4 participants respectively), followed by Gemini (3) and Cursor (2). This prior familiarity with Claude largely predicted baseline tool choices during the study: six of seven participants used Claude in some form, with approaches ranging from the web interface alone, to Claude paired with a Colab notebook, to Claude Code, to combinations of Claude and Cursor. The sole exception used Cursor exclusively with auto model selection. One participant switched tools between every task (Gemini in Colab, then Antigravity with Opus, then Claude web), driven by dissatisfaction with the results of each. All tools used represented the highest-tier models available at the time of the study.

Despite access to ChatGPT, Codex, VS Code, and other tools on the provided workstation, participants almost universally defaulted to familiar tools, reflecting a strong familiarity bias even among experts.

\subsection{Pre-Study Survey}
\label{app:quant-feedback}

Seven participants completed a pre-study survey prior to the study 
covering demographics, visualization expertise, library familiarity, 
and workflow habits. All seven reported regularly using data 
visualization in their work. Tables~\ref{tab:survey-demo}, 
\ref{tab:survey-libs}, and~\ref{tab:survey-workflow} summarize 
the responses.

\begin{table}[h]
\centering
\caption{Participant demographics and self-reported visualization 
expertise (1--5 scale).}
\label{tab:survey-demo}
\small
\begin{tabular}{lcccccc}
\toprule
\textbf{P} & \textbf{Age} & \textbf{Gender} & \textbf{Position} 
& \textbf{Overall} & \textbf{InfoVis} & \textbf{SciVis} \\
\midrule
P1 & 27 & M & PhD      & 4 & 4 & 3 \\
P2 & 34 & M & Post-Doc & 5 & 4 & 5 \\
P3 & 22 & M & PhD      & 4 & 4 & 4 \\
P4 & 28 & M & Post-Doc & 4 & 4 & 3 \\
P5 & 32 & M & Post-Doc & 5 & 5 & 5 \\
P6 & 30 & F & Post-Doc & 4 & 5 & 4 \\
P7 & 29 & F & PhD      & 5 & 5 & 4 \\
\bottomrule
\end{tabular}
\end{table}

\begin{table}[h]
\centering
\caption{Self-reported familiarity with visualization libraries 
(1~=~Unfamiliar, 2~=~Vaguely Familiar, 3~=~Familiar, 
4~=~Proficient, 5~=~Very Proficient). Top group: InfoVis-oriented 
libraries. Bottom group: SciVis/WebGL-oriented libraries.}
\label{tab:survey-libs}
\small
\begin{tabular}{lccccccc}
\toprule
\textbf{P} & \textbf{D3} & \textbf{Vega} & \textbf{Mpl} 
& \textbf{Sns} & \textbf{ggplot2} & \textbf{Altair} 
& \textbf{MATLAB} \\
\midrule
P1 & 4 & 5 & 4 & 4 & 4 & 5 & 2 \\
P2 & 3 & 4 & 3 & 2 & 2 & 3 & 2 \\
P3 & 3 & 2 & 5 & 4 & 2 & 2 & 2 \\
P4 & 4 & 4 & 4 & 3 & 3 & 4 & 2 \\
P5 & 5 & 4 & 4 & 4 & 4 & 4 & 3 \\
P6 & 4 & 4 & 4 & 4 & 4 & 4 & 2 \\
P7 & 4 & 4 & 4 & 4 & 5 & 4 & 2 \\
\midrule
\textbf{P} & \textbf{VTK} & \textbf{ParaView} & \textbf{Neuro.} 
& \textbf{Three.js} & \textbf{Deck.gl} & \textbf{PyVista} & \\
\midrule
P1 & 1 & 1 & 1 & 3 & 1 & 1 & \\
P2 & 2 & 1 & 1 & 5 & 2 & 1 & \\
P3 & 1 & 1 & 1 & 2 & 1 & 1 & \\
P4 & 2 & 2 & 3 & 2 & 5 & 1 & \\
P5 & 5 & 5 & 5 & 5 & 4 & 4 & \\
P6 & 1 & 1 & 1 & 1 & 1 & 1 & \\
P7 & 1 & 1 & 1 & 2 & 1 & 1 & \\
\bottomrule
\end{tabular}
\end{table}

\begin{table}[h]
\centering
\caption{Participant workflow habits. All participants primarily 
create visualizations by writing code; P1 and P5 also reported 
regularly using AI tools as part of their workflow.}
\label{tab:survey-workflow}
\small
\begin{tabular}{lll}
\toprule
\textbf{P} & \textbf{AI Systems Used} & \textbf{Other Libraries} \\
\midrule
P1 & ChatGPT, Claude, Gemini, Cursor & --- \\
P2 & ChatGPT, Claude                 & Observable Plot \\
P3 & Claude                          & --- \\
P4 & Claude, Gemini, Cursor          & --- \\
P5 & ChatGPT, Claude, Gemini         & --- \\
P6 & ChatGPT, Claude                 & plot.js, gosling.js \\
P7 & None                            & p5.js, Illustrator, ggplot2 ext. \\
\bottomrule
\end{tabular}
\end{table}

\end{document}